\documentclass[review,onefignum,onetabnum]{siamart190516}

\usepackage{comment}
\usepackage{caption}
\usepackage{float}
\usepackage{placeins}
\usepackage{subcaption}

\usepackage{lipsum}
\usepackage{amsfonts}
\usepackage{graphicx}
\usepackage{epstopdf}
\usepackage{algorithmic}
\ifpdf
  \DeclareGraphicsExtensions{.eps,.pdf,.png,.jpg}
\else
  \DeclareGraphicsExtensions{.eps}
\fi
\usepackage{soul}

\newsiamremark{remark}{Remark}
\newsiamremark{hypothesis}{Hypothesis}
\crefname{hypothesis}{Hypothesis}{Hypotheses}
\newsiamthm{claim}{Claim}

\headers{PeleLM-FDF Solver for LES of Turbulent Reacting Flows}{A. Aitzhan, S. Sammak, P. Givi and A. G. Nouri}

\title{PeleLM-FDF Large Eddy Simulator of Turbulent Combustion\thanks{Submitted to the editors December 27, 2021.
\funding{This work was funded by the NSF under Grant CBET-2042918.}}}

\author{Aidyn Aitzhan\thanks{Department of 
Mechanical Engineering and Materials Science, University of Pittsburgh, Pittsburgh, PA 
  (\email{aia29@pitt.edu}).}
\and Shervin Sammak \thanks{Department of 
Mechanical Engineering and Materials Science, University of Pittsburgh, Pittsburgh, PA; Center for Research Computing University of Pittsburgh, Pittsburgh, PA
  (\email{ shervin.sammak@pitt.edu}).}
\and Peyman Givi \thanks{Department of 
Mechanical Engineering and Materials Science, University of Pittsburgh, Pittsburgh, PA
  (\email{ pgivi@pitt.edu}).}
\and Arash G. Nouri \thanks{Department of 
Mechanical Engineering and Materials Science, University of Pittsburgh, Pittsburgh, PA
  (\email{ arash.nouri@pitt.edu}).}
}

\usepackage{amsopn}

\newcommand{\be}{\begin{equation}}
\newcommand{\ee}{\end{equation}}
\newcommand{\prt}{\partial}
\newcommand{\non}{\nonumber}
\newcommand{\ov}{\overline}

\newcommand{\la}{\left<}
\newcommand{\ra}{\right>}
\newcommand{\lp}{\left(}
\newcommand{\rp}{\right)}
\newcommand{\bm}{\mathbf}

\newcommand{\bx}{\mbox{${\bf x}$}}

\newcommand{\cond}{\mbox{$\Big{\arrowvert}$}}

\newcommand{\rhob}{\mbox{$\la \rho \ra_{\ell}$}}
\newcommand{\rrb}{\mbox{$\la S_\alpha \ra_L$}}

\newcommand{\Pb}{\mbox{$\la p \ra_\ell$}}

\newcommand{\uib}{\mbox{$\la u_i \ra_L$}}
\newcommand{\ujb}{\mbox{$\la u_j \ra_L$}}

\newcommand{\fab}{\mbox{$\la \phi_\alpha \ra_L$}}

\newcommand{\taub}{\mbox{${\tau}$}}

\newcommand{\bphi}{\mbox{\boldmath $\phi$}}
\newcommand{\bpsi}{\mbox{\boldmath $\psi$}}

\def\la{\langle}
\def\ra{\rangle}
\def\ov{\overline}

\def\exa{\expandafter}
\ifpdf
\hypersetup{
  pdftitle={title},
  pdfauthor={author}
}
\fi

\begin{document}
\nolinenumbers

\maketitle

\begin{abstract}

A new computational methodology, termed ``PeleLM-FDF'' is developed and utilized  for high fidelity large eddy simulation (LES) of complex turbulent combustion systems. This methodology is constructed via a hybrid scheme combining the Eulerian PeleLM base flow solver with the Lagrangian Monte Carlo simulator of the filtered density function (FDF) for the subgrid scale reactive scalars.  The resulting  methodology is  capable of simulating  some of the most intricate physics of complex turbulence-combustion interactions. This is demonstrated by LES of a non-premixed CO/H$_2$ temporally evolving jet flame.  The chemistry is modelled via a skeletal kinetics model, and the results are appraised  via detail {\it a posteriori} comparisons against direct numerical simulation (DNS) data of the same flame.  Excellent agreements are observed for the time evolution of various statistics of the  thermo-chemical quantities, including the manifolds of the multi-scalar mixing.  The new methodology is capable of capturing the complex  phenomena of flame-extinction and re-ignition at a 1/512 of the computational cost of the DNS.  The high fidelity and the computational affordability of the new PeleLM-FDF  solver warrants its consideration for LES of practical turbulent combustion systems.
  
\end{abstract}



\section{Introduction}
Since its original proof of concept \cite{CJGP98,JCJGP99}, the filtered density function (FDF) has become very popular  for   large eddy simulation (LES) of turbulent flows. This popularity  is due to inherent capability of the FDF to account full statistics of the subgrid scale (SGS) quantities; and thus it is more accurate than conventional SGS models which are  based on  low order SGS moments.  This superior performance comes at a price. The FDF transport equation is somewhat more difficult and computationally more expensive to solve, as compared to traditional LES schemes. The last decade has witnessed significant progress to improve FDF simulations, as evidenced by a rather large number of publications; \textit{e.g.}\ Ref.\ \cite{almeida2018les,zhang2018ecples,Damasceno2018Simulation,Castro2021Implementation,mejia2016scalar,MCP14,SBGM16,sewerin2018lespbepdf,Zhou2019AIAA,ASGG15,APSG12,AGSG11,YNSSG11,PYSG13,NYGSP10,NYSG10,SAMG20,inkarbekov2020gpu,komperda2020hybrid,natarajan2021high,SNBMG17,NSPG17,NNGLP17,NGL19}. Parallel with these developments, there have also been extensive studies regarding the FDF  accuracy \& reliability \cite{TP16a,klimenko2010convergence,NYSG10,chibbaro2014particle,SBGM16},  and sensitivity analysis of its simulated results \cite{zhou2017investigation,zhao2019transported}. For comprehensive reviews of progress within the last decade, see Refs.\ \cite{SRG20,Yang2021}. 

Despite the remarkable  progress as noted, there is still a continuing demand for further improvements of  LES-FDF for prediction of complex turbulent combustion systems.  In particular, it is desirable to develop FDF tools which are of high fidelity, and  are also computationally  affordable. In this work, the   PeleLM \cite{Nonaka2018Conservative} base flow solver is combined with the parallel Monte Carlo FDF simulator \cite{APSG12,AJSG11} in a hybrid manner that takes full advantage of modern developments in both   strategies. PeleLM is a massively parallel simulator of reactive flows at low Mach numbers. These flows are of significant interest in  several industries such as gas turbines, IC engines, furnaces and many others.  The  solver is based on block-structured AMR  algorithm \cite{bell2005amr}  through the AMReX numerical software library \cite{AMReX_JOSS} (formerly called BoxLib \cite{Zhang2016BoxLib}). This solver uses a variable density projection method \cite{Almgren1998Conservative, Pember1998Adaptive, bell2002parallel} for solving three-dimensional Navier-Stokes and reaction-diffusion equations. The computational  discretization is based on structured finite volume for spatial discretization, and a modified spectral deferred correction (SDC) algorithm \cite{Nonaka2012Deferred, Pazner2016HighOrder, Hamon2019Concurrent, Esclapez2020Spectral} for temporal integration. The solver is capable of dealing with  complex geometries via the embedded boundary method \cite{Pember1995AnEB, Dragojlovic2006embedded}, and runs  on modern platforms for parallel computing such as MPI + OpenMP for CPUs and MPI + CUDA or MPI + HIP for GPUs. The fidelity of PeleLM has been demonstrated to be effective for DNS of a variety  of reactive turbulent flows \cite{Aspden2016ThreeD, Dasgupta2017Effect, Aspden2019Towards, Dasgupta2019Analysis, Wimer2021Numerical}.  Here, the PeleLM is augmented  to include LES capabilities by hybridizing it with the FDF-SGS closure. The resulting solver is shown to be computationally efficient, and to produce results consistent with those generated by high-fidelity, and much more expensive  DNS.

\section{Formulation}
We consider a variable density turbulent reacting flow involving $N_s$ species in which the flow velocity is much less than the speed of sound.  In this flow, the primary
transport variables are the fluid density $\rho(\bm{x}, t) $, the velocity vector $u_i(\bm{x}, t),\
i=1,2,3$ along the $x_i$ direction, the total specific enthalpy $h (\bm{x}, t)$,
the pressure $p(\bm{x}, t)$, and the species mass fractions $Y_{\alpha}(\bm{x}, t) \
(\alpha=1,2,\dots,N_s)$. The conservation equations governing these
variables are the continuity, momentum, enthalpy (energy) and species
mass fraction equations, along with an equation of state
\cite{Williams85}:
\begin{equation}
{\prt \rho \over \prt t}+ {\prt \rho u_i \over \prt
x_i} = 0,
\label{1}
\end{equation}
\begin{equation}
{\prt \rho u_j \over \prt t}+ {\prt \rho u_i u_j \over
\prt x_i} = - {\prt p \over \prt x_j} + {\prt \tau_{ij}
\over \prt x_i},
\label{2}
\end{equation}
\begin{equation}
{\prt \rho \phi_\alpha \over \prt t}+ {\prt \rho u_i
\phi_\alpha \over \prt x_i} = -{\prt J_i^\alpha \over \prt
x_i} + \rho S_{\alpha},
\ \ \  \alpha=1,2,\dots,\sigma=N_s +1,
\label{3}
\end{equation}
\begin{equation}
p=\rho R^0 T \sum_{\alpha=1}^{N_s} Y_{\alpha} / {\cal M}_{\alpha},
\label{4}
\end{equation}
where $t$ represents time, $R^0$ is the universal gas constant and
${\cal M}_{\alpha}$ denotes the molecular weight of species $\alpha$.
  Equation \cref{3} represents  transport of the species' mass
fractions and enthalpy in a common form with:
\begin{equation}
\phi_{\alpha} \equiv Y_{\alpha}, \ \alpha=1,2,\dots,N_s,\ \
\phi_{\sigma} \equiv h =
\sum_{\alpha=1}^{N_s}h_{\alpha} \phi_{\alpha}.
\label{4b}
\end{equation}
With the low Mach number approximation, the  chemical source terms $S_{\alpha}$ are functions of the composition variables $\bphi=[Y_1,Y_2,\dots,Y_{N_s}, h]$ only; \textit{i.e.} $S_{\alpha}= S_{\alpha}(\bphi)$.  For a Newtonian
fluid with zero bulk viscosity and Fickian diffusion, the viscous
stress tensor $\tau_{ij}$ and the mass and the heat fluxes $J^{\alpha}_i \ 
(\alpha=1,2,\dots,\sigma$) are given by:
\begin{equation}
\tau_{ij}=\mu \left( {\prt u_i \over \prt x_j} + {\prt u_j
\over \prt x_i} - {2 \over 3} {\prt u_k \over \prt x_k}
\delta_{ij} \right),\ \ \ 
J_i^\alpha=- \gamma {\prt \phi_\alpha \over \prt x_i},
\label{4e}
\end{equation}
where $\mu$ is the dynamic viscosity and $\gamma$ denotes
the thermal and the mass molecular diffusivity coefficients.  Both
$\mu$ and $\gamma$ are assumed temperature dependent, and the molecular  Lewis number is
assumed to be unity.  

Large eddy simulation involves the use of the spatial filtering
operation \cite{Aldama90}:
\begin{equation}
\la Q(\bx,t) \ra_{\ell}=\int_{-\infty}^{+\infty}
Q(\bx^\prime,t) {\cal G} (\bx^\prime, \bx)d \bx^\prime,
\label{5}
\end{equation}
where ${\cal G}$ denotes the filter function of width $\Delta_G$, $\la
Q(\bx,t) \ra_{\ell}$ represents the filtered value of the transport
variable $Q(\bx,t)$.  In variable density
flows it is convenient to consider the Favre filtered quantity
$ {\la Q(\bx,t) \ra_L =}  {\la \rho Q \ra_{\ell}} / {\rhob }
$.   The application of the filtering operation to the transport equations
yields:
\begin{equation}
{\prt \rhob \over \prt t} + {\prt
\rhob  \uib \over \prt x_i}=0,
\label{6}
\end{equation}
\begin{equation}
{\prt \rhob \ujb \over \prt t} + {\prt \rhob \uib \ujb
\over \prt x_i} = - { \prt \Pb \over \prt x_j } + {
{\prt \la \tau_{ij} \ra_{\ell} } \over \prt x_i } - {\prt
T_{ij} \over \prt x_i},
\label{7}
\end{equation}
\begin{equation}
{\prt \rhob \fab \over \prt t} + {\prt \rhob \uib \fab
\over \prt x_i} = -{ {\prt \la J_i^\alpha \ra_{\ell} } \over
\prt x_i} -{\prt M^\alpha_i \over \prt x_i} + \la \rho
S_{\alpha} \ra_{\ell}, \ \ \ \alpha=1,2,\dots,\sigma,
\label{8}
\end{equation}
where $ T_{ij} = \rhob ( \la u_i u_j \ra_L - \uib \ujb)$, and $ \
M^\alpha_i = \rhob ( \la u_i \phi_\alpha \ra_L - \uib \fab) $ denote
the subgrid stress and the subgrid mass fluxes, respectively. The
filtered reaction source terms are denoted by $\la \rho S_{\alpha}
\ra_{\ell}= \rhob \rrb \ (\alpha=1,2,\dots,N_s)$.

\section{Filtered Density Function} 
\label{SECTION:FDF}
The complete SGS statistical information pertaining to the scalar field, is contained within the FDF, defined as \cite{Pope90}: 

\begin{equation}
\label{EQ:VSFMDF_FDFdef} F_L\lp\bpsi; \bx,t\rp =
\int^{+\infty}_{-\infty} \rho(\bx',t) \zeta \lp
\bpsi, \bphi(\bx',t) \rp G(\bx', \bx)d\bx',
\end{equation}
where
\begin{equation}
\label{EQ:VSFMDF_FDFdef2} \zeta\lp
\bpsi, \bphi(\bx,t) \rp = \prod^{\sigma}_{\alpha=1} \delta\lp
\psi_{\alpha}-\phi_{\alpha}(\bx,t)\rp.
\end{equation}
In this equation, $\delta$ denotes the Dirac delta function and
$\bpsi$ is the scalar array in the sample space. The term $\zeta $ is the ``fine-grained'' density
\cite{OBrien80}. 
With the condition of a positive filter kernel \cite{VGK94}, $F_L$ has
all of the properties of a mass density function \cite{Pope00}.
Defining the ``conditional filtered value'' of the variable $Q(\bx,t)$
as: \be \label{EQ:VSFMDF_QC} \la Q \cond\bpsi
\ra_\ell \equiv \frac{\int^{+\infty}_{-\infty} Q\lp \bx',t \rp
\rho(\bx',t)\zeta\lp \bpsi, \bphi(\bx',t) \rp G\lp
\bx',\bx \rp d\bx' }{F_L\lp \bpsi; \bx,t\rp}, \ee
the FDF is governed by \cite{AJSG11}:
\begin{align}
\label{19} \frac{\prt F_L}{\prt t} +
\frac{\prt [\la u_i (\bx,t) \vert \bpsi \ra_\ell F_L]}
{\prt x_i} &= \frac{\prt}{\prt \psi_\alpha}\left[\left<\frac{1}{\rho
(\bphi)} \left. \frac{\prt J_i^\alpha}{\prt x_i} \right| \bpsi \right>_\ell F_L \right] \non \\
&- \frac{\prt}{\prt \psi_\alpha}[S_\alpha
(\bpsi) F_L].
\end{align}
This is the exact transport equation for the FDF, in which the effects of chemical reaction (the last term on the right hand side) appear in a closed form.  The unclosed terms due to convection and molecular mixing are identified by the conditional averages (identified by a vertical bar). The gradient diffusion model, and the 
linear mean square estimation (LMSE) approximations are employed for closure of these terms. With these assumptions, the modelled transport equation for the FDF becomes \cite{Givi06}:

\begin{align}
\frac{\prt F_L}{\prt t} + \frac{\prt [ \uib F_L]}{\prt
x_i} &= \frac{\prt}{\prt x_i} \left[(\gamma + \gamma_t)
\frac{\prt (F_L / \rhob)}
{\prt x_i}\right] \non \\ 
&+ \frac{\prt}{\prt
\psi_{\alpha} }\left[ \Omega (\psi_{\alpha}  -
\la \phi_{\alpha} \ra_{L}) F_L \right] - \frac{\prt}{\prt
\psi_{\alpha} }\left[ S_{\alpha}\left( \bpsi \right) F_L \right],
\label{EQ:29}
\end{align}

where $\gamma_t$ is the turbulent viscosity model \cite{Vreman2004Eddy}, $\Omega =C_{\phi}\lp \gamma +\gamma_{t}\rp / \lp \la \rho \ra_{\ell}\Delta_G^{2} \rp$ is the SGS mixing frequency \cite{Pope00,OBrien80}, and $C_\phi$ is a model constant.

\section{Hybrid PeleLM-FDF Solver}
Equation \cref{EQ:29} may be integrated to obtain the modeled transport equations for the SGS moments, \textit{e.g.}  the filtered mean, $\la \phi_{k} \ra_{L}$ and the SGS variance $\taub_{k} \equiv \la \phi_{k}^2  \ra_L- \la \phi_{k} \ra^2_L$. A convenient means of solving this equation is via the Lagrangian Monte Carlo (MC) procedure \cite{Yang2021}.  In this procedure, each of the MC  elements (particles) undergo motion in physical space by convection due to the filtered mean flow velocity and diffusion due to
molecular and subgrid diffusivities. These are determined by viewing  Eq.\ \cref{EQ:29} as a Fokker-Planck equation, for which the corresponding Langevin equations describing  transport of the MC particles are \cite{Risken89,Gardiner90}: 
\begin{equation}
dX_i(t)=\left[\uib + \frac{1}{\rhob} \frac{\prt (\gamma+\gamma_t)}
{\prt x_i} \right] dt+\sqrt{ 2(\gamma+\gamma_t) / \rhob }\ dW_i(t),
\label{EQ:MC1}
\end{equation}
with the change in the compositional makeup according to: 
\be
\frac{d \phi_{k}^{+}}{dt} = -\Omega \left( \phi_{k}^{+} - \la \phi_{k}\ra_{L} \right) + S_{k}\left( \bphi^{+} \right) \; \left(k = 1,2, \dots N_s+1 \right).
\label{EQ:MC2}
\ee
In these equations,  ${W}_i$ denotes the Wiener-Levy process, $\phi_{k}^{+} = \phi_{k}\left(\mathbf X, t\right)$ is the scalar value of the particle with the Lagrangian position $X_i$. 

\begin{figure}
    \centering
    \begin{subfigure}[b]{0.45\textwidth}
    \centering
    \includegraphics[width=1\textwidth]{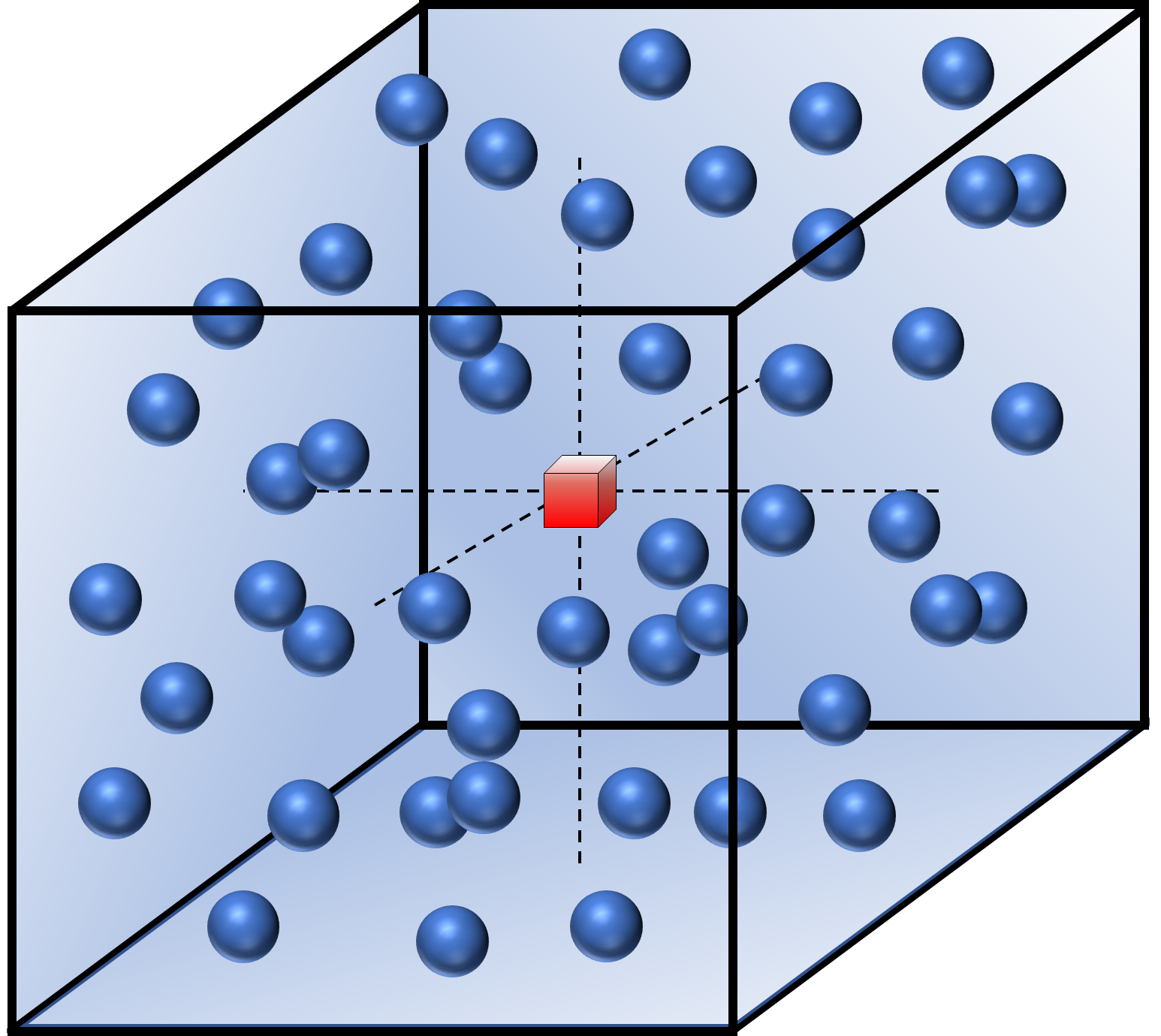}
    \caption{}
    \label{fig:ens_domain}
   \end{subfigure}
     \hfill
     \begin{subfigure}[b]{0.45\textwidth}
         \centering
    \includegraphics[width=1\textwidth]{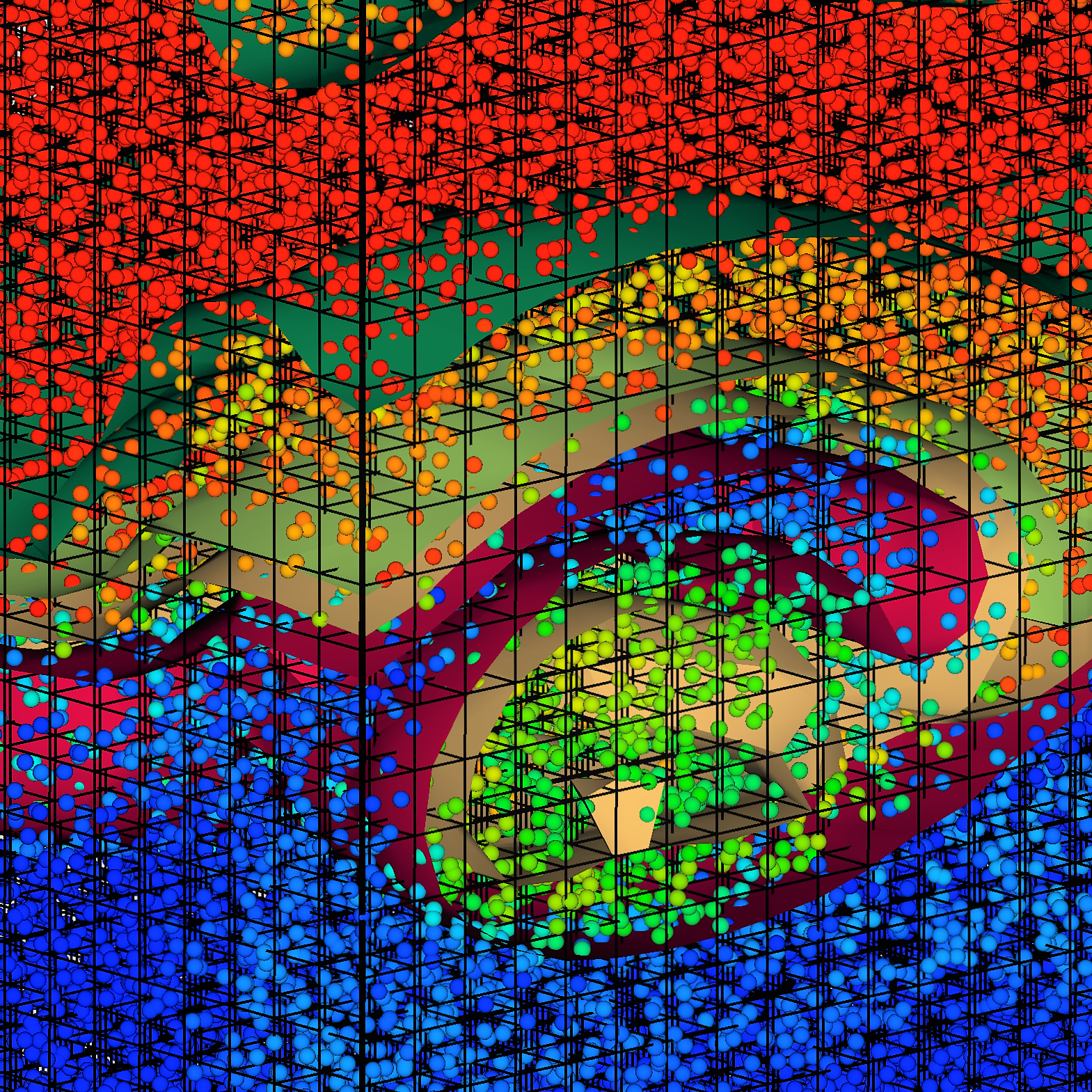}
    \caption{}
    \label{fig:ParticeGrid}
   \end{subfigure}
        \caption{(a) Ensemble averaging in MC simulations.
The red cube denotes the finite volume cell center, and the blue spheres denote the MC
particles. (b) Example of MC particles within the Eulerian field identified by PeleLM.  The colors of the MC particles provide a measure of the particle's scalar values.}
        \label{fig:doubleFigs}
\end{figure}

AMReX library is a very powerful computational software with many useful functions, templates and classes including linear solvers \cite{Williams2014Krylov} and particle containers \cite{Musser2021MFIX}. The latter one, is especially useful for our purpose.  The principal algorithm is based on a variable density projection method for low Mach number flows is  described in Ref.\ \cite{day2000numerical}. The domain is discretized by an ensemble of finite-volume cells and the particles are free move within the domain  (\cref{fig:doubleFigs}).  The MC  procedure is implemented  by deriving a new class from the particle container of the AMReX library, adding all the required functions. The  particle transport as given by the  SDEs \cref{EQ:MC1} is tracked via  Euler-Maruyama method \cite{KPS97},
The compositional makeup (Eq.\ \cref{EQ:MC2}) is implemented with variety of methods involving third-party solvers like VODE \cite{Brown1989VODE}, CVODE \cite{SUNDIALS}, and our in-house adaptive Runge-Kutta solver.  

With the hybrid scheme as developed, some of the quantities are obtained by MC-FDF, some by the base flow solver (PeleLM) and some by {\it both}.  So, there is a ``redundancy'' in determination of some of the quantities. In general, all of the equations for the filtered quantities can be
solved by PeleLM, in which all of the unclosed terms are evaluated by the MC-FDF solver. This process can be done at any filtered SGS moment level \cite{AGSG11}. With the hydrodynamic solver given by PeleLM, the scalar transport is implemented via both of these ways. In doing so, the filtered source terms are evaluated by the ensemble values over the MC particles:
\be
\la S_\alpha\left(\bx,t \right) \ra_L  \approx \frac{1}{N_E} \sum_{n \in \Delta_E} S_\alpha(\bphi^{(n)}),
\label{EQ:Ensavg}
\ee 
where $N_E$ is number of particles within the $\Delta_E$ neighborhood of point $\bx$. The choice of $\Delta_E$ is independent of the grid size $\Delta x$, and the LES filter size $\Delta_G$. It is desirable to set $\Delta_E$ as small as possible. The particle-grid interaction is schematically illustrated in \cref{fig:ens_domain}, while the example of an actual hybrid Eulerian-Lagrangian simulation is shown in \cref{fig:ParticeGrid}. The transfer of information from the grid points to the MC particles is accomplished via a linear interpolation.

\section{Flow Configuration and Model Specifications} 
\begin{figure}
    \centering
    \includegraphics[width=0.75\textwidth]{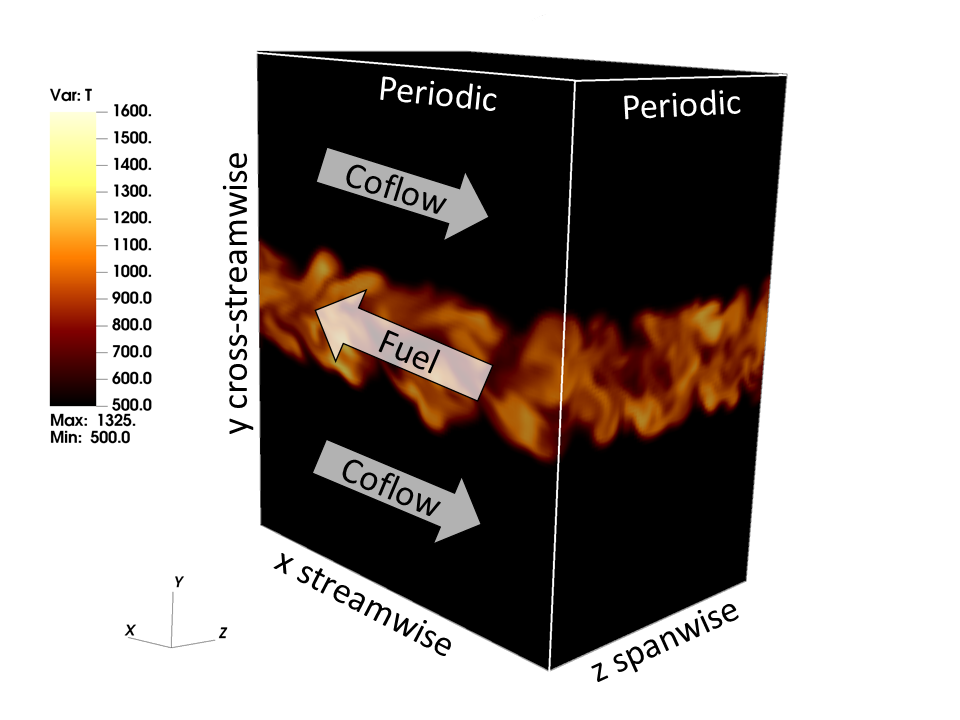}
    \caption{Schematics of the temporally developing turbulent jet flame. The jet consists of a central fuel stream surrounded by two counter-flowing oxidizer streams. The fuel stream is comprised of 50$\%$ of CO, 10$\%$ H$_2$ and 40$\%$ N$_2$ by volume, while oxidizer streams contain 75$\%$ N$_2$ and 25$\%$ O$_2$. The initial temperature of both streams is 500K and the thermodynamic pressure is set to 1 atm.}
    \label{fig:FLOW}
\end{figure}

The performance of the PeleLM-FDF solver is assessed by conducting LES of a temporally evolving planar turbulent CO/H$_2$ jet flame.  This flame has been the subject of detailed  DNS  \cite{Hawkes2007Scalar}, and several subsequent modeling and simulations \cite{YANG2013Large, Punati2011Evaluation, Vo2018MMC, Yang2017Sensitivity, Sen2010Large}. The flame is rich with strong flame–turbulence interactions resulting in local extinction followed by re-ignition. The flow configuration is the same as that considered in DNS and is depicted in \cref{fig:FLOW}.  The jet consists of a central fuel stream of width $H$  surrounded by counter-flowing oxidizer streams. The fuel stream is comprised of 50$\%$ of CO, 10$\%$ H\textsubscript{2} and 40$\%$ N\textsubscript{2} by volume, while oxidizer streams contain 75$\%$ N\textsubscript{2} and 25$\%$ O\textsubscript{2}. The initial temperature of both streams is 500K and thermodynamic pressure  is set to 1 atm. The  velocity difference between the two streams is $U = 276$m/s. The fuel stream velocity and the oxidiser stream velocity are $U/2$ and $-U/2$, respectively. The Reynolds number, based on $U$ and $H$ is $Re=9,079$. The sound speeds in the fuel and oxidizer streams denoted as $C_1$ and $C_2$, respectively and the Mach number $Ma = U / \left(C_1 + C_2 \right) = 0.3$ is small enough to justify a low Mach number approximation. The combustion chemistry is modelled via the skeletal kinetics, containing 11 species with 21 reaction steps \cite{Hawkes2007Scalar}. The initial conditions are taken directly from DNS. The boundary conditions are periodic in stream wise ($x$) and spanwise ($z$) directions, and the outflow boundary conditions imposed at $y = \pm L_y/2$. The models in FDF are the same as those in previous LES-FDF \cite{AJSG11}, with minor upgrades. The SGS stresses and mass fluxes are modeled by the standard Boussinesq approximation \cite{Smagorinsky63,garnier2009large}, and the gradient diffusion approximation, respectively. The SGS viscosity coefficient $\mu_t$ is calculated using the Vreman's model~\cite{Vreman2004Eddy}. The model parameters are: $C_{\phi} = 5$, $c = 0.05625$ and  $Sc_t = Pr_t = 0.7$.

The size of the computational domain is $L_x \times L_y \times L_z = 12H \times 14H \times 8H$. The time is normalized by $t_j = H/U$. The domain is discretized into equally spaced structured fixed grids of size $N_x \times N_y \times N_z = 108 \times 126 \times 72$. The resolution, as selected, is the largest that was conveniently available to us, and kept the SGS energy within the allowable $15\% \sim 20\%$ of the total energy. This resolution should be compared with $N_{x, DNS} \times N_{y, DNS} \times N_{z, DNS} = 864 \times 1008 \times 576$ grids as utilized in DNS  \cite{Hawkes2007Scalar}. The sizes of the ensemble domain, the subgrid filter and the finite-volume cell are taken to be equal $\Delta_E = \Delta_G = \Delta x = \Delta y = \Delta z = L_x / N_x$, and the timestep for temporal integration is $\Delta t = 10^{-7} s$. The number of MC  particles per grid point is set to 64; so there are over  62.7 million MC particles portraying the FDF at all time. With a factor of 512 times smaller number of grids, the total computational time for the simulations is around 400 CPU hours on 2 nodes of 28-core Intel Xeon E5-2690 2.60 GHz (Broadwell) totalling 56 processors.

The simulated results are analyzed both instantaneously and statistically.  In the former, the instantaneous contours (snap-shots) and the scatter plots of the reactive scalar fields are considered.
This pertains to the temperature and mass fractions of all of the species.  In the
latter, the ``Reynolds-averaged'' statistics are constructed.  With the assumption of a temporally developing layer, the flow is homogeneous in the $z-$ and the $x-$ directions.  Therefore, all of the  Reynolds averaged values, denoted by an overline, are temporally evolving and determined by ensemble averaging over the  $x-z$ planes.  The   resolved stresses are denoted by $R\lp a,b \rp=\ov{\la a \ra_{L} \la b \ra_{L}} - \lp \ov{\la a \ra_{L}} \rp \lp \ov{\la b \ra_{L}}\rp $, and the total stresses are denoted by  $ r\lp a,b \rp= \ov{\lp ab \rp} - \ov{a} \ov{b}$.  The latter can be evaluated directly  from the fine-grid DNS data $r_{\text{DNS}}\lp a,b \rp$.  In LES with the assumption of a {\sl generic} filter, \textit{i.e.}\ $\ov{\la Q \ra_{L}} = \ov{Q}$, the total stresses are  approximated by $r_{\text{LES}} \lp a,b \rp =R\lp a,b \rp + \ov{\tau \lp a,b \rp}$ \cite{Germano96,VNPM13}. 
The root mean square (RMS) values are square roots of these stresses.  
To analyze the compositional flame structure, the ``mixture fraction'' field  $Z (\bx ,t)$ is also constructed.   Bilger's formulation \cite{Bilger1976Structure, Peters00} is employed for this purpose. 

\section{Presentation of Results}  
\begin{figure}
    \centering

    \includegraphics[width=\textwidth]{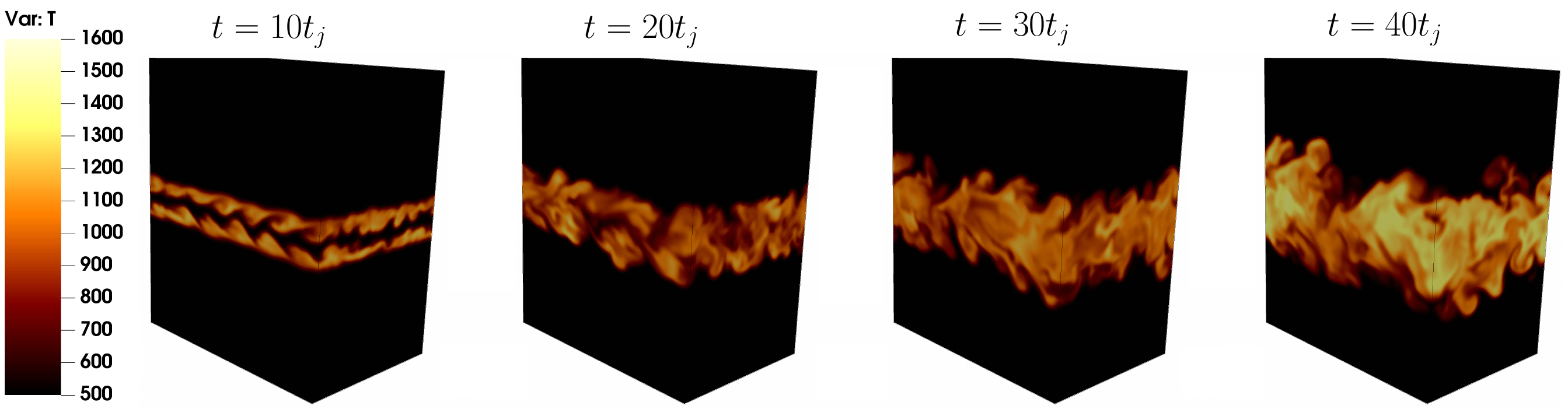}

    \caption{Temporal evolution of the temperature field.}
    \label{fig:MT}
\end{figure}

\begin{figure}
    \centering
     \begin{subfigure}[b]{\textwidth}
         \centering
         \includegraphics[width=0.3\textwidth]{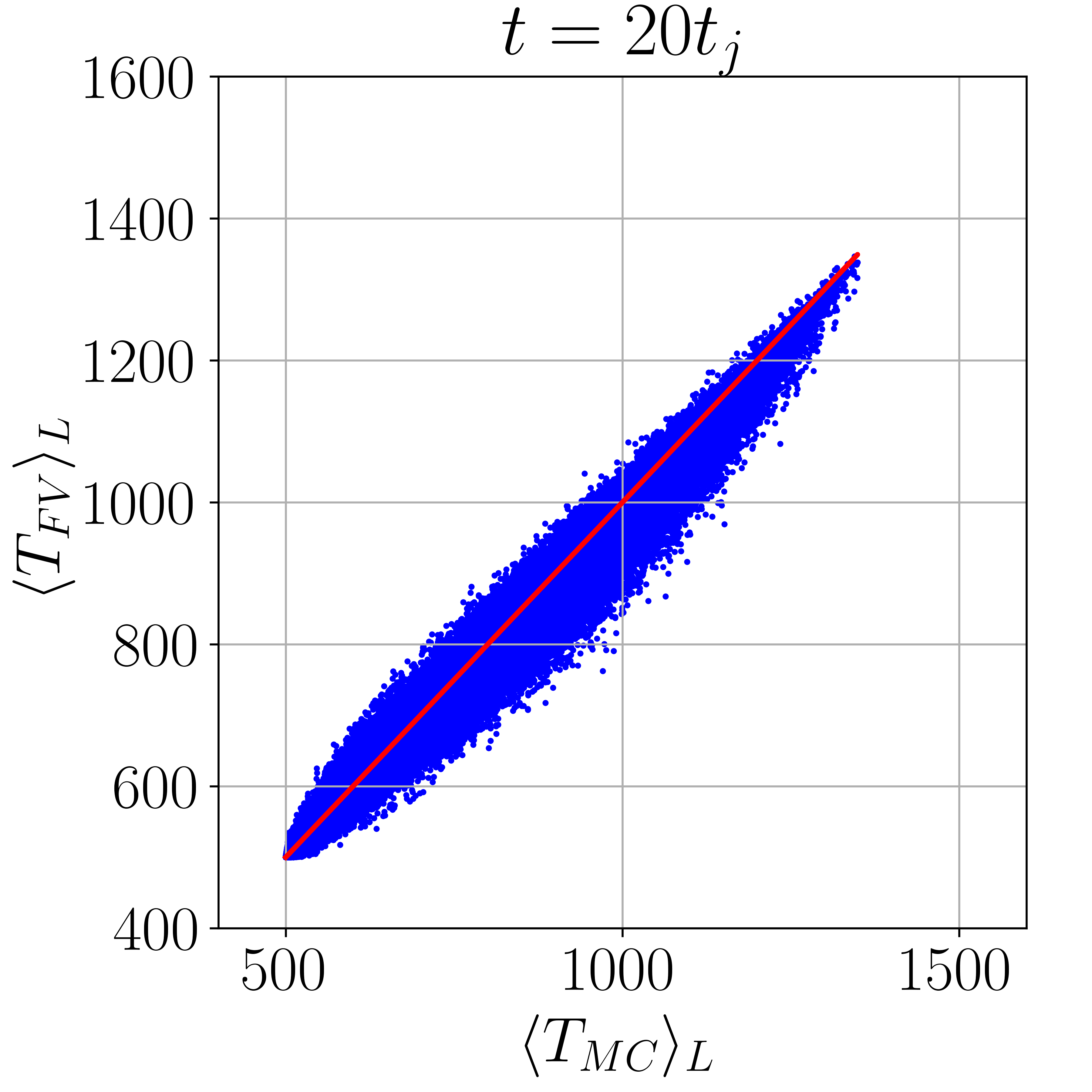}
         \includegraphics[width=0.3\textwidth]{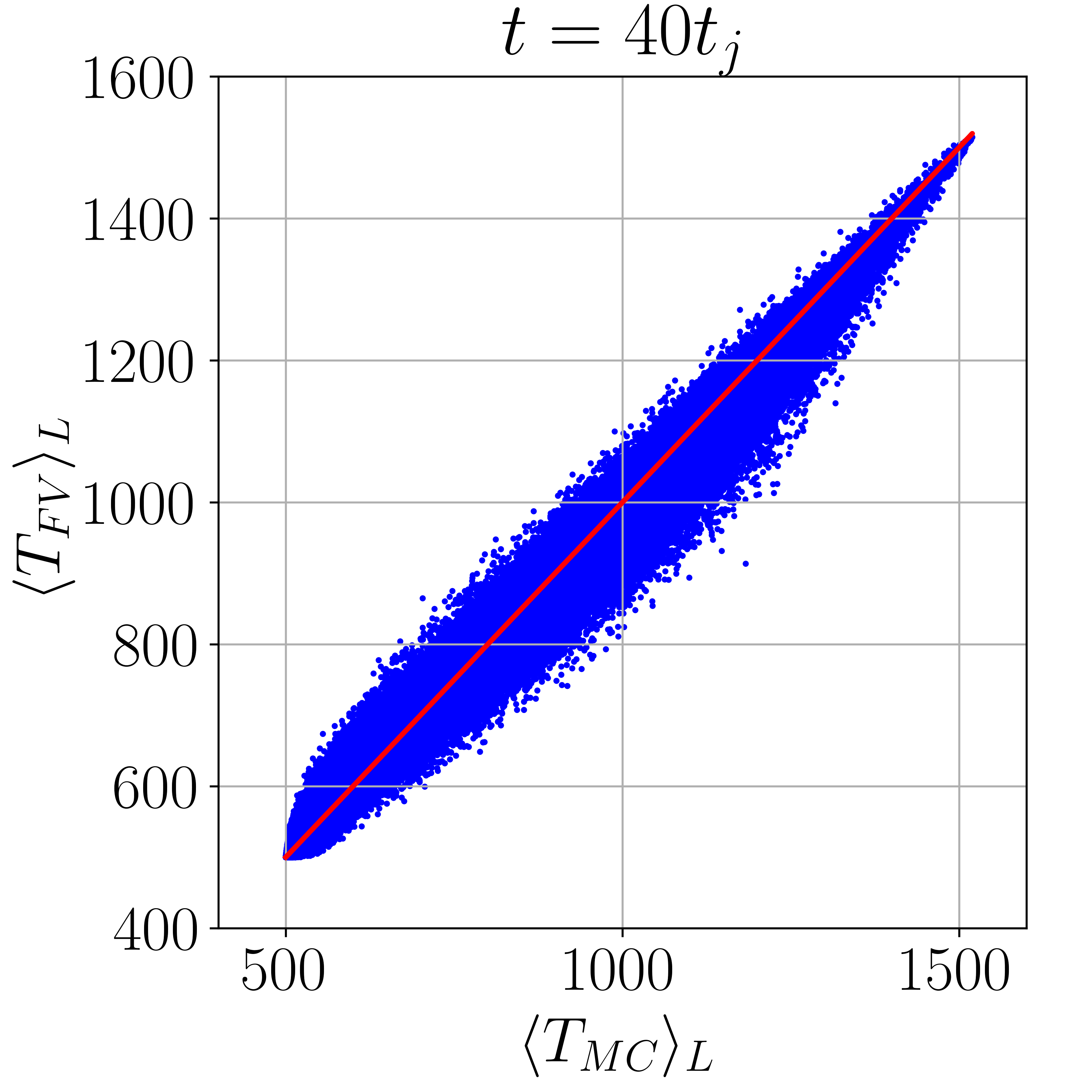}
         \caption{Temperature}
         \label{fig:ScatterT}
     \end{subfigure}

     \begin{subfigure}[b]{\textwidth}
         \centering
         \includegraphics[width=0.3\textwidth]{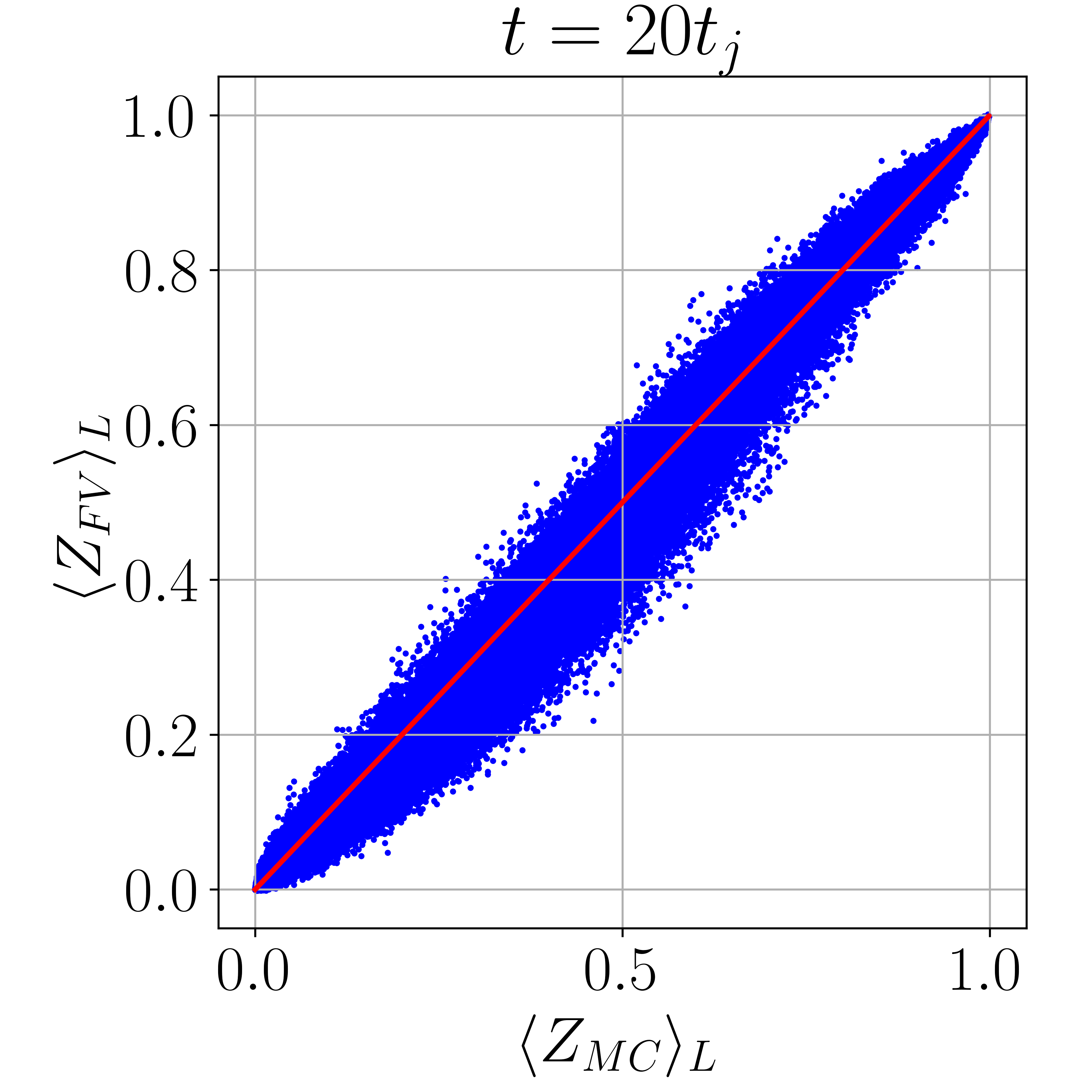}
         \includegraphics[width=0.3\textwidth]{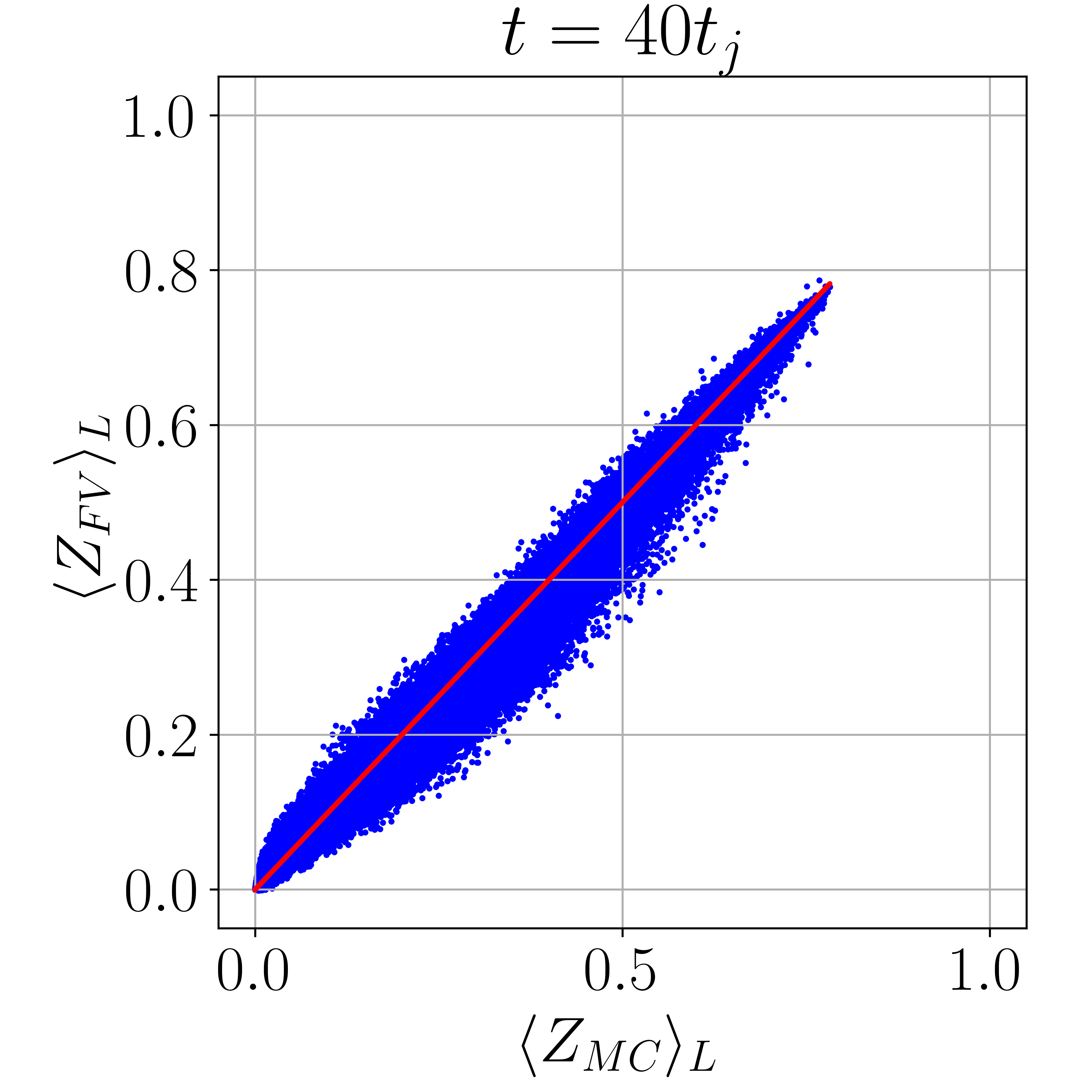}
         \caption{Mixture fraction}
         \label{fig:ScatterZ}
     \end{subfigure}

    \caption{Scatter plots of the  Eulerian vs.\ the  Lagrangian filtered values.}
    \label{fig:ScatterTZ}
\end{figure}

\begin{figure}
    \centering
     \begin{subfigure}[b]{\textwidth}
         \centering
         \includegraphics[width=0.3\textwidth]{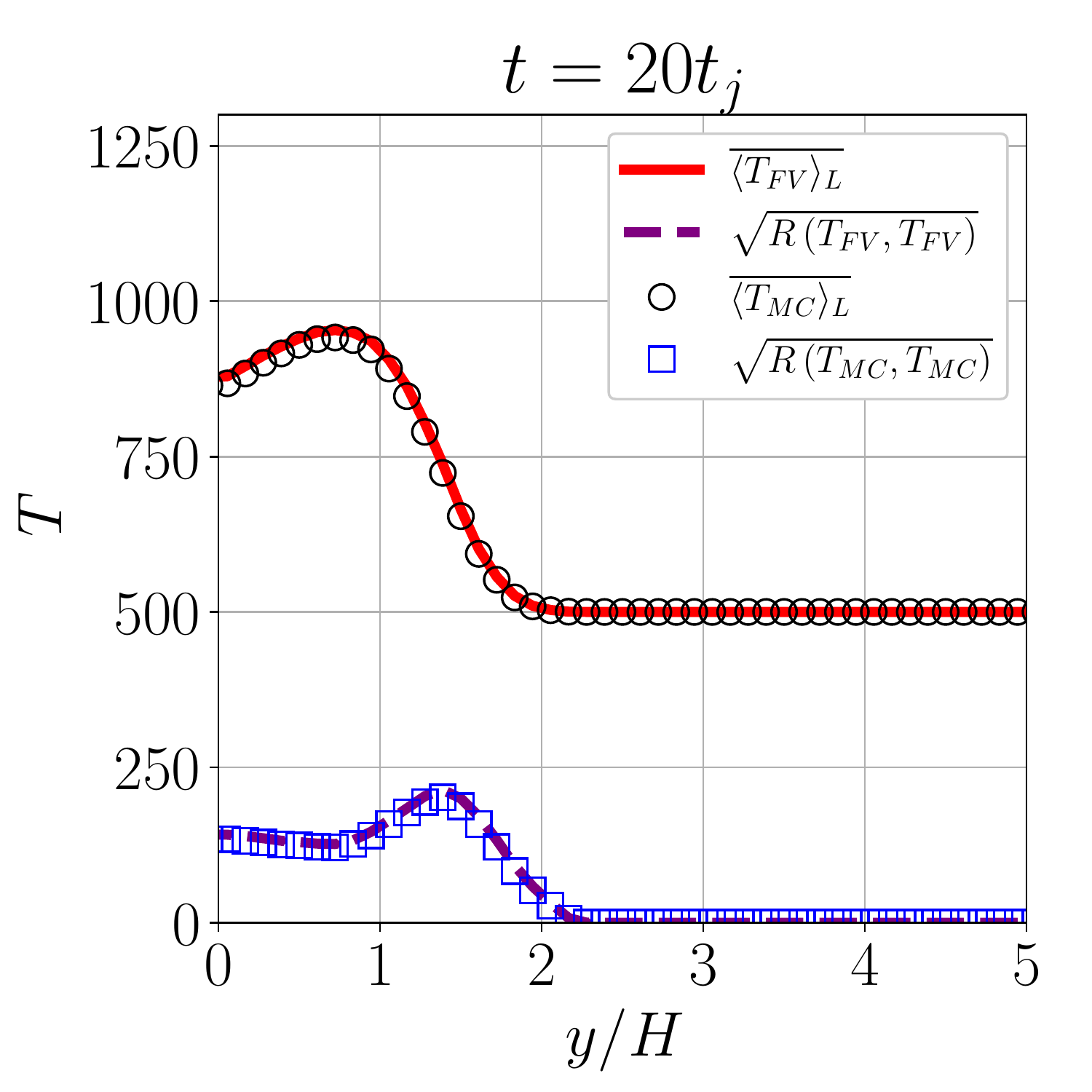}
         \includegraphics[width=0.3\textwidth]{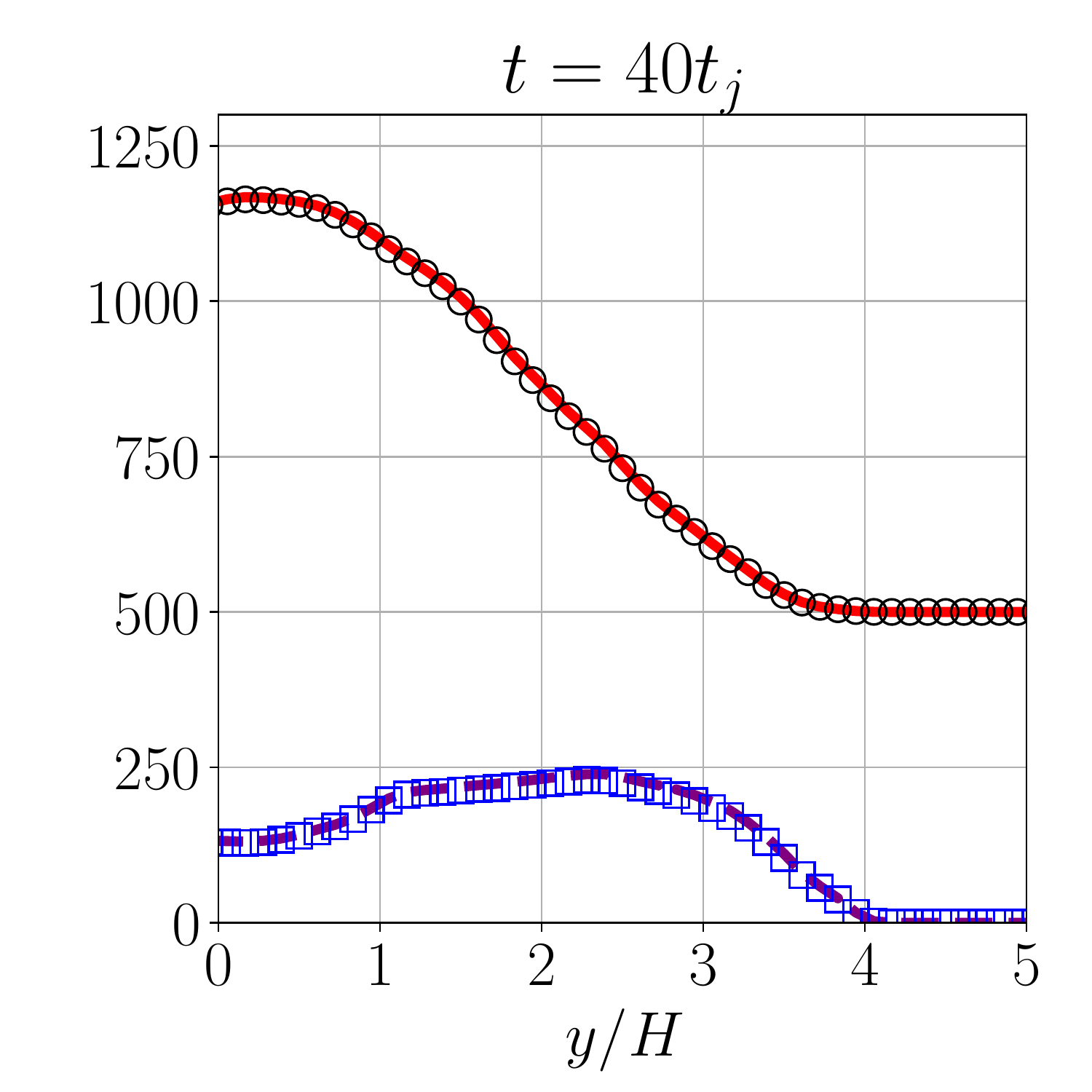}
         \caption{Temperature}
         \label{fig:RFDMCT}
     \end{subfigure}

     \begin{subfigure}[b]{\textwidth}
         \centering
         \includegraphics[width=0.3\textwidth]{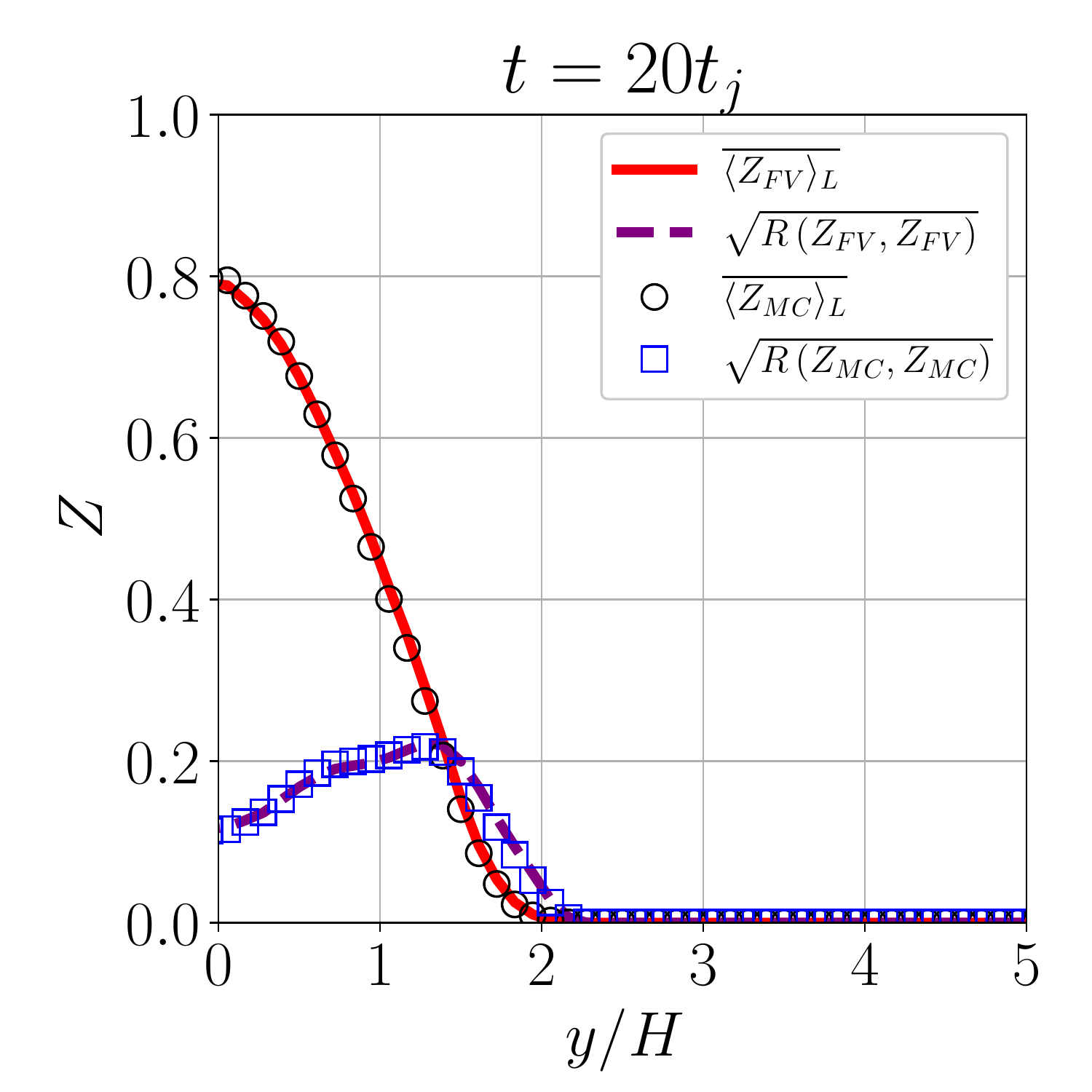}
         \includegraphics[width=0.3\textwidth]{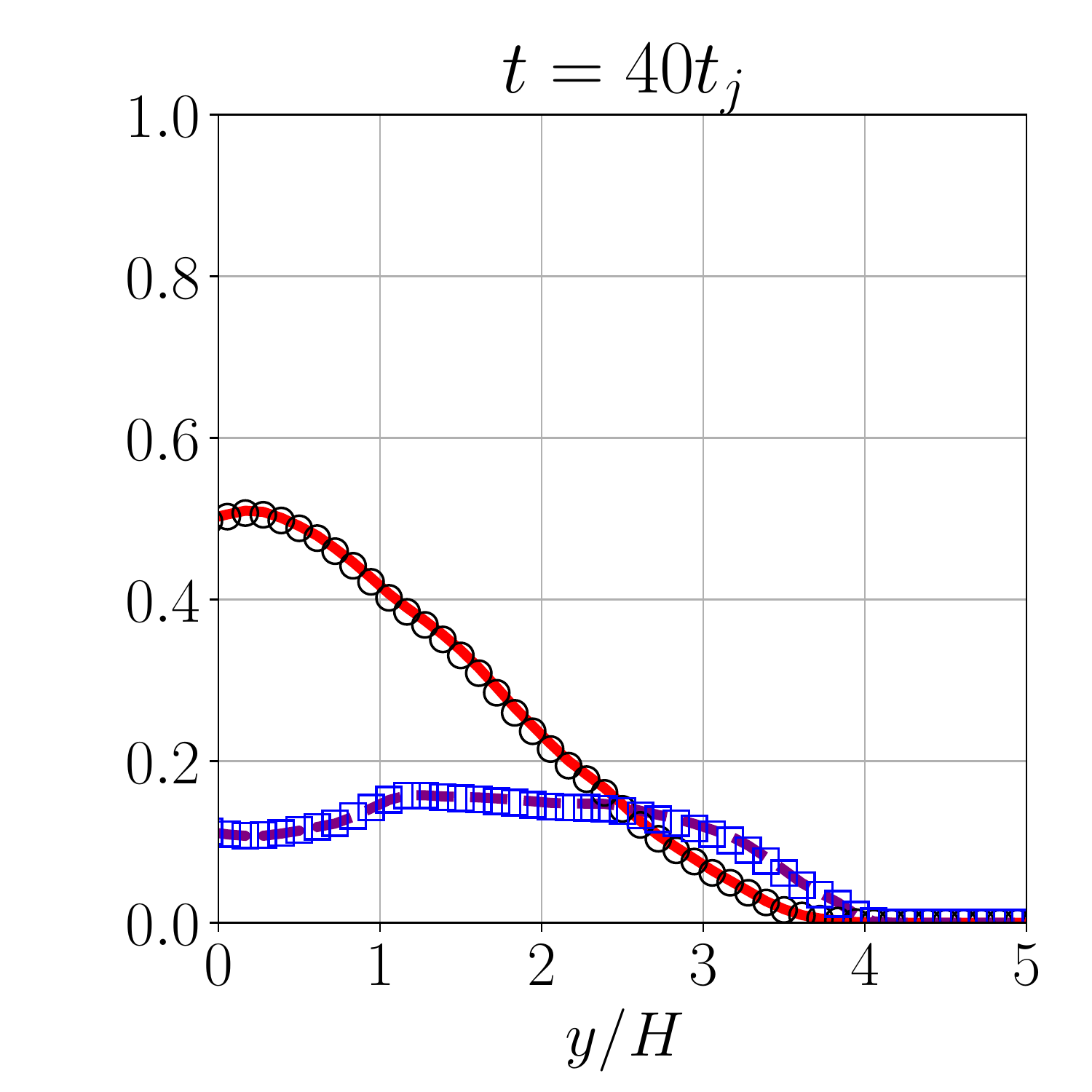}
         \caption{Mixture fraction}
         \label{fig:RFDMCZ}
     \end{subfigure}

    \caption{Reynolds-averaged Eulerian (lines) vs.\ the  Lagrangian filtered (symbols) values in the cross-stream direction.}
    \label{fig:RFDMC}
\end{figure}

For the purpose of  flow visualization, the contour plots of the temperate field are presented in \cref{fig:MT} for several consecutive time-instances.  These contours show the formation of structures within the flow, and the growth of the layer from the initial laminar to a highly three-dimensional turbulent flow. To demonstrate the consistency, comparisons are made between the filtered values as obtained by the Lagrangian and Eulerian simulators.  \cref{fig:ScatterTZ} shows the instantaneous scatter plots of the temperature and mixture fraction, and \cref{fig:RFDMC} shows the Reynolds averaged values of these variables.  The similarity of FDF and PeleLM results is evident. 

\begin{figure}
    \centering

    \includegraphics[trim=0.0cm 1.25cm 0.75cm 0cm, clip=true, height=3.4cm]{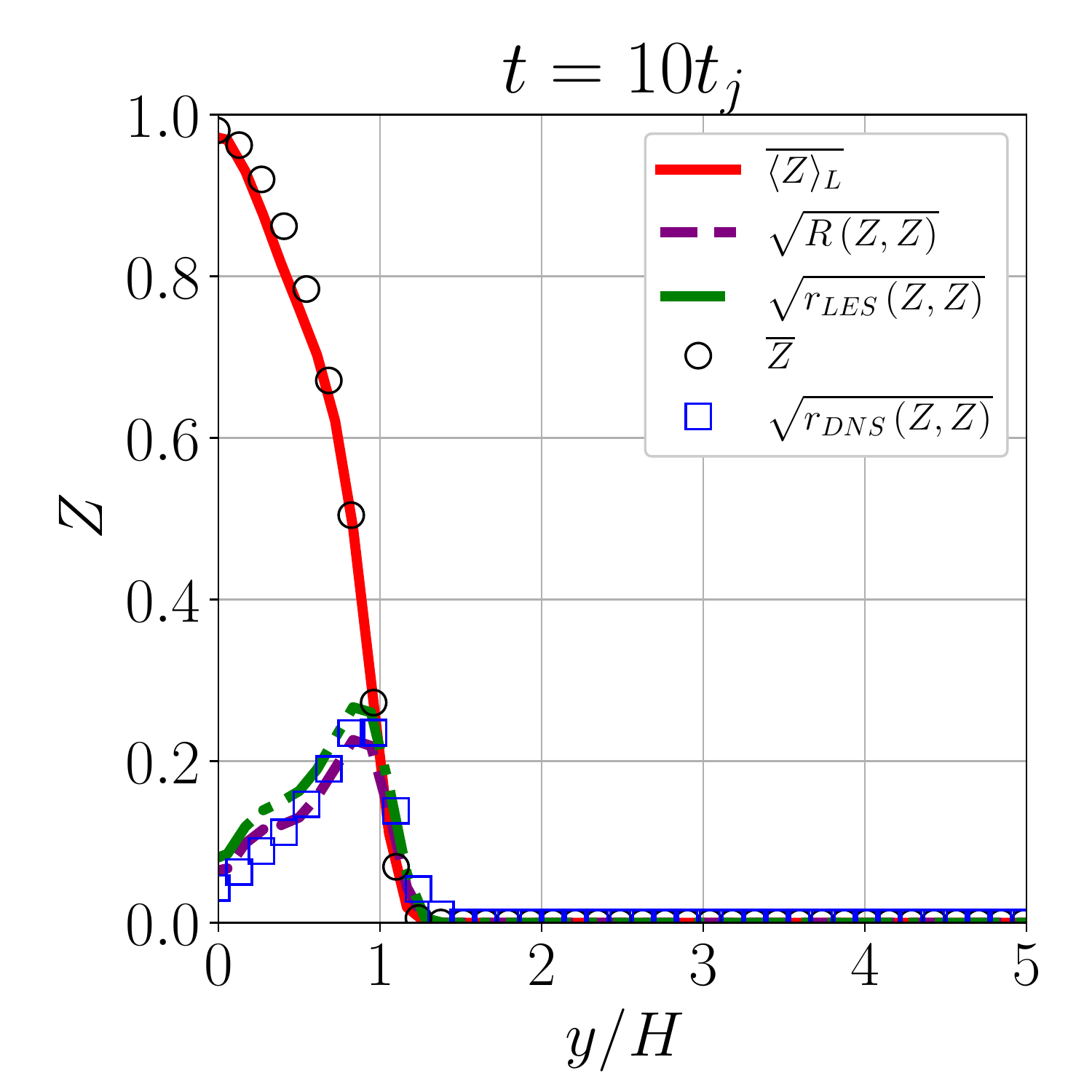} 
    \includegraphics[trim=2.8cm 1.25cm 0.75cm 0cm, clip=true, height=3.4cm]{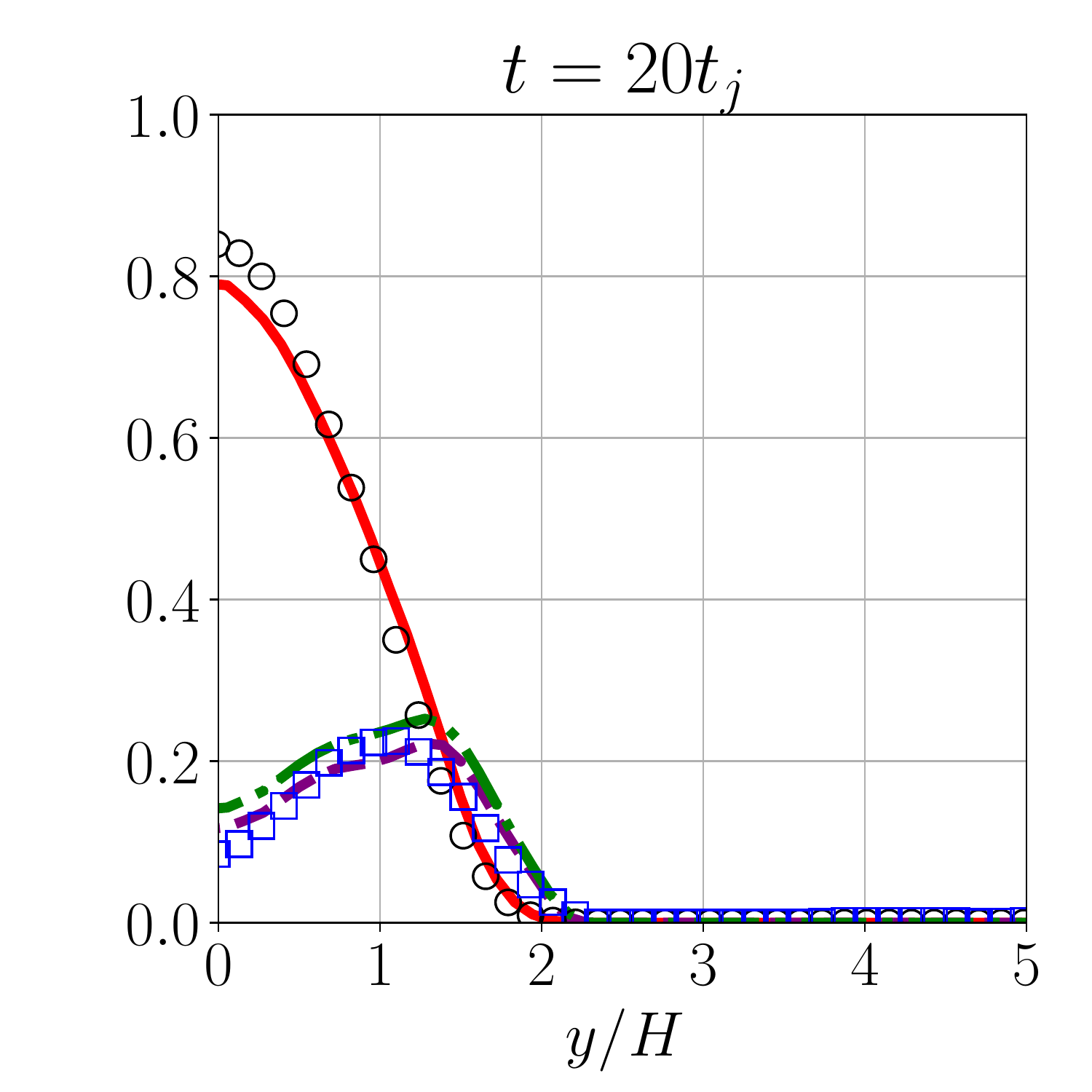}
    \includegraphics[trim=2.8cm 1.25cm 0.75cm 0cm, clip=true, height=3.4cm]{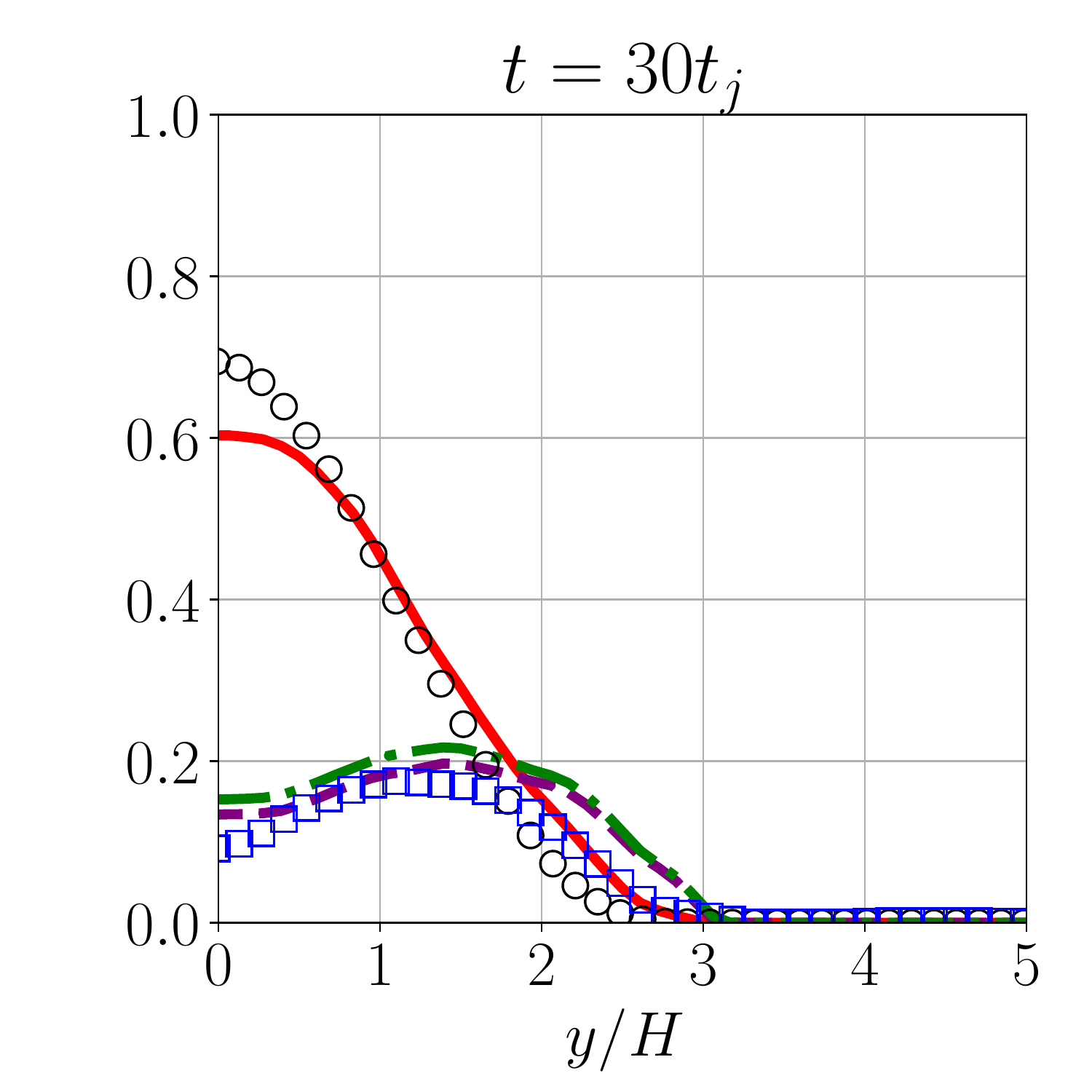}
    \includegraphics[trim=2.8cm 1.25cm 0.75cm 0cm, clip=true, height=3.4cm]{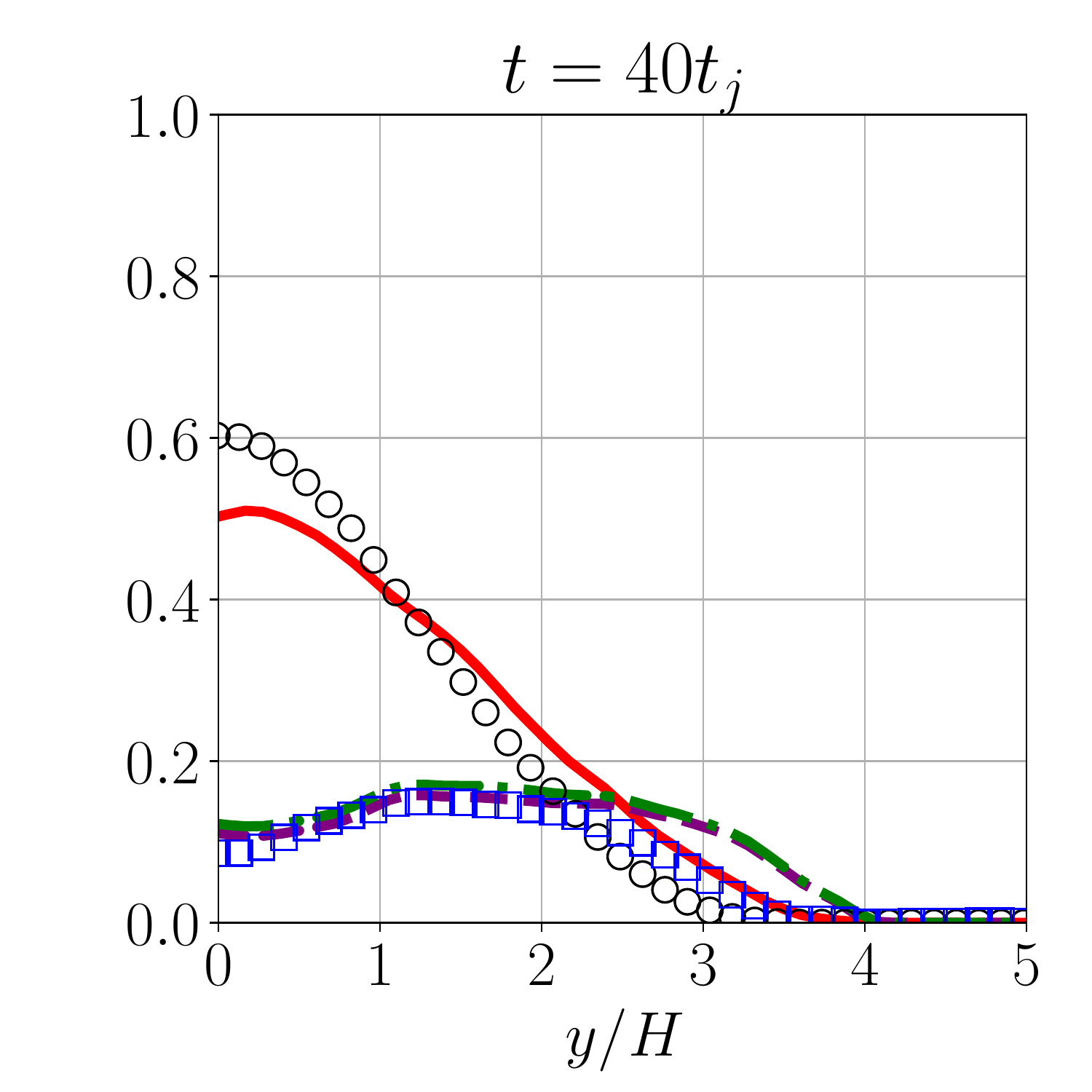}

    \includegraphics[trim=0.0cm 1.25cm 0.75cm 0cm, clip=true, height=3.4cm]{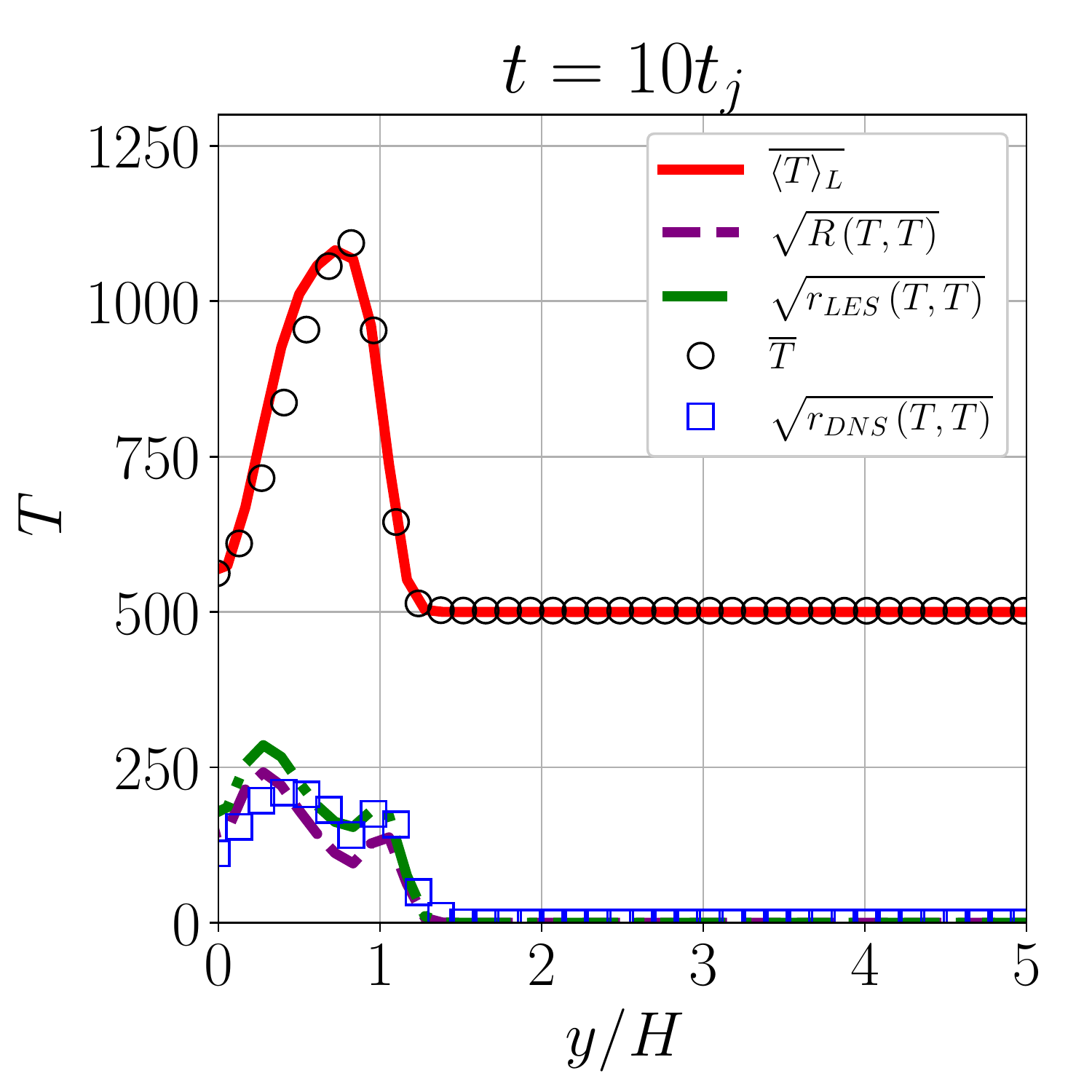}
    \includegraphics[trim=2.8cm 1.25cm 0.75cm 0cm, clip=true, height=3.4cm]{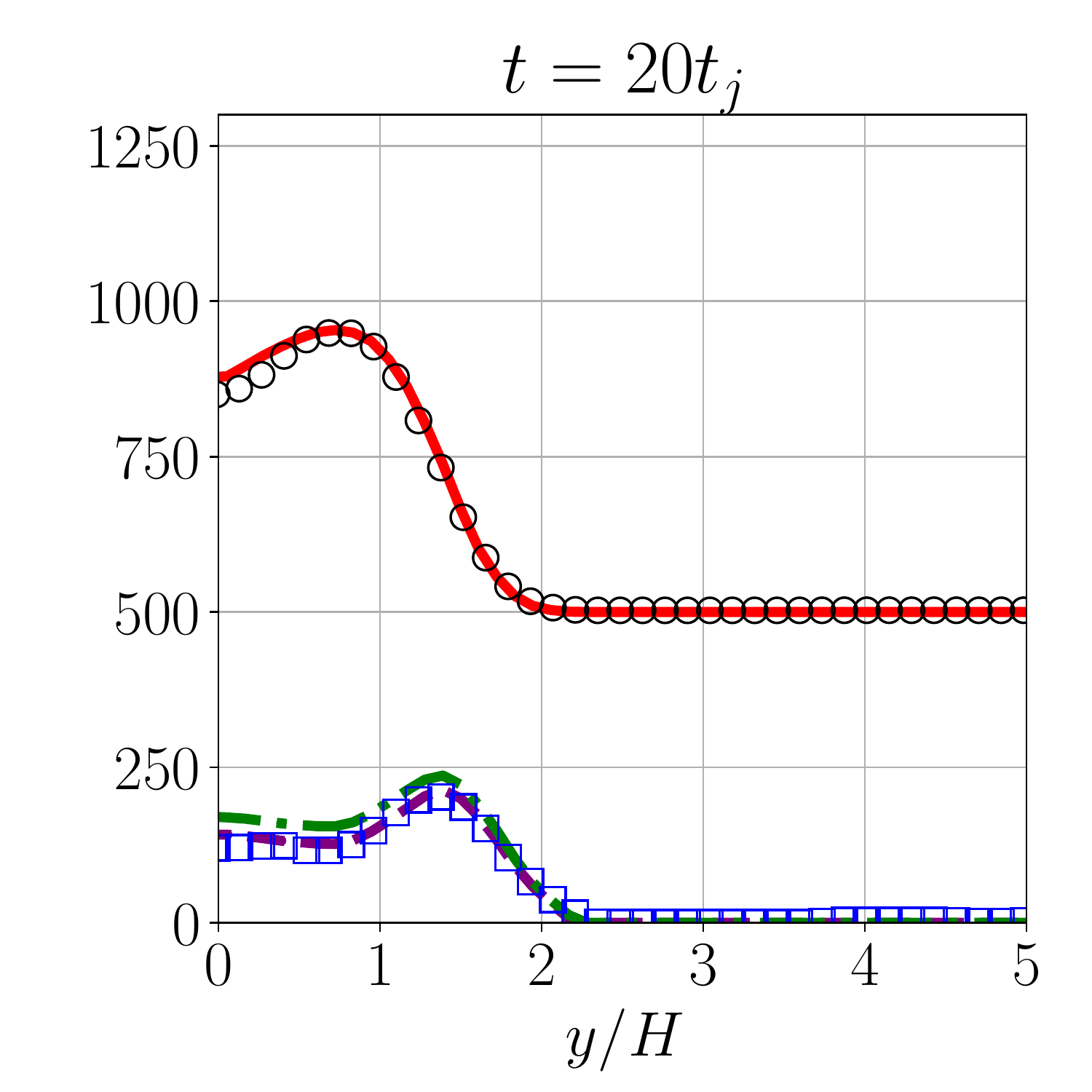}
    \includegraphics[trim=2.8cm 1.25cm 0.75cm 0cm, clip=true, height=3.4cm]{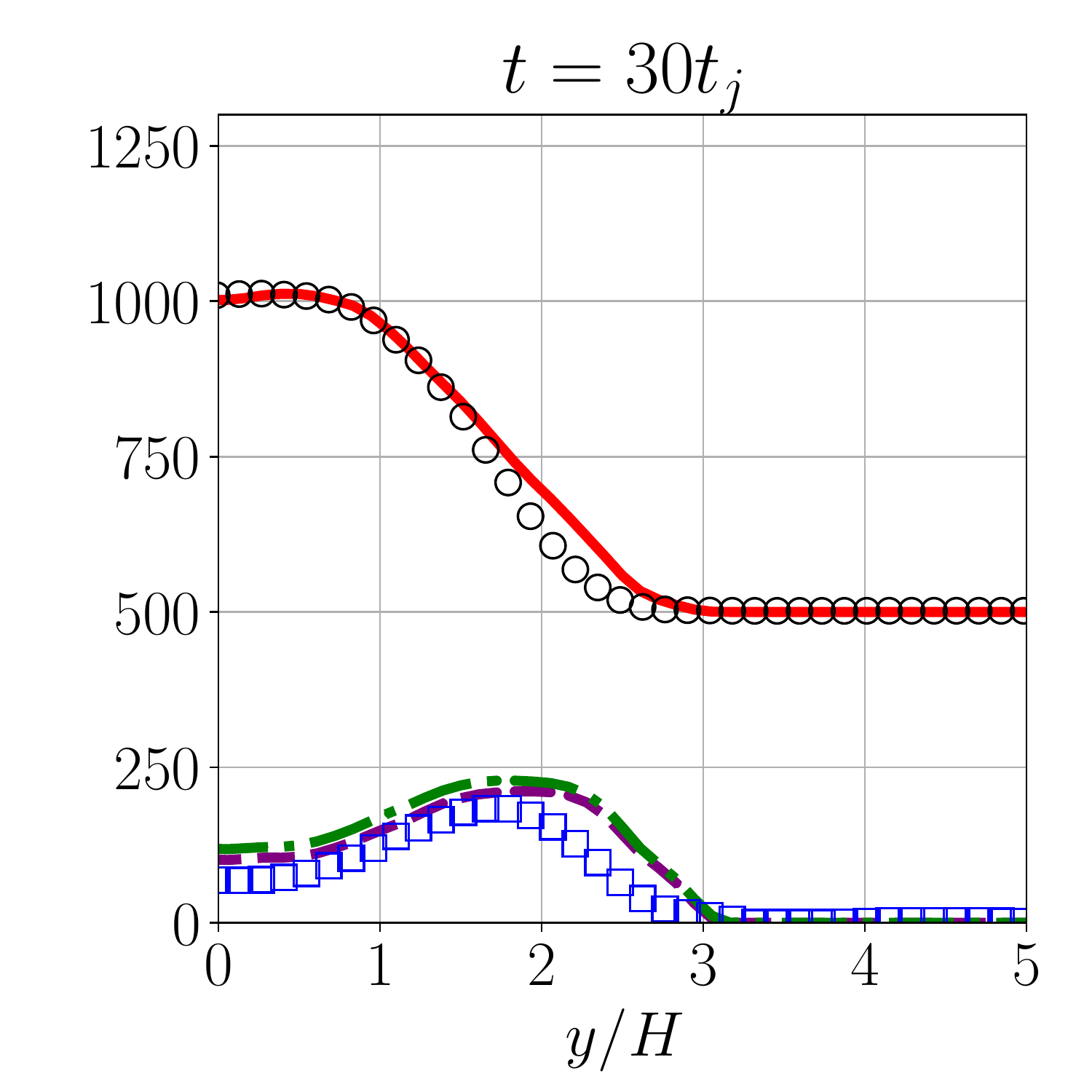}
    \includegraphics[trim=2.8cm 1.25cm 0.75cm 0cm, clip=true, height=3.4cm]{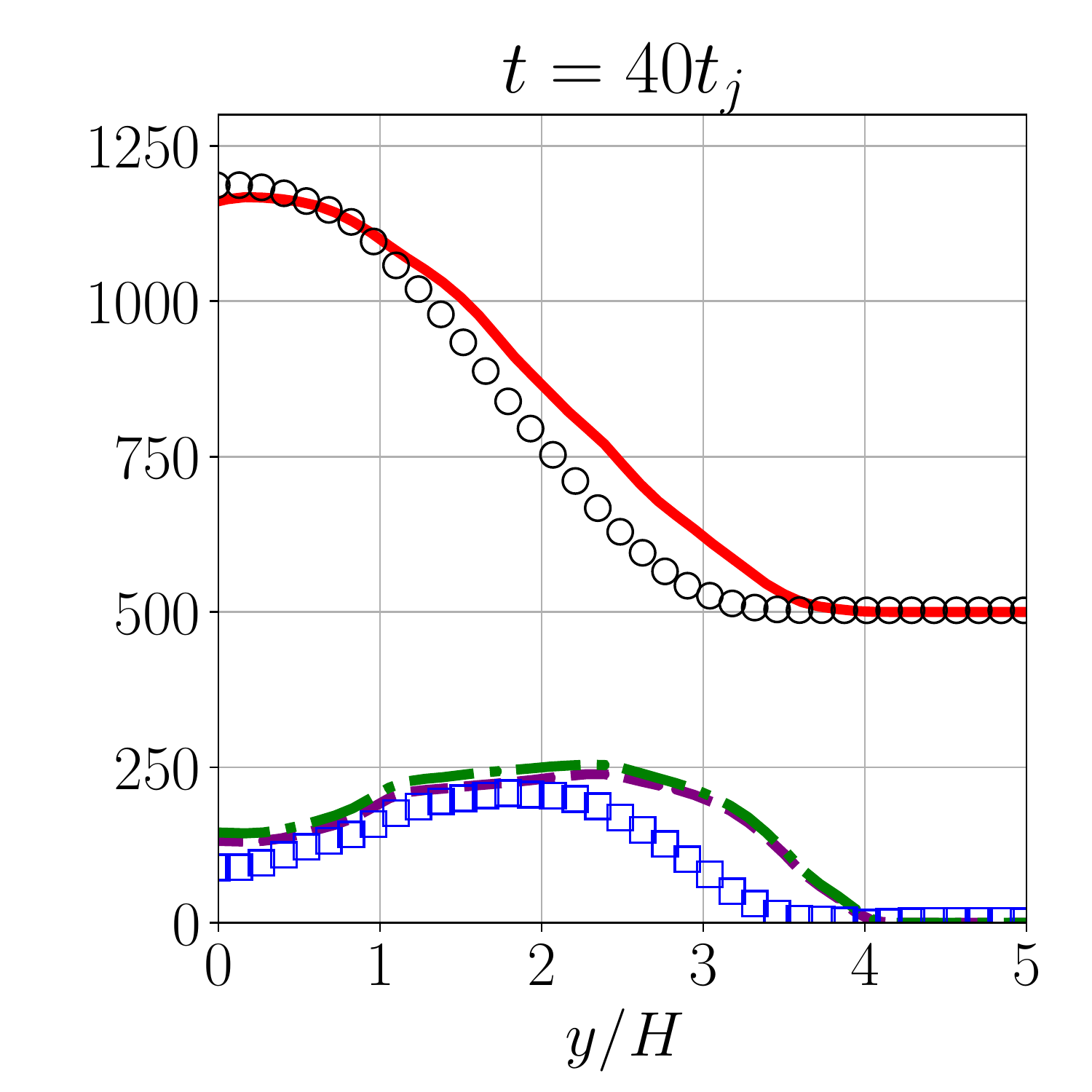}

    \includegraphics[trim=0.0cm 1.25cm 0.75cm 0cm, clip=true, height=3.4cm]{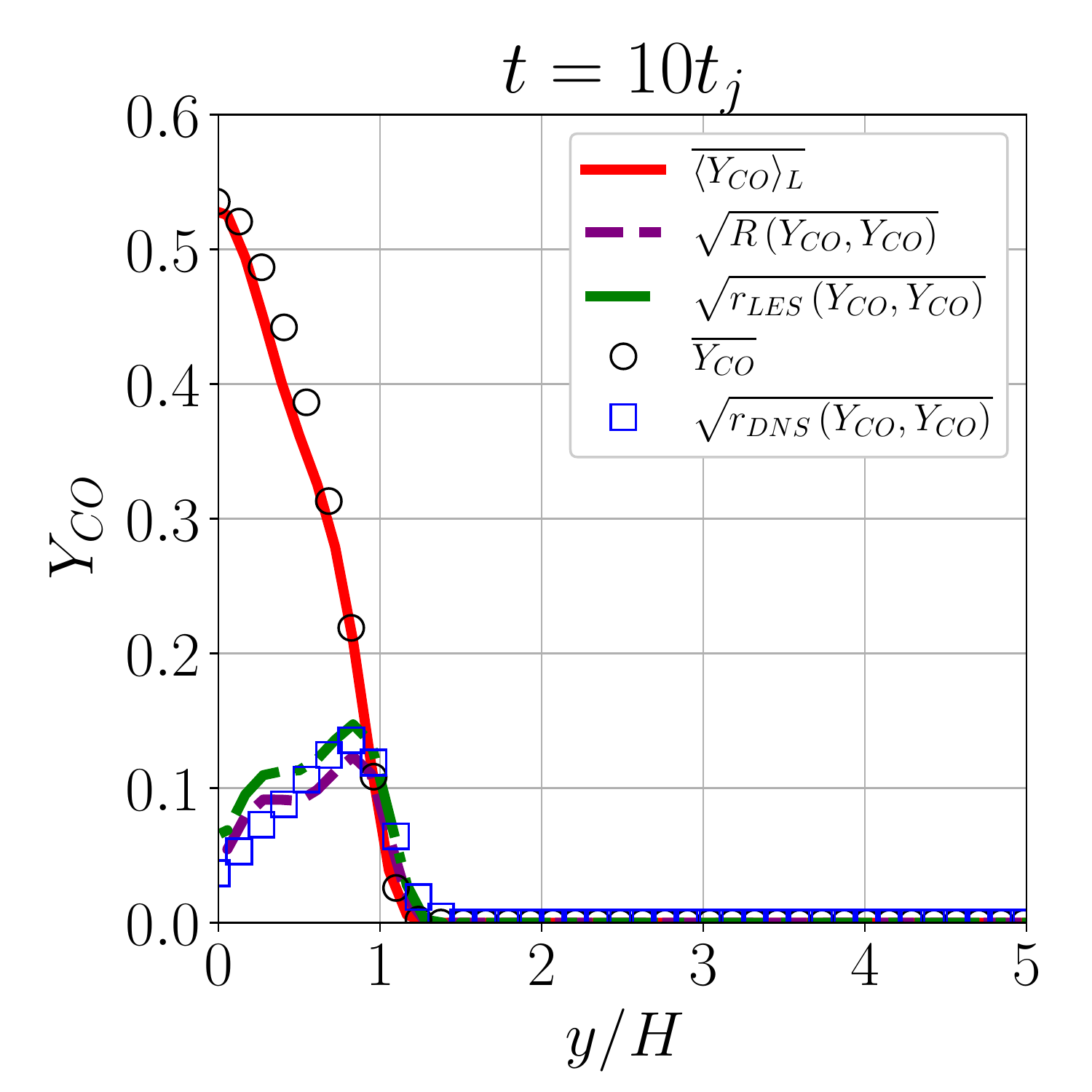}
    \includegraphics[trim=2.8cm 1.25cm 0.75cm 0cm, clip=true, height=3.4cm]{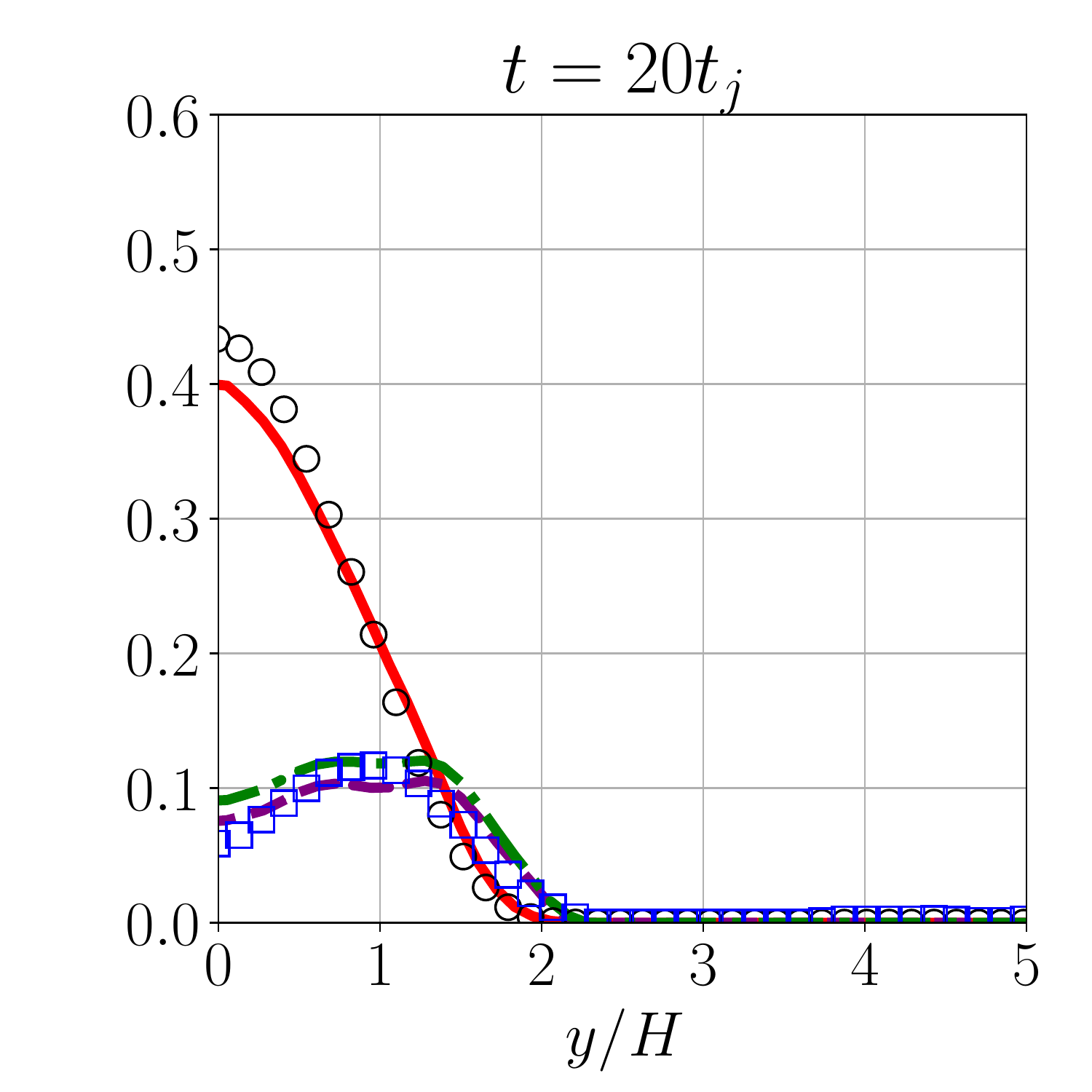}
    \includegraphics[trim=2.8cm 1.25cm 0.75cm 0cm, clip=true, height=3.4cm]{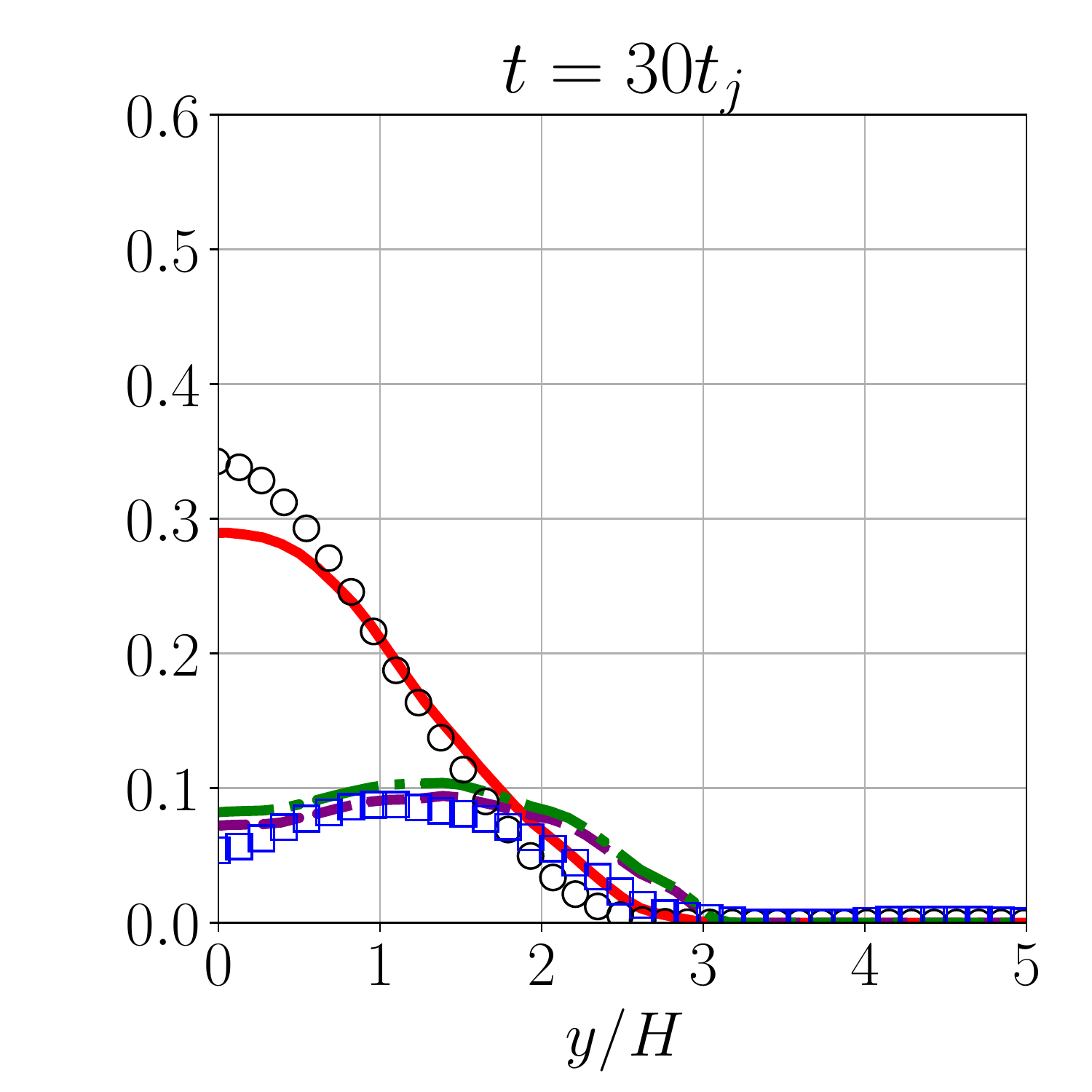}
    \includegraphics[trim=2.8cm 1.25cm 0.75cm 0cm, clip=true, height=3.4cm]{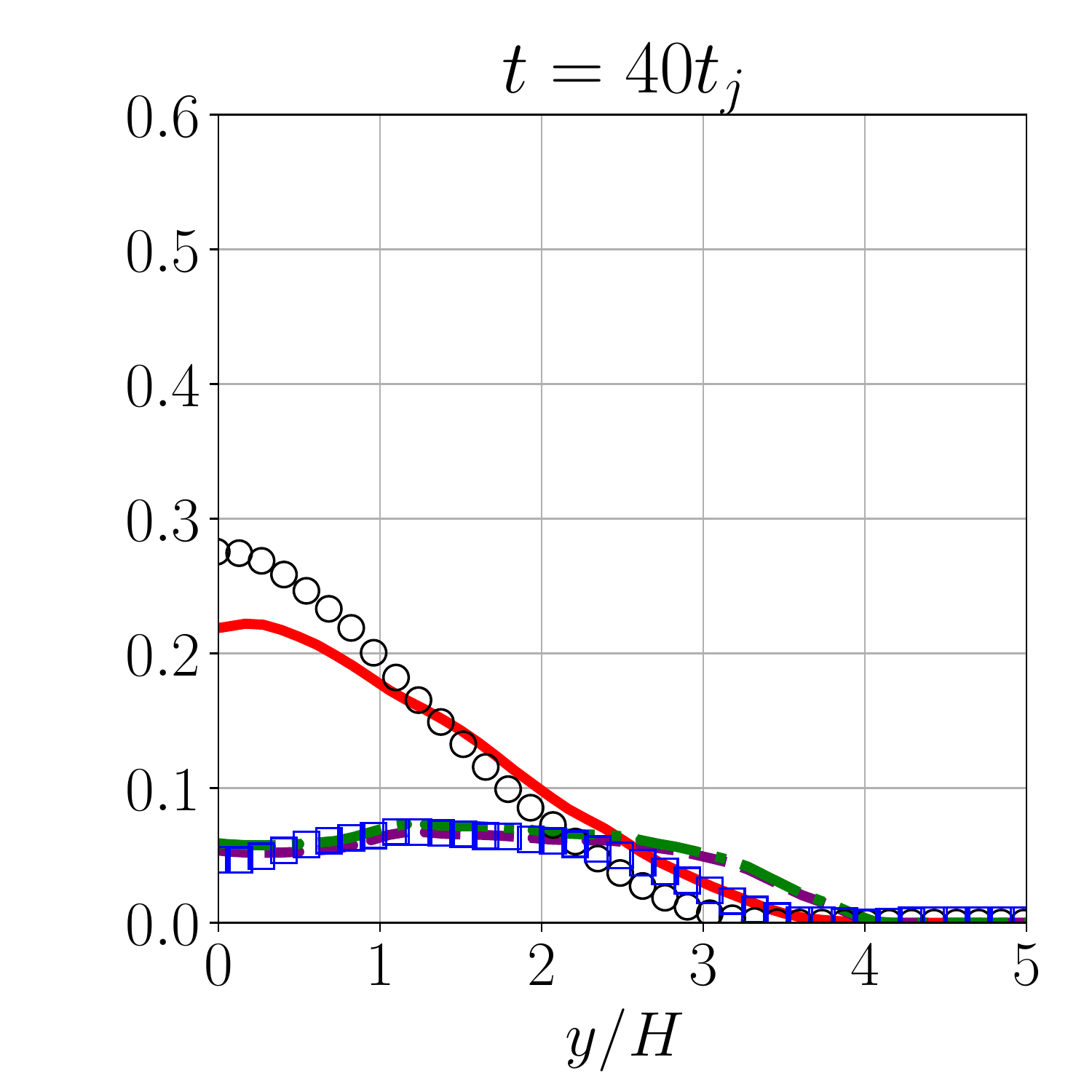}

    \includegraphics[trim=0.0cm 1.25cm 0.75cm 0cm, clip=true, height=3.4cm]{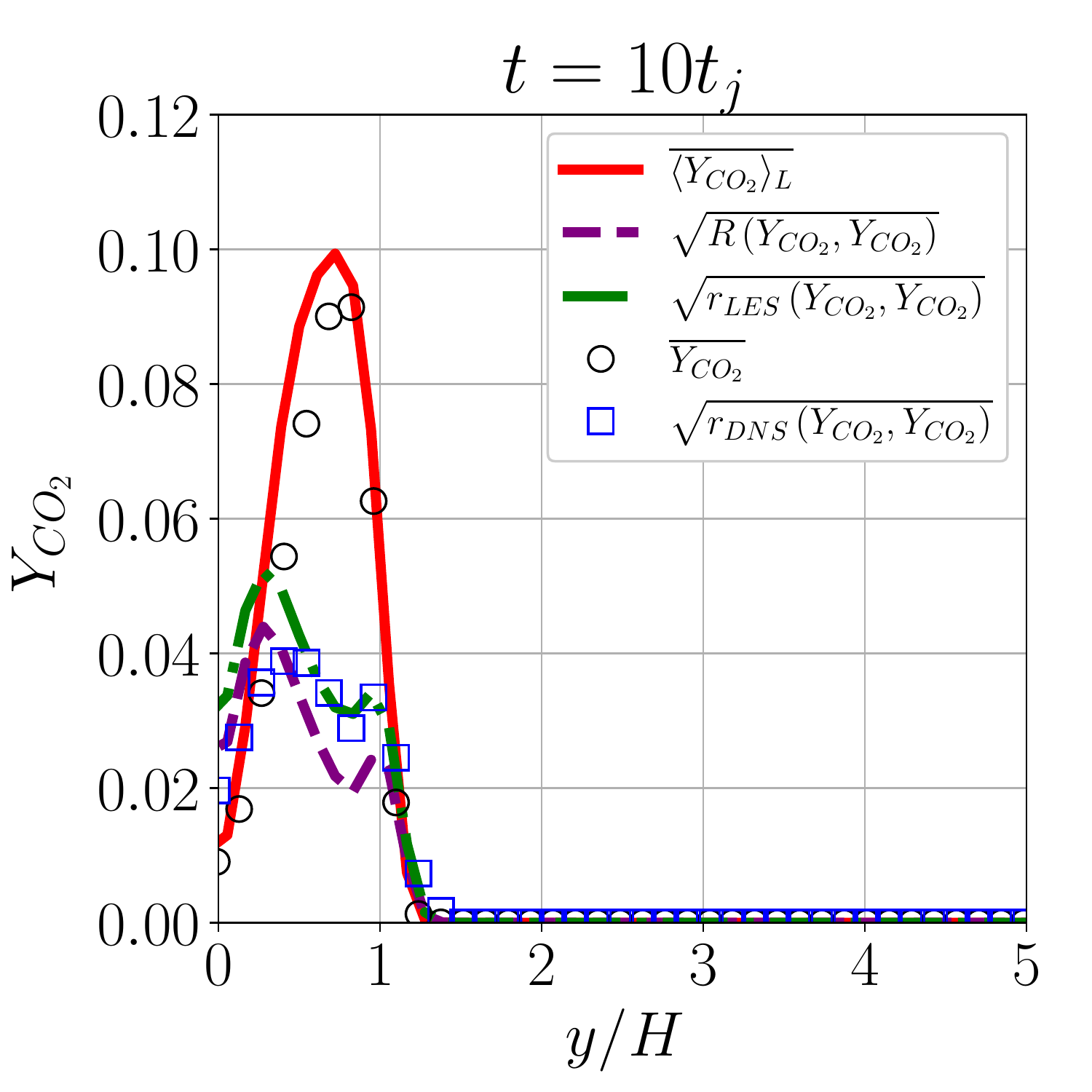}
    \includegraphics[trim=2.8cm 1.25cm 0.75cm 0cm, clip=true, height=3.4cm]{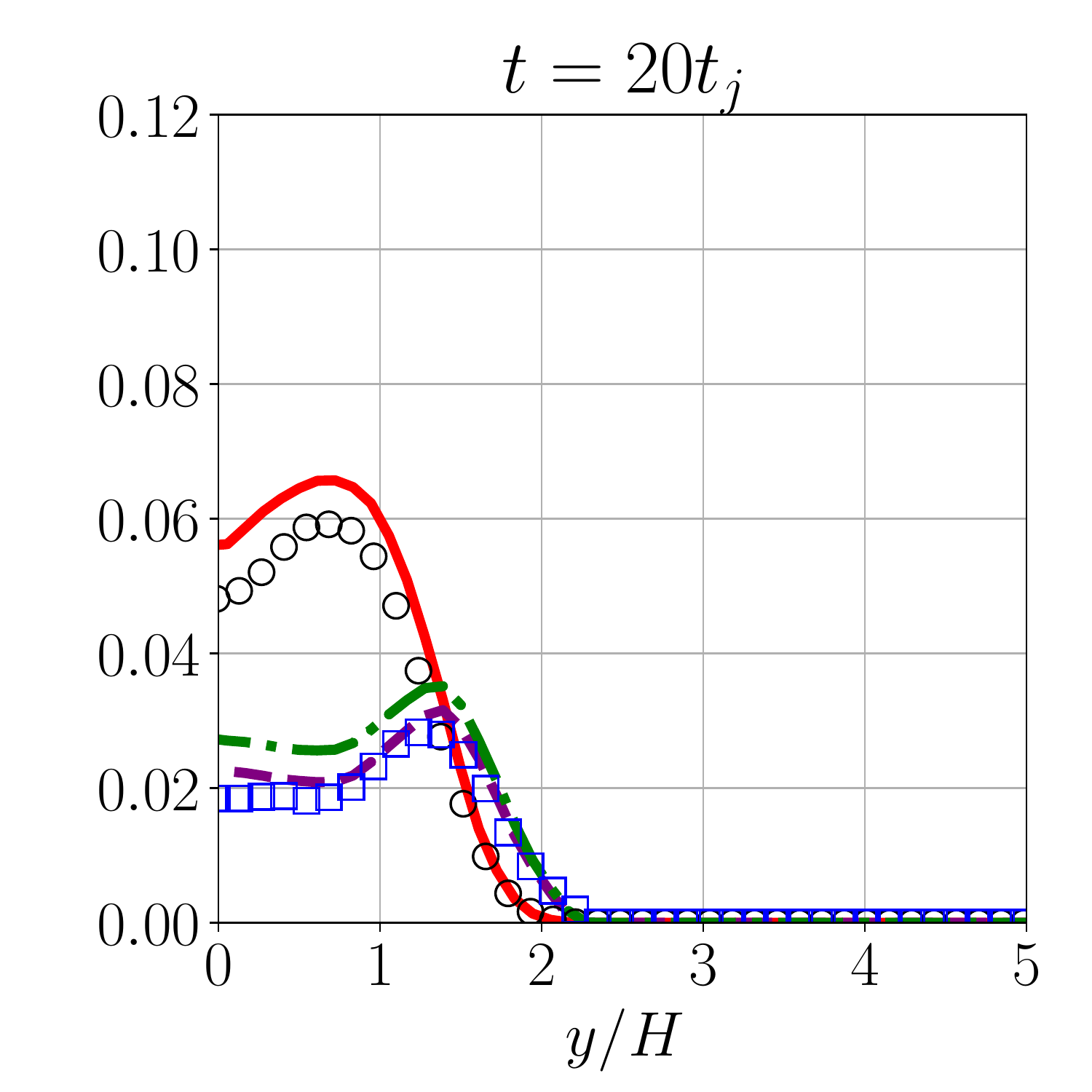}
    \includegraphics[trim=2.8cm 1.25cm 0.75cm 0cm, clip=true, height=3.4cm]{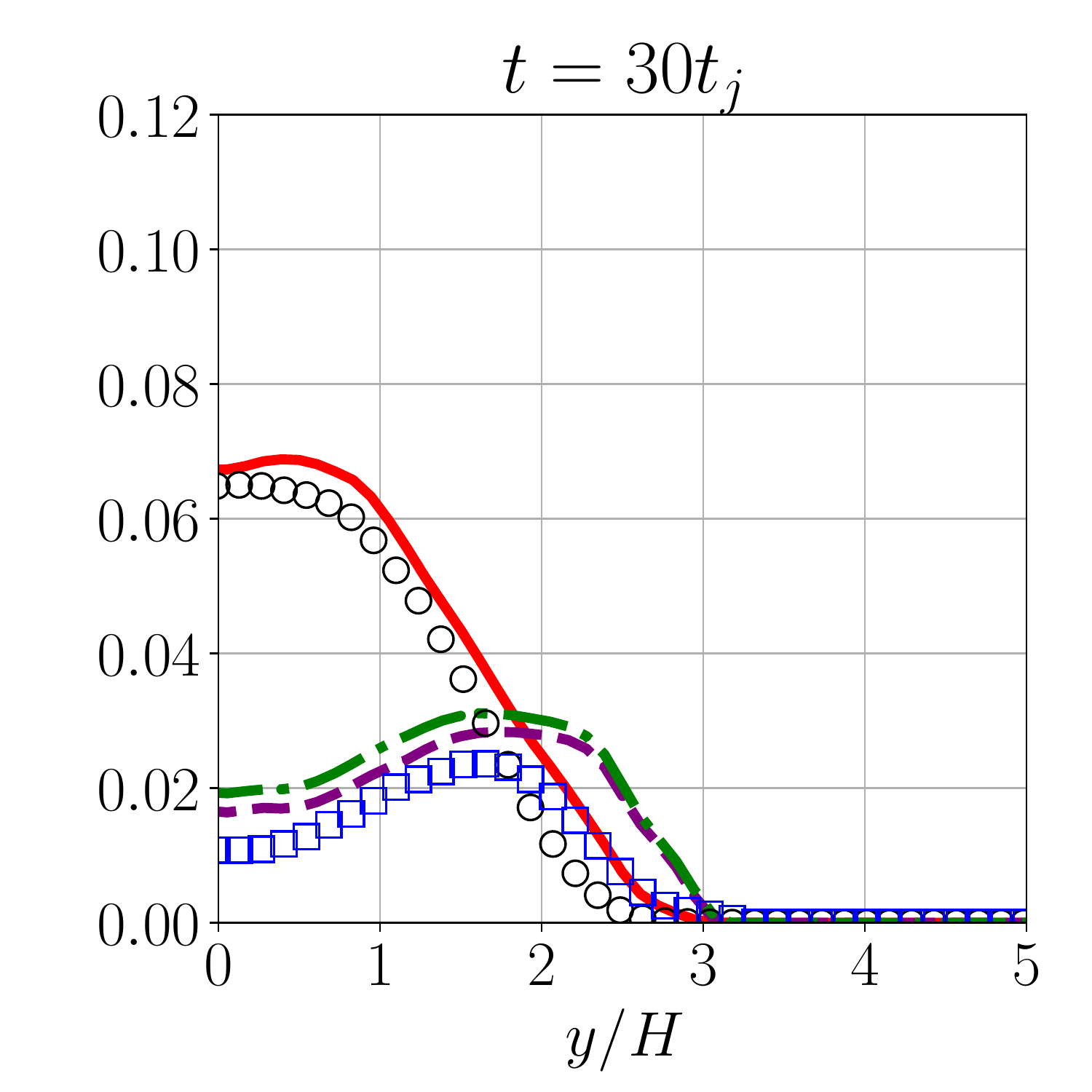}
    \includegraphics[trim=2.8cm 1.25cm 0.75cm 0cm, clip=true, height=3.4cm]{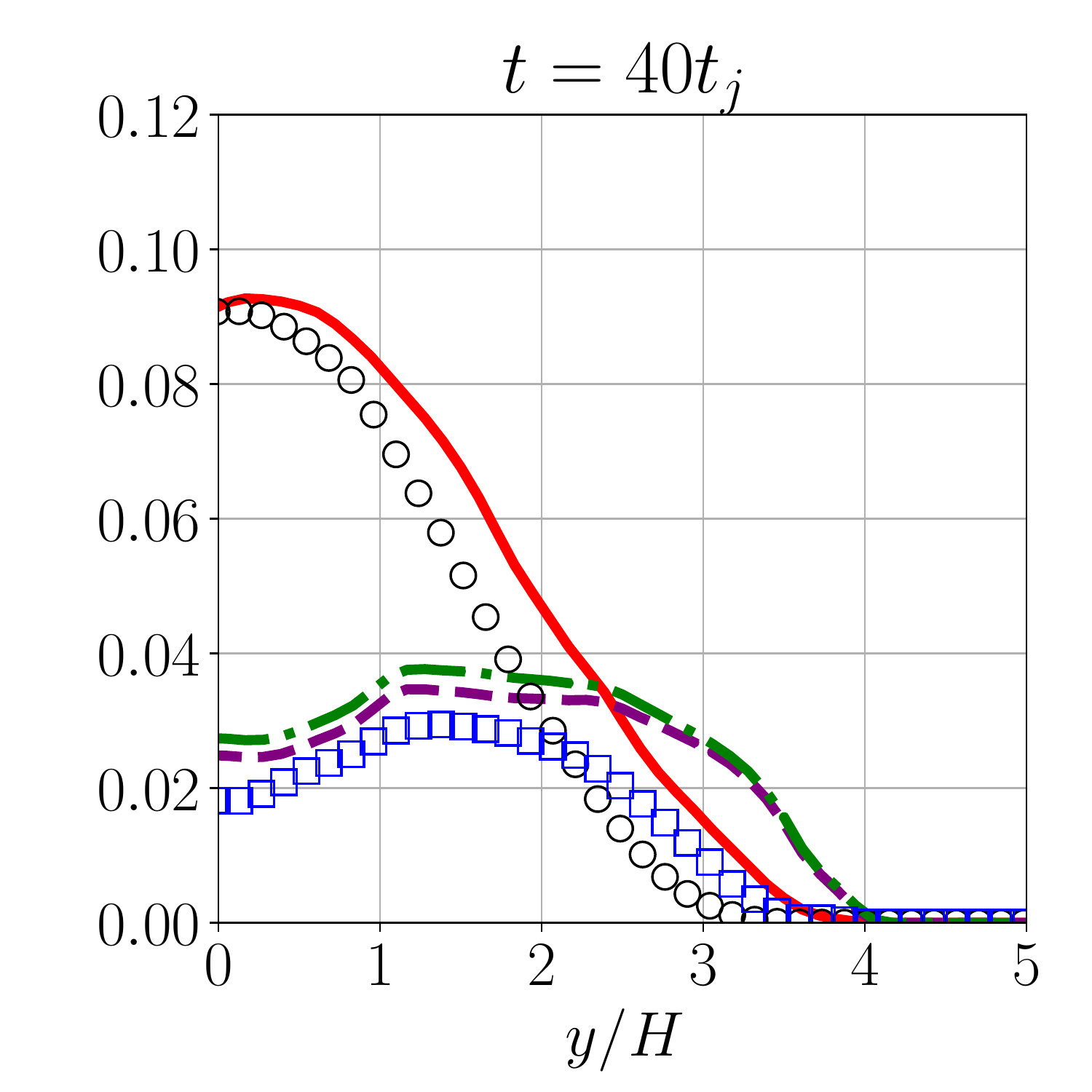}
    
    \vspace{-0.125cm}
    
    \includegraphics[trim=0.0cm 0cm 0.75cm 14cm, clip=true, width=3.52cm]{Figures/bar_CO2_05000.pdf}
    \includegraphics[trim=2.8cm 0cm 0.75cm 14cm, clip=true, width=2.84cm]{Figures/bar_CO2_10000.pdf}
    \includegraphics[trim=2.8cm 0cm 0.75cm 14cm, clip=true, width=2.84cm]{Figures/bar_CO2_15000.pdf}
    \includegraphics[trim=2.8cm 0cm 0.75cm 14cm, clip=true, width=2.84cm]{Figures/bar_CO2_20000.pdf}

    \caption{Reynolds-averaged mean and RMS values of the mixture fraction ($Z$), temperature ($T$), CO mass fraction ($Y_{CO}$), and CO\textsubscript{2} mass fraction ($Y_{CO_2}$). Lines and symbols denote LES and  DNS results, respectively.}
    \label{fig:RZTC}
\end{figure}

\begin{figure}
    \centering
     \begin{subfigure}[b]{\textwidth}
         \centering
         \includegraphics[width=0.3\textwidth]{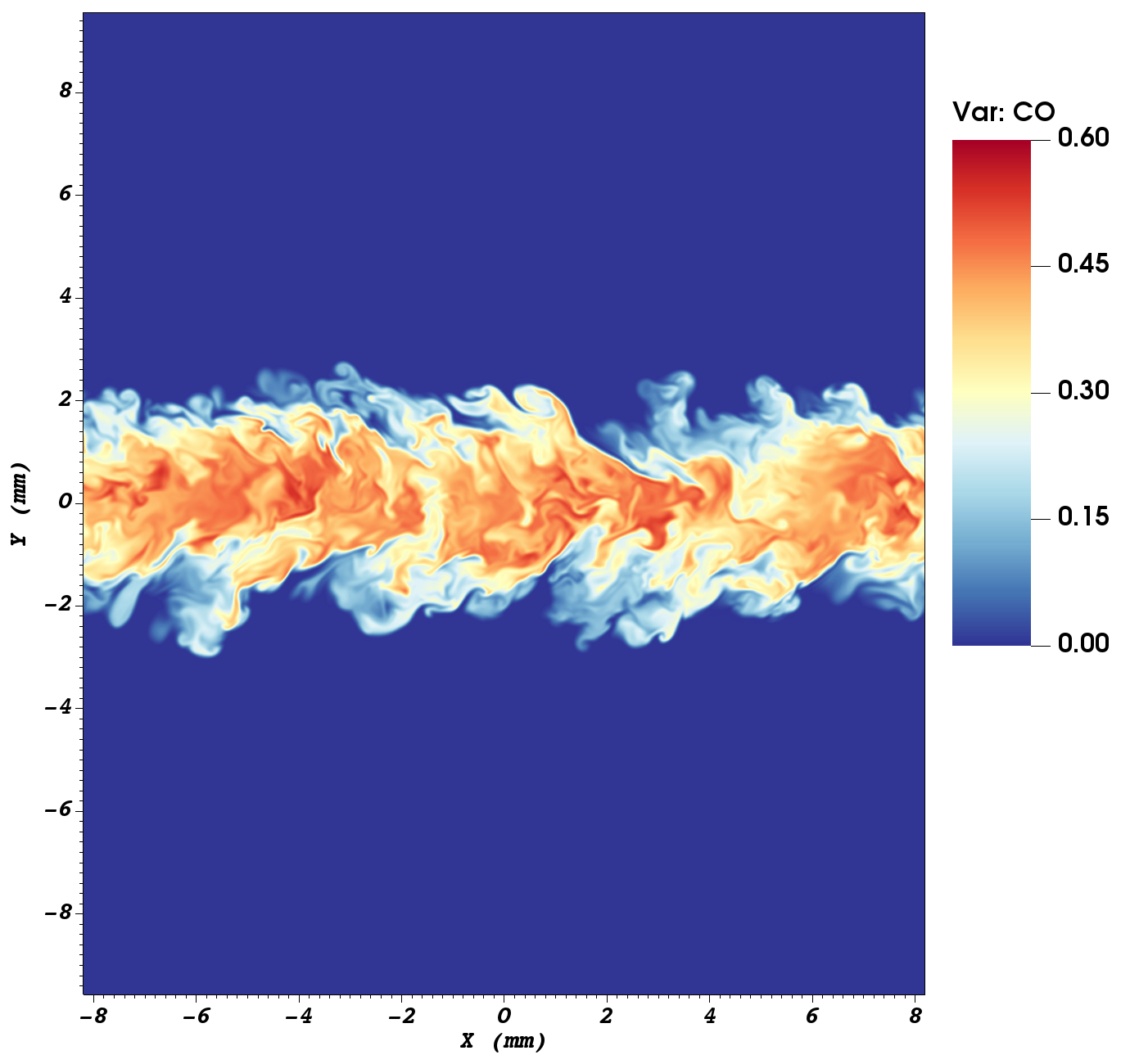}
         \includegraphics[width=0.3\textwidth]{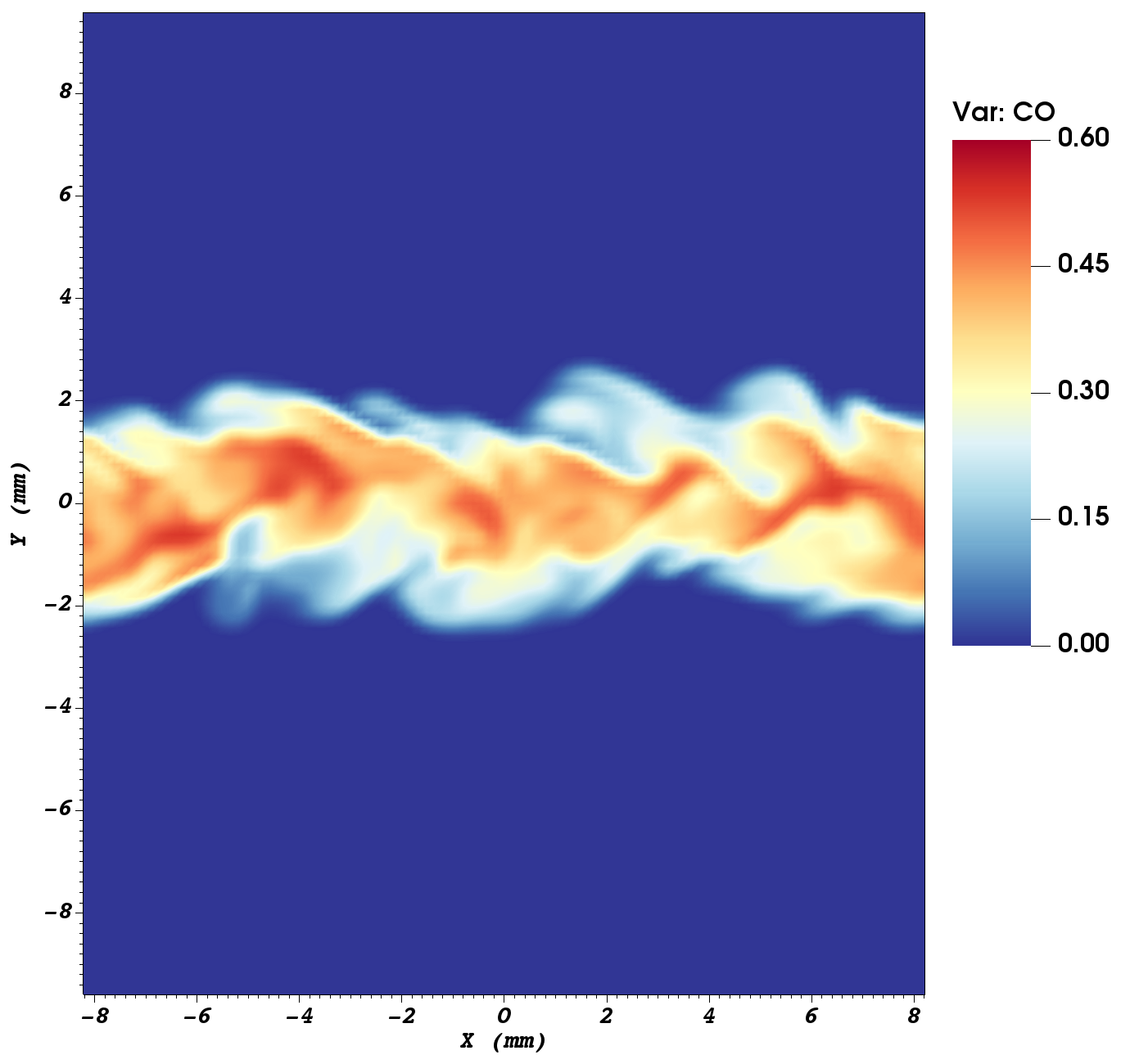}
         \caption{$t=20t_j$}
         \label{fig:COSlices20}
     \end{subfigure}
     
     \begin{subfigure}[b]{\textwidth}
         \centering
         \includegraphics[width=0.3\textwidth]{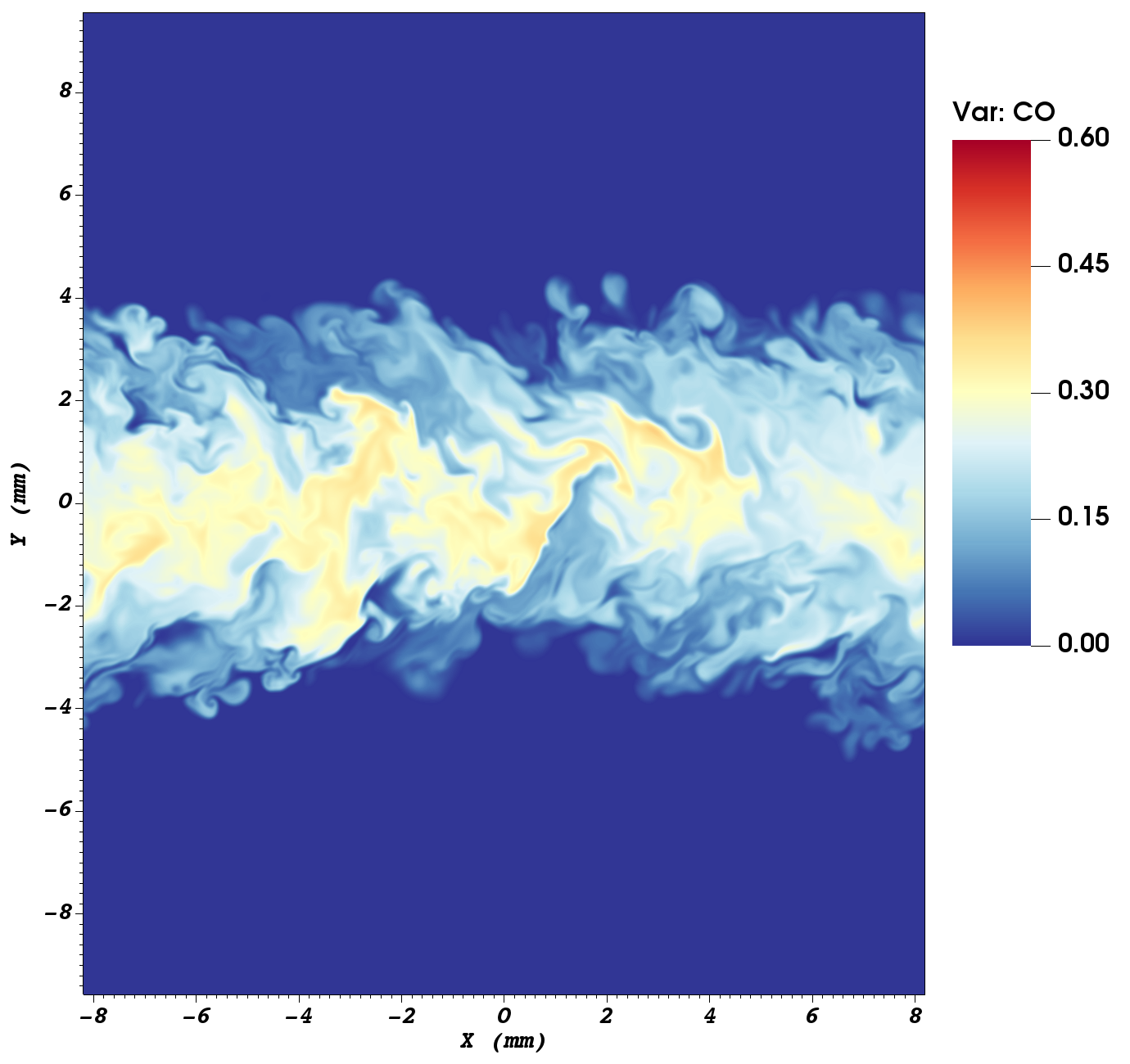}
         \includegraphics[width=0.3\textwidth]{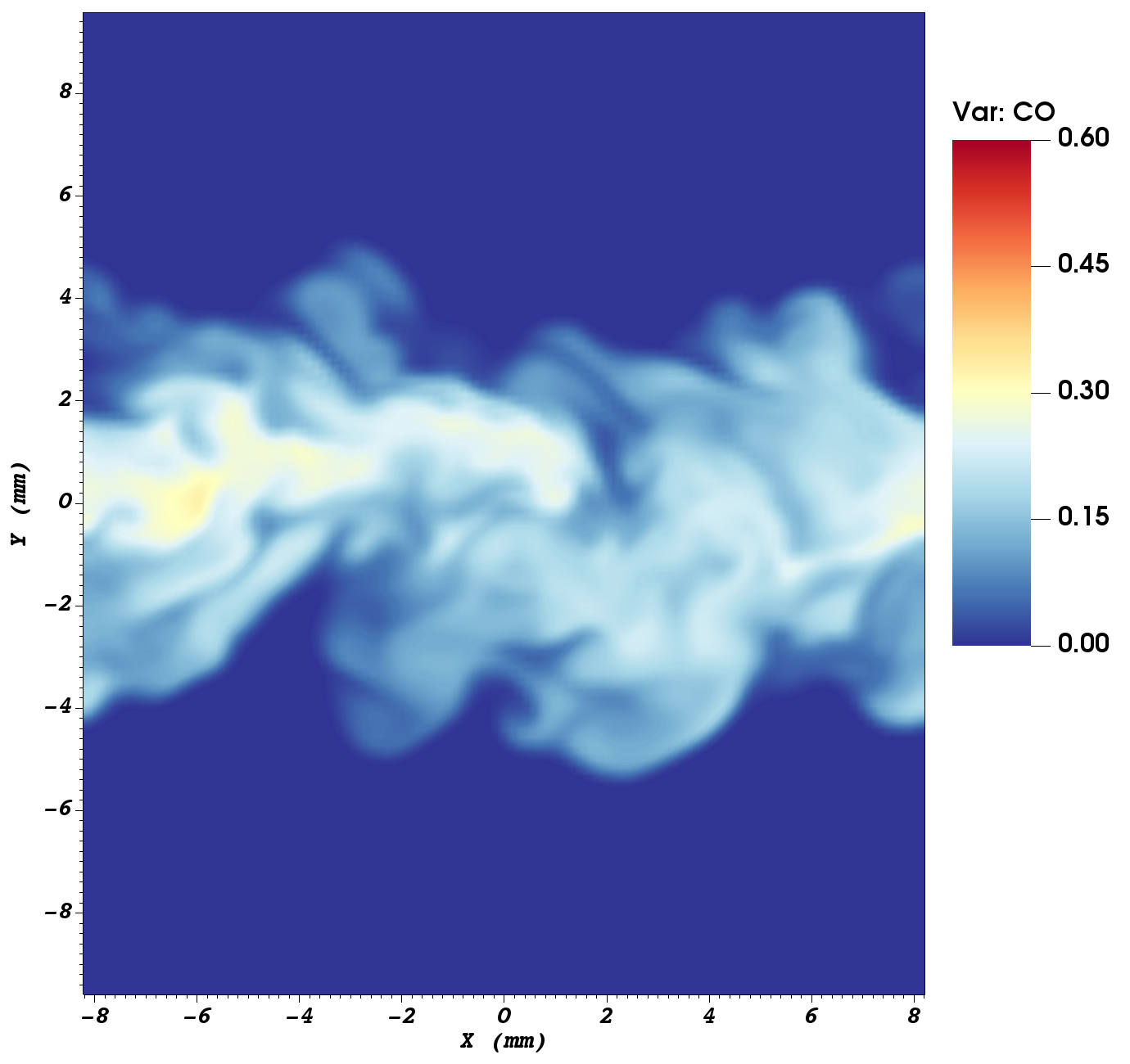}
         \caption{$t=40t_j$}
         \label{fig:COSlices40}
     \end{subfigure}
    
    \caption{Instantaneous slice plots at $z=0$ of CO mass fraction obtained from DNS (left) and LES-FDF (right).}
    \label{fig:COSlices}
\end{figure}

The fidelity of LES predictions are assessed via comparisons with DNS.  This is shown for the first and second Reynolds-moments of the  mixture fraction, the temperature. and mass fractions of major species (CO, CO\textsubscript{2}) at several time levels in \cref{fig:RZTC}. Additionally, 2D slice plots of LES-FDF and DNS are shown in \cref{fig:COSlices} for more detailed view. In all of these cases, the DNS captures more of the small scale features which are filtered out by LES. Therefore, the spreading rate as predicted by LES is somewhat larger than that in DNS. The initial decrease of the  temperature at $t \approx 20t_j$ is an indication of  flame extinction, and its increase at later times $(t \approx 40 t_j)$ signals re-ignition.

\begin{figure}
    \centering
    \includegraphics[width=0.3\textwidth]{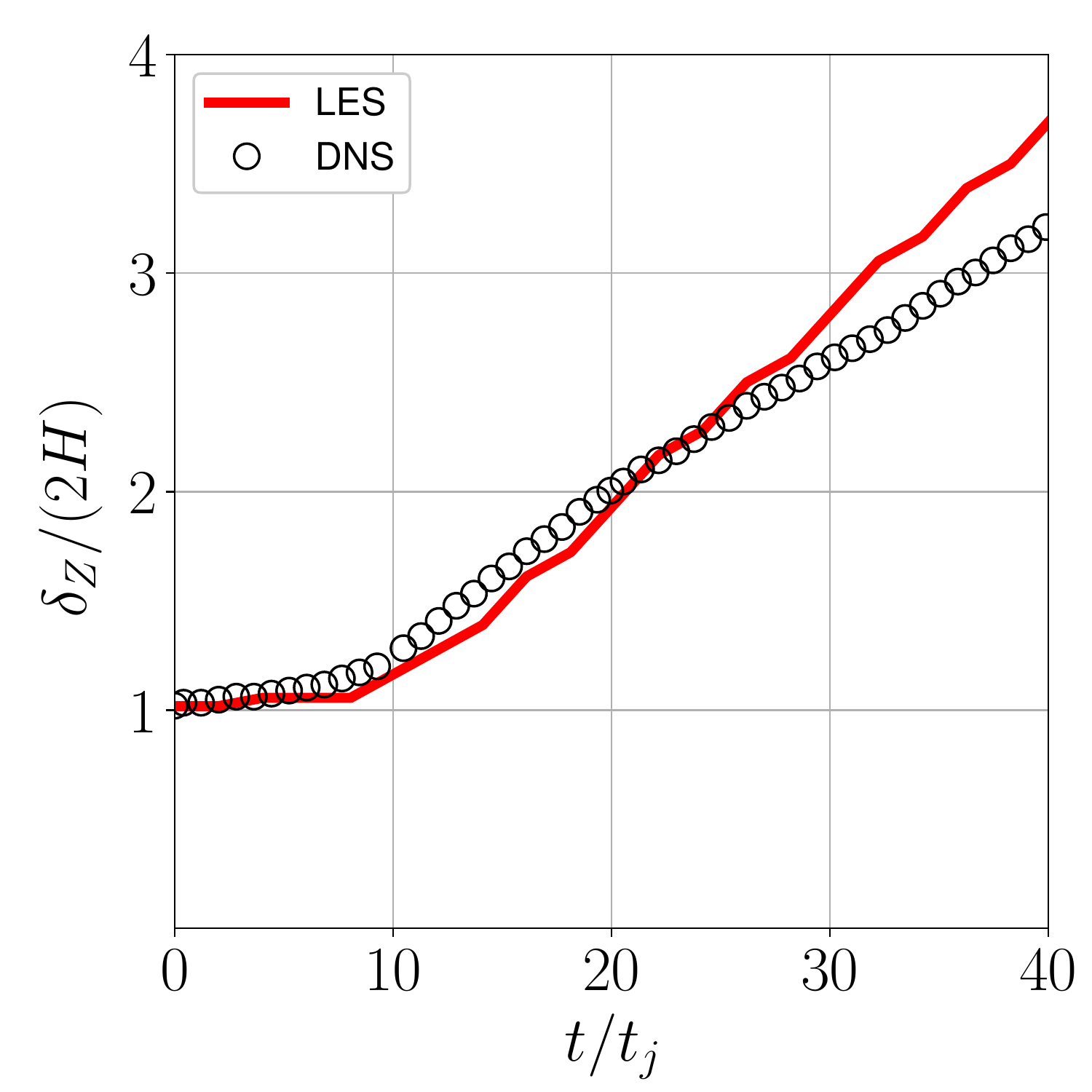}
    \caption{Temporal evolution of mixture fraction thickness. Lines and symbols denote the LES and DNS values, respectively.}
    \label{fig:Zthickness}
\end{figure}

As an evidence of overall layer growth, the mixture fraction thickness is constructed. This thickness in defined as $
\delta_Z = 2 \ {\rm arg \ min} \lp \left| \overline{Z} \lp y \rp - \epsilon \right| \rp$ for $y > 0$, where $\epsilon$ is small positive number. The temporal evolution of this thickness is shown in  \cref{fig:Zthickness}, and indicates that the growth of a turbulent layer predicted by LES is close to that obtained by DNS at initial times. However, as the flow develops the LES predicts a larger spreading of the layer.

\begin{figure}
    \centering

    \includegraphics[trim=0.0cm 1.25cm 0.75cm 0cm, clip=true, height=3.425cm]{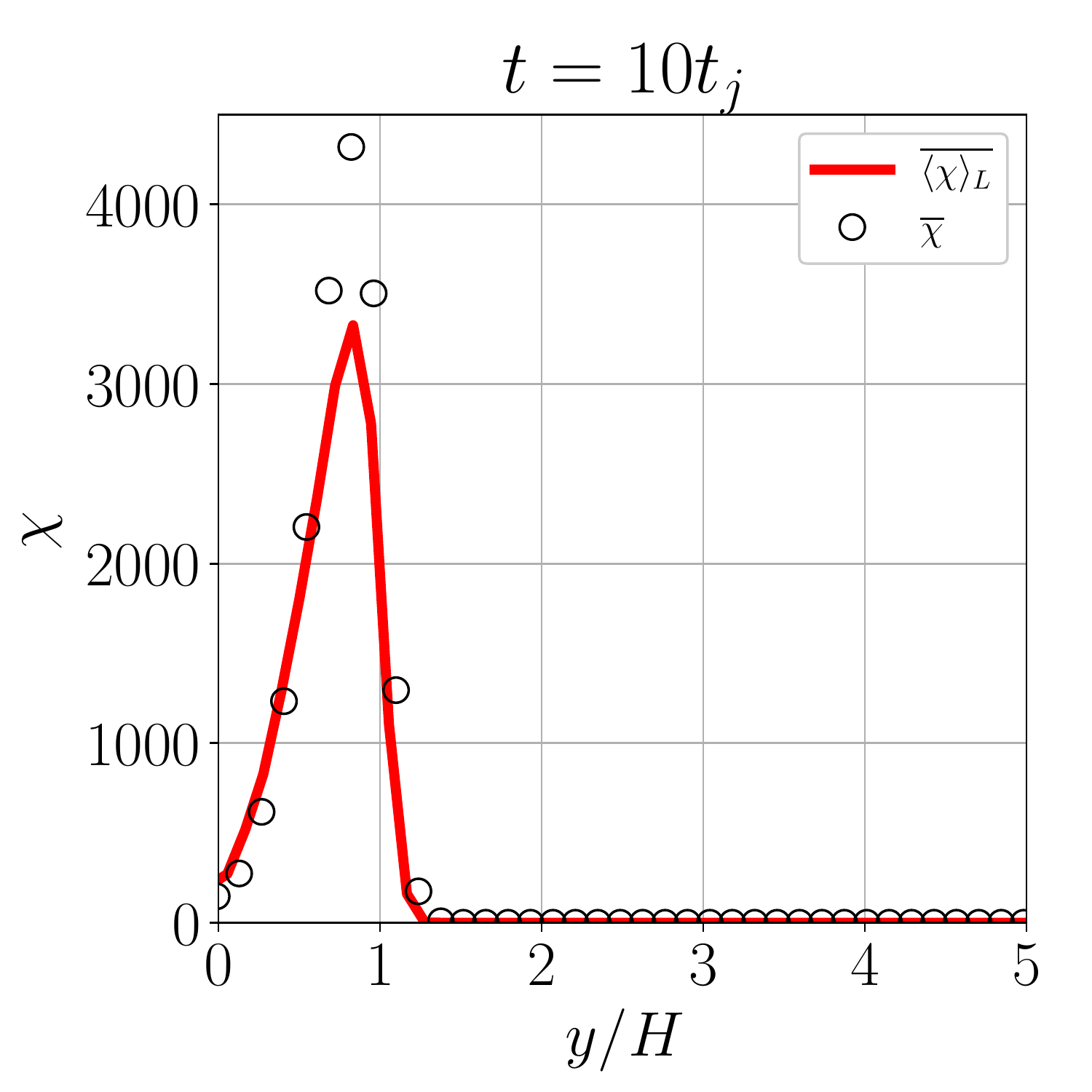} 
    \includegraphics[trim=2.8cm 1.25cm 0.75cm 0cm, clip=true, height=3.425cm]{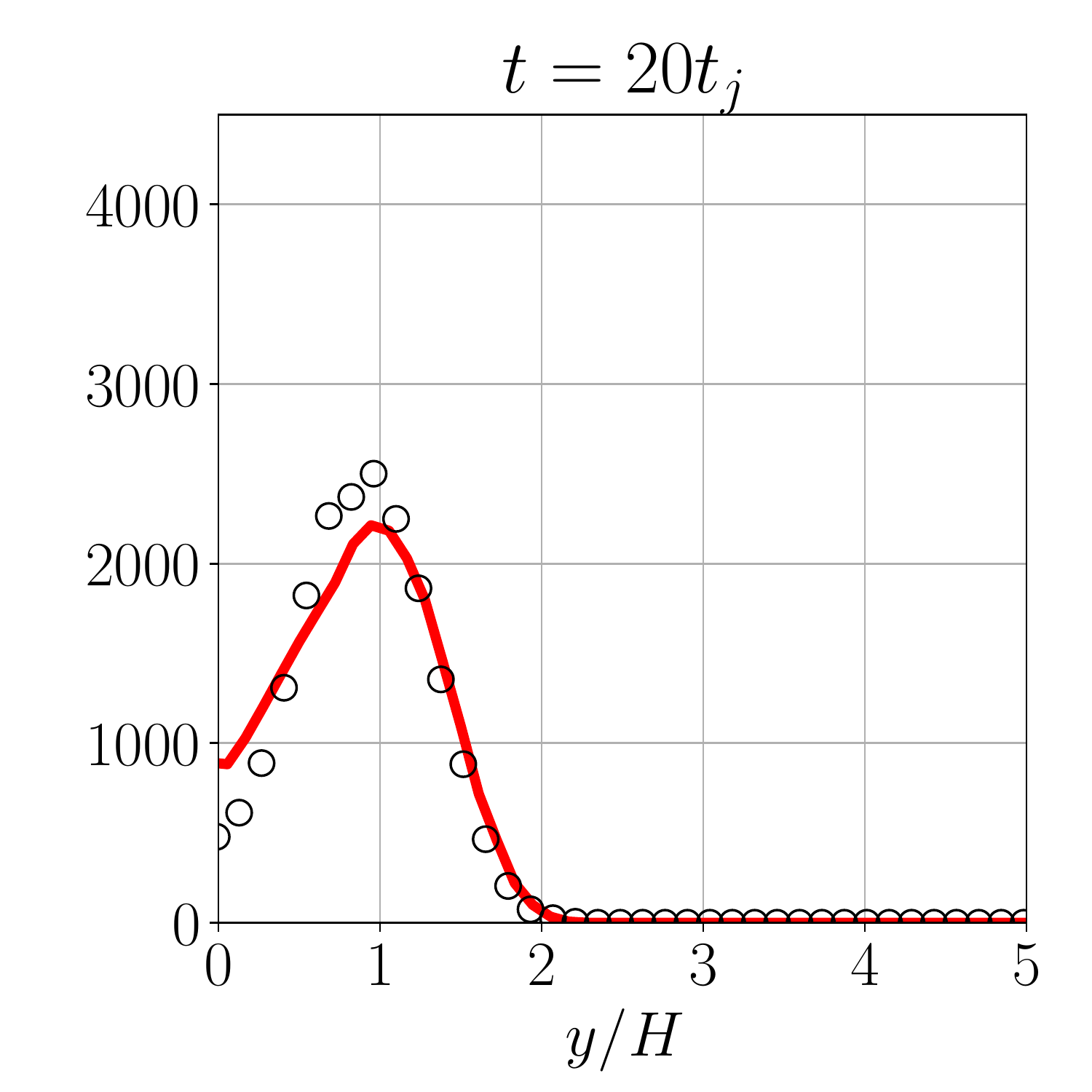}
    \includegraphics[trim=2.8cm 1.25cm 0.75cm 0cm, clip=true, height=3.425cm]{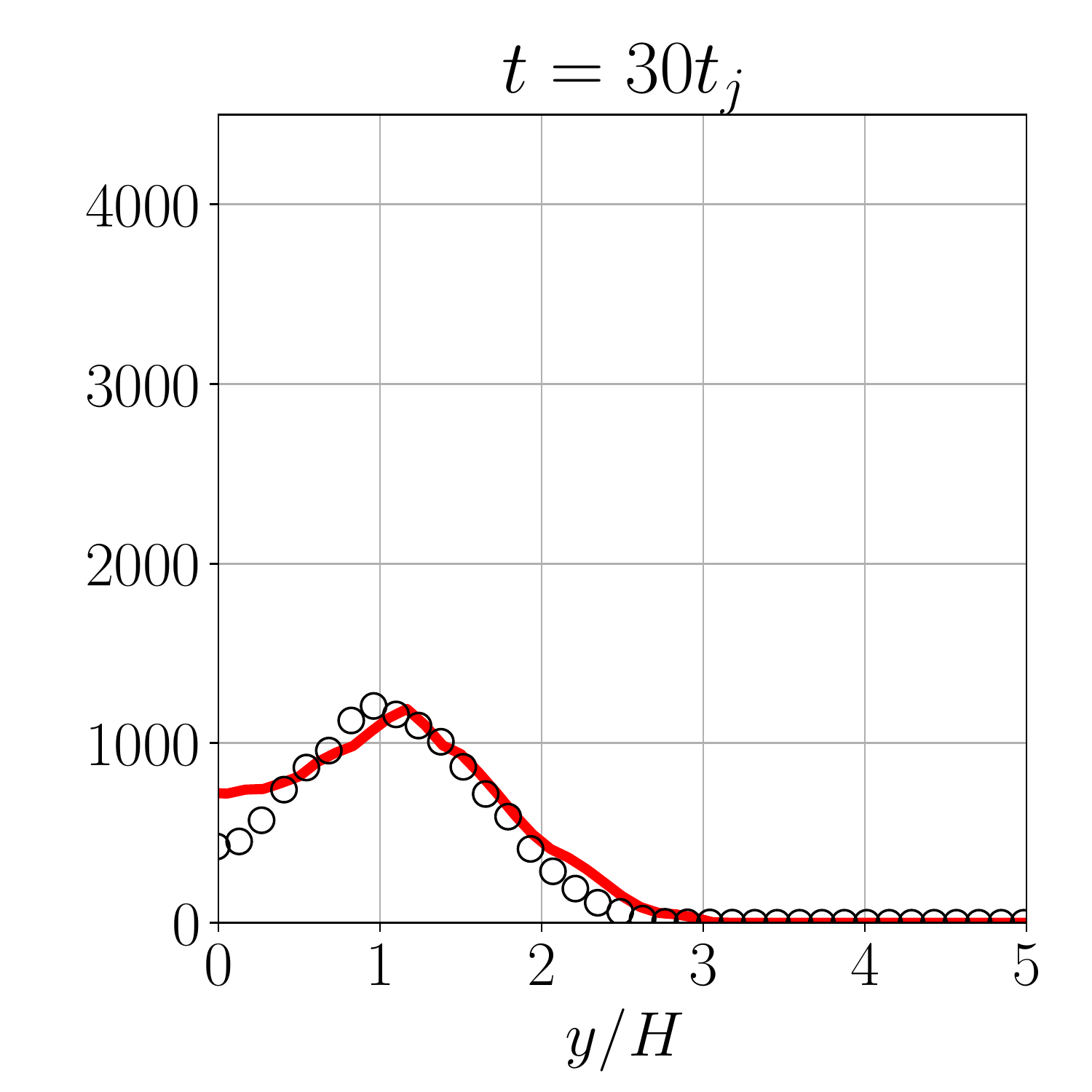}
    \includegraphics[trim=2.8cm 1.25cm 0.75cm 0cm, clip=true, height=3.425cm]{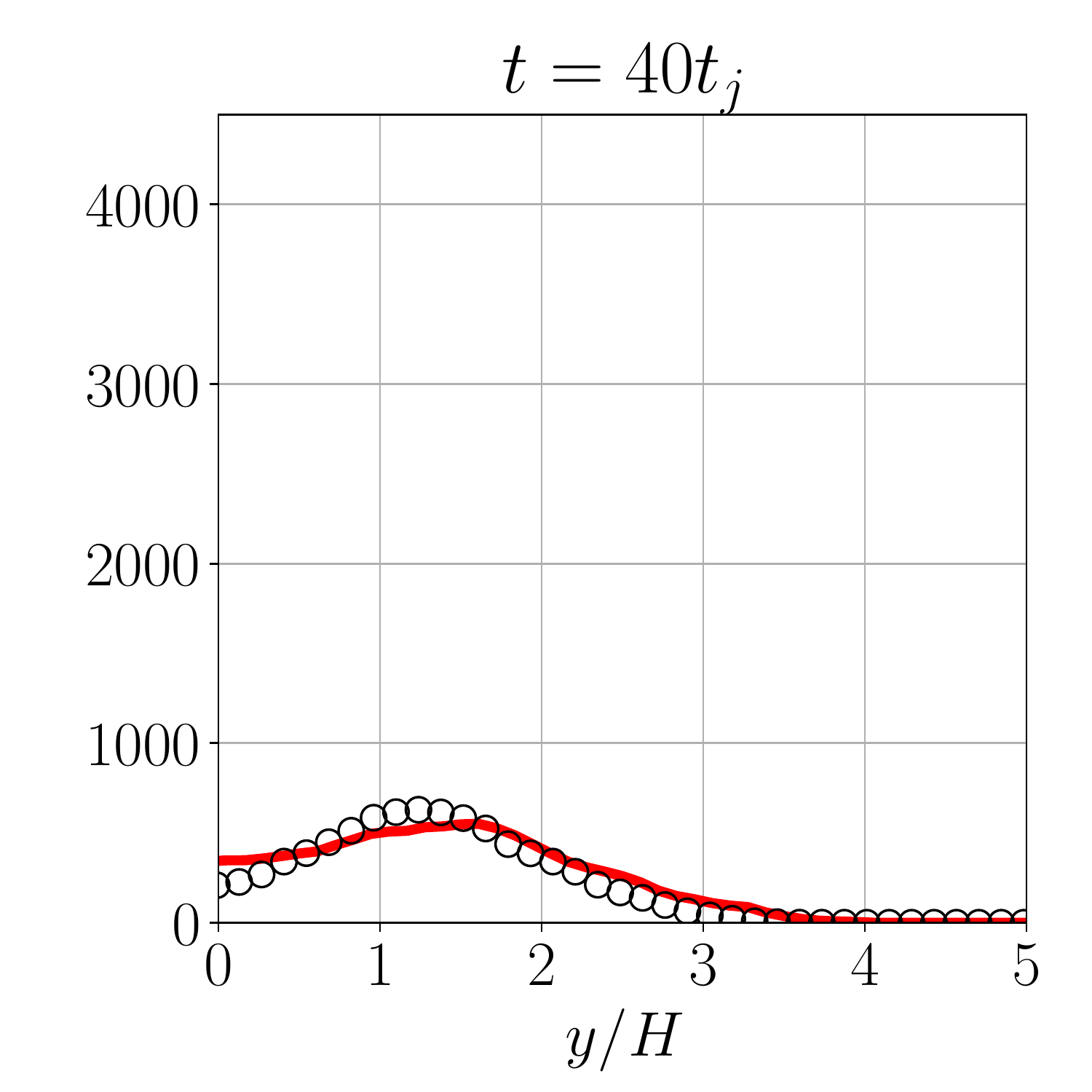}

    \vspace{-0.125cm}
    
    \includegraphics[trim=0.0cm 0cm 0.75cm 14cm, clip=true, width=3.52cm]{Figures/bar_chi_05000.pdf}
    \includegraphics[trim=2.8cm 0cm 0.75cm 14cm, clip=true, width=2.84cm]{Figures/bar_chi_05000.pdf}
    \includegraphics[trim=2.8cm 0cm 0.75cm 14cm, clip=true, width=2.84cm]{Figures/bar_chi_05000.pdf}
    \includegraphics[trim=2.8cm 0cm 0.75cm 14cm, clip=true, width=2.84cm]{Figures/bar_chi_05000.pdf}

    \caption{Reynolds-averaged values of scalar dissipation rate. Lines and symbols denote the LES and DNS values, respectively.}
    \label{fig:RChi}
\end{figure}

\begin{figure}
    \centering

    \includegraphics[trim=0.0cm 1.25cm 0.75cm 0cm, clip=true, height=3.425cm]{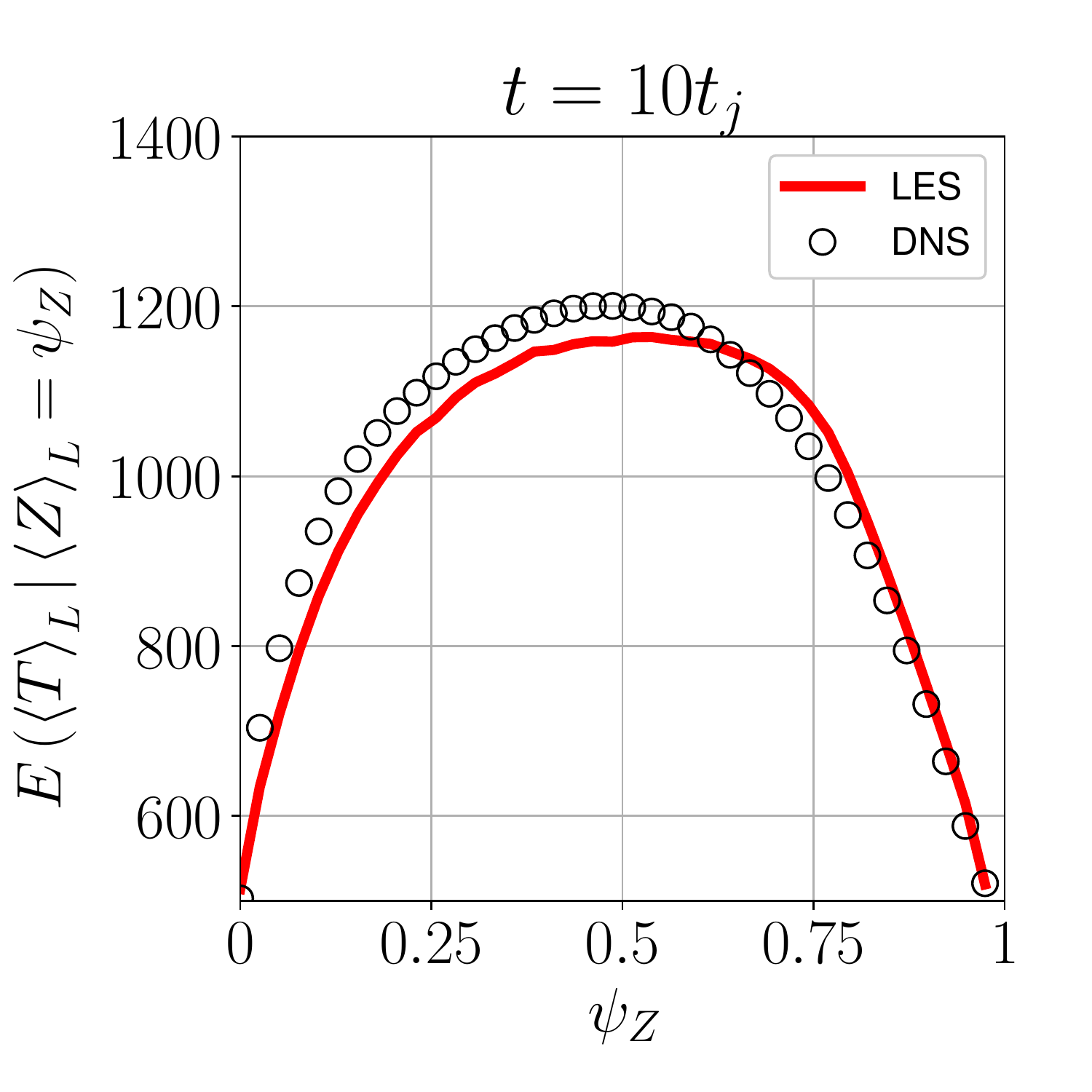} 
    \includegraphics[trim=3.1cm 1.25cm 0.75cm 0cm, clip=true, height=3.425cm]{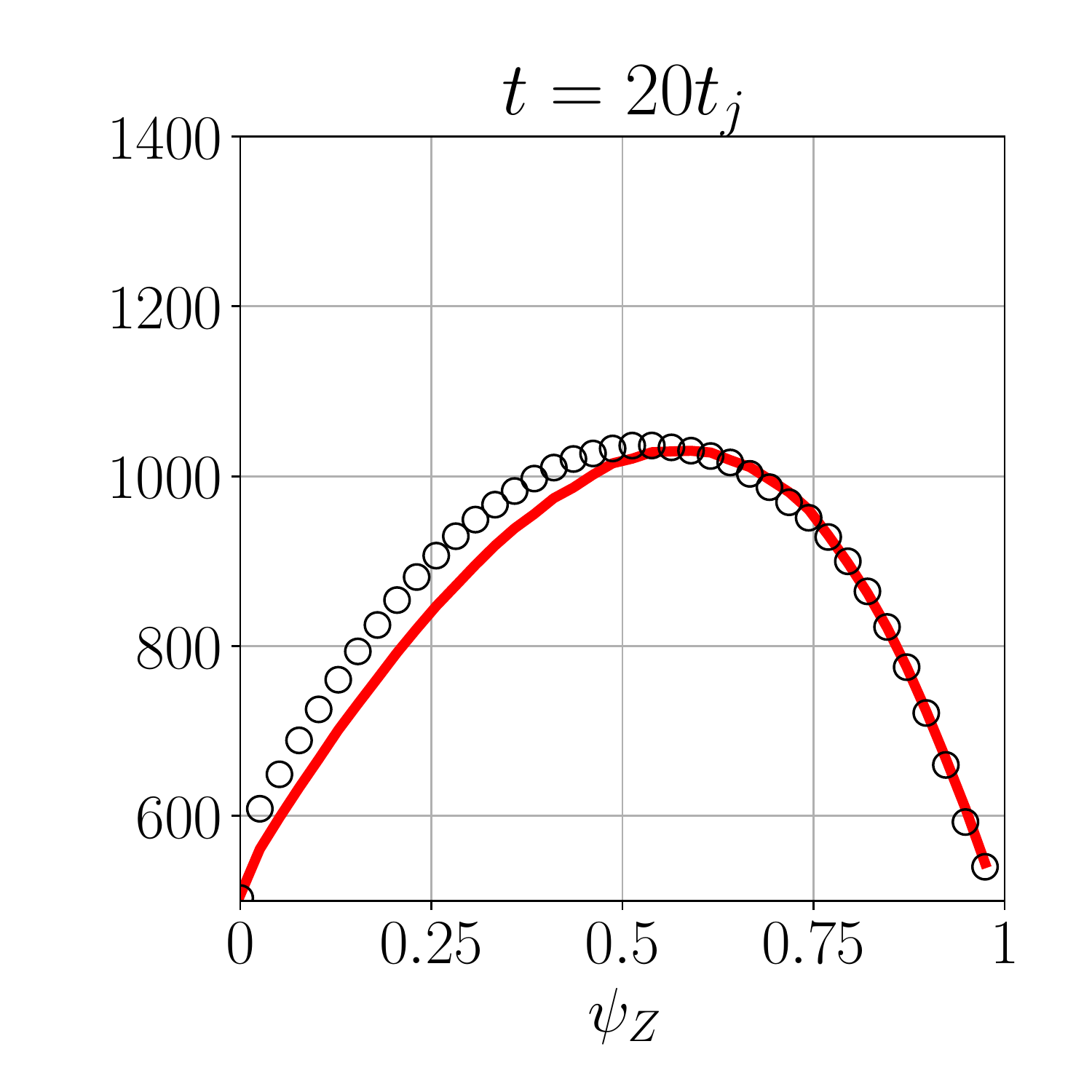}
    \includegraphics[trim=3.1cm 1.25cm 0.75cm 0cm, clip=true, height=3.425cm]{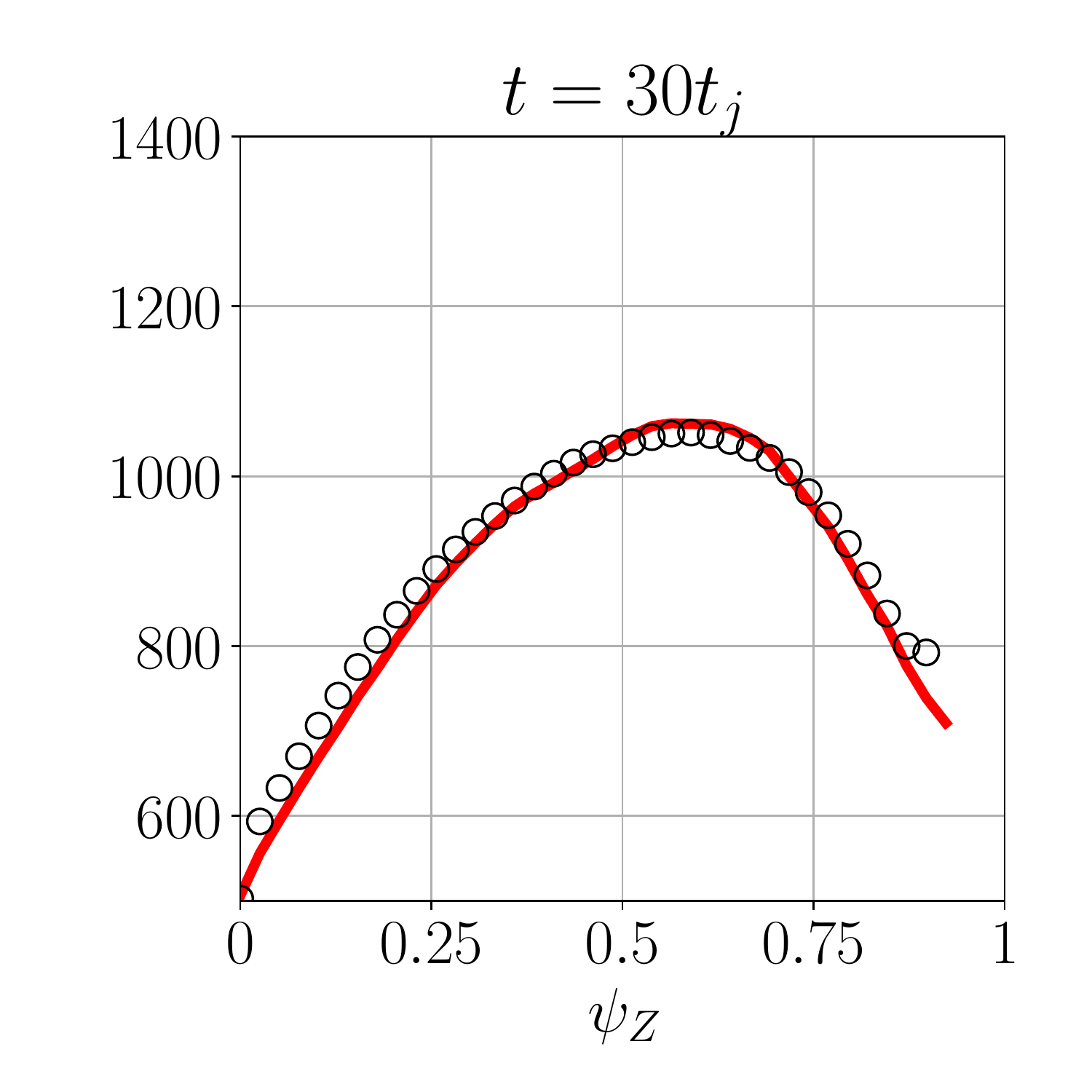}
    \includegraphics[trim=3.1cm 1.25cm 0.75cm 0cm, clip=true, height=3.425cm]{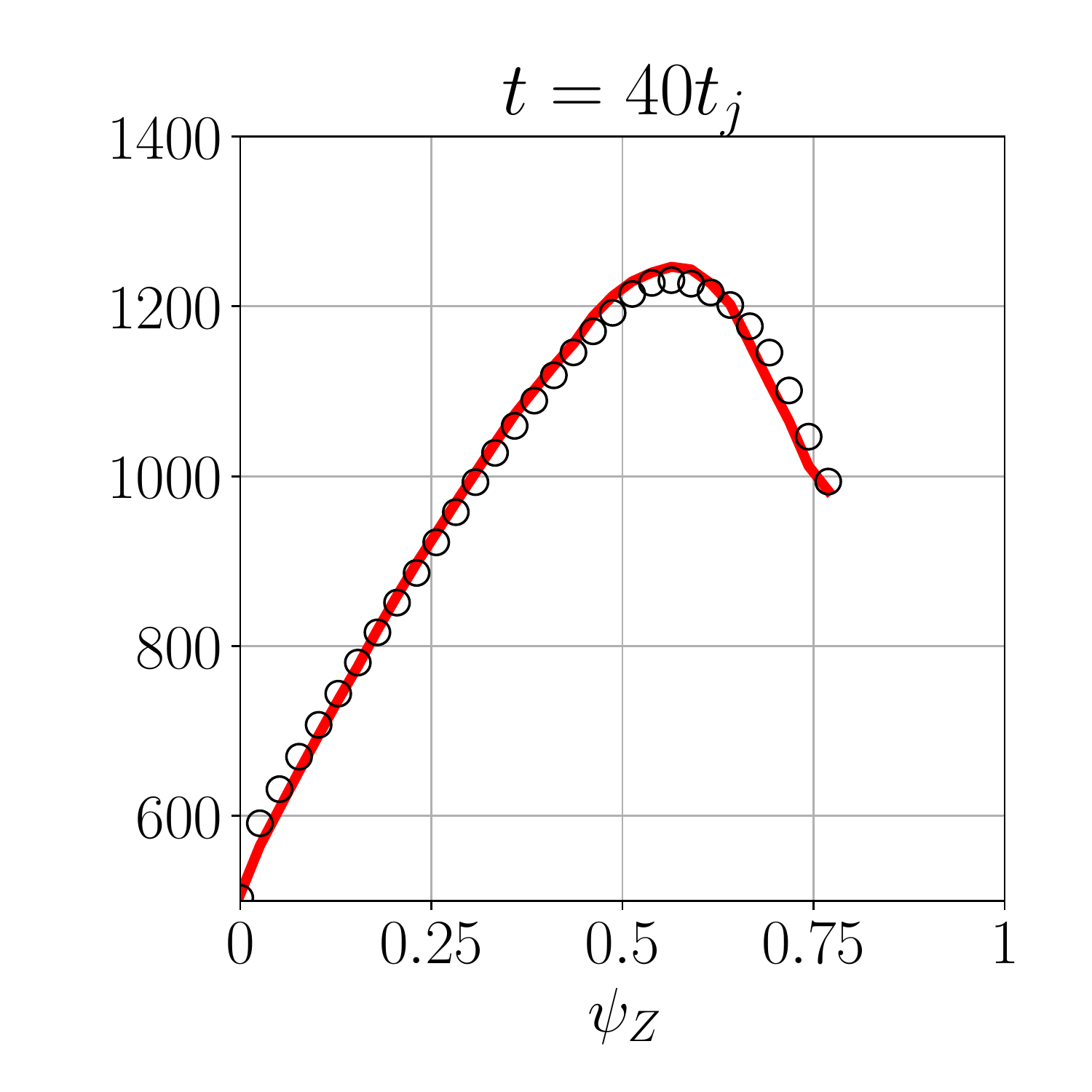}

    \vspace{-0.125cm}
    
    \includegraphics[trim=0.0cm 0cm 0.75cm 14cm, clip=true, width=3.52cm]{Figures/COND_05000.pdf}
    \includegraphics[trim=3.1cm 0cm 0.75cm 14cm, clip=true, width=2.84cm]{Figures/COND_05000.pdf}
    \includegraphics[trim=3.1cm 0cm 0.75cm 14cm, clip=true, width=2.84cm]{Figures/COND_05000.pdf}
    \includegraphics[trim=3.1cm 0cm 0.75cm 14cm, clip=true, width=2.84cm]{Figures/COND_05000.pdf}

    \caption{Volume averaged temperature conditioned on mixture fraction. Lines and symbols denote the LES and DNS values, respectively}
    \label{fig:COND}
\end{figure}

The flame extinction phenomenon and its subsequent re-ignition is explained in terms of the  dissipation of the mixture fraction: $\chi = 2 \gamma / \rho \nabla Z \cdot \nabla Z$ \cite{Peters00, Givi1986FlameExtinction}.  The Reynolds-averaged values of this dissipation,  implicitly modelled here as: $\la \chi \ra_{L} = 2 \Omega \tau\left(Z,Z\right)$  are shown in \cref{fig:RChi}. All of the predicted results agree very  well with DNS measured data.  At initial times, when the dissipation rates are   large, the flame  cannot be sustained  and is locally extinguished.  At later times, when the dissipation values are lowered, the flame is re-ignired and the temperature  increases.  This dynamic is more clearly depicted in \cref{fig:COND}, where the  expected temperature values conditioned on the mixture fraction   are shown.  By $t=20t_j$ the temperature at the stoichimetric   mixture fraction ($Z_{st} = 0.42$) decreases from $T = 1400 K$,  stays below extinction limit for a while, and then rises after $t=30 t_j$. The agreement with DNS data for this conditional expected value is very good.  

\begin{figure}
    \centering
     \begin{subfigure}[b]{0.3\textwidth}
         \centering
         \includegraphics[width=\textwidth]{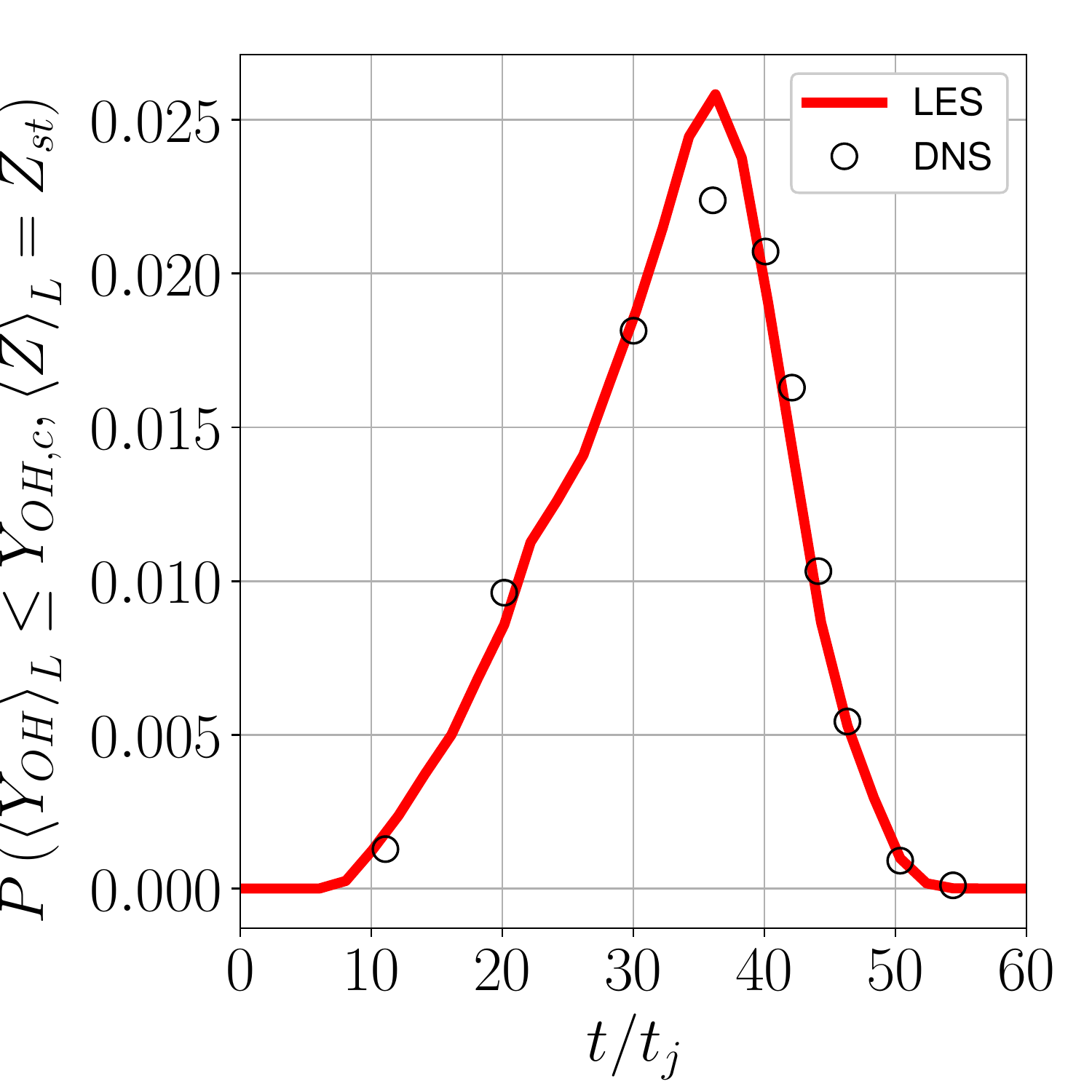}
         \caption{Volume averaged extinction marker}
         \label{fig:PEXT}
     \end{subfigure}
     \hspace{1cm}
     \begin{subfigure}[b]{0.3\textwidth}
         \centering
         \includegraphics[width=\textwidth]{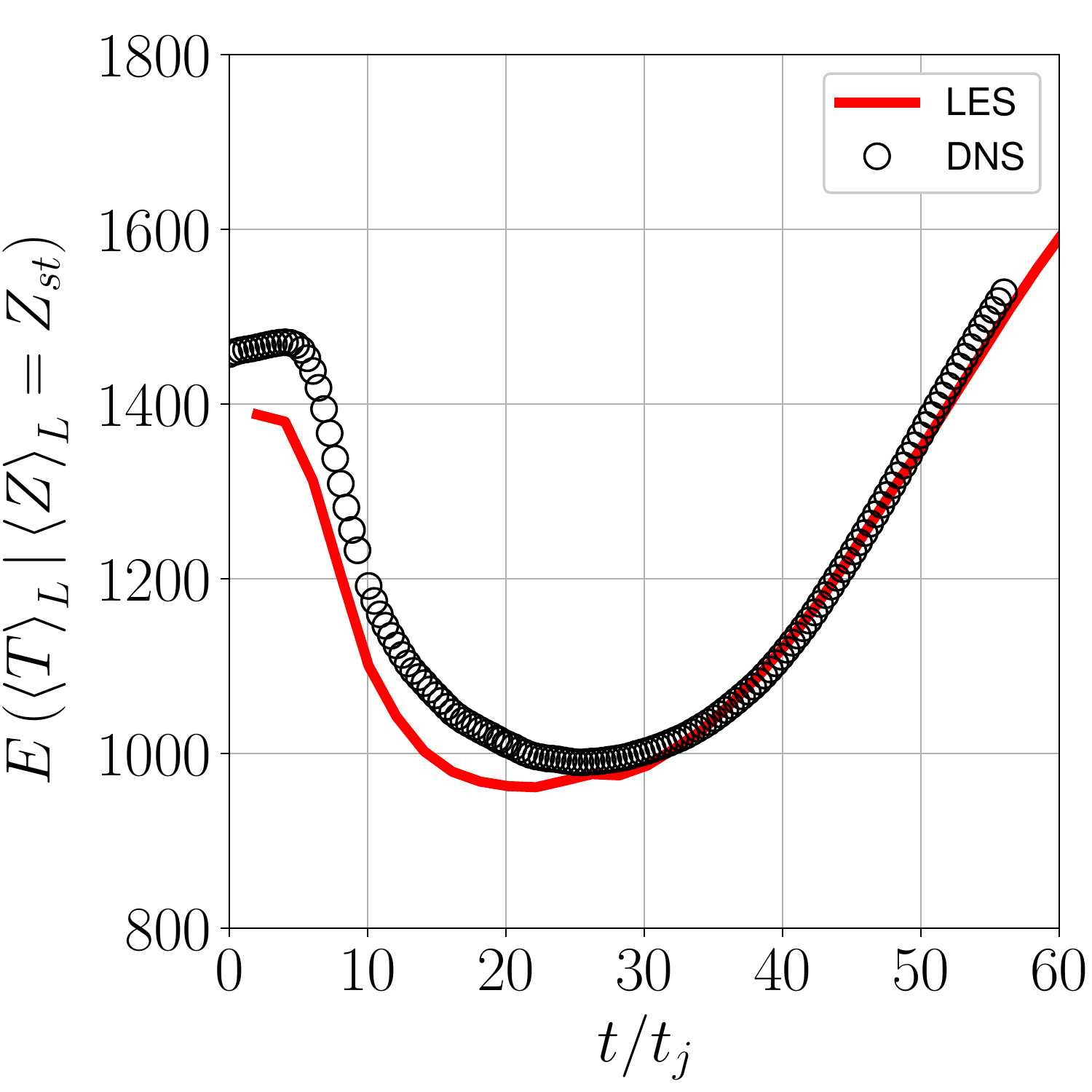}
         \caption{Mean temperature on the stoichiometric surface}
         \label{fig:CONDZST}
     \end{subfigure}
     \caption{Temporal evolution of average extinction and re-ignition. Lines and symbols denote the LES and DNS values, respectively.}

    \label{fig:TEMPORAL}
\end{figure}

To provide a more quantitative assessment of the flame structure within the entire domain,  an  ``extinction marker'' is defined: $M_{ext} = \left( H\left(Y_{OH} - Y_{OH, c} \right) | Z = Z_{st}\right) $ \cite{hawkes2007reignition}. Here  $Y_{OH, c} = 0.0007$ is a cut-off mass fraction of hydroxyl radical and $H\left(x\right)$ denotes the Heaviside  function. The choice of OH mass fraction as a main scalar used in marker is made upon a visual inspection of the fields of heat release. (A video-clip is provided in Supplementary Materials.) The volume averaged extinction marker defines the probability of a point experiencing extinction$\frac{1}{V} \int_{V} M_{ext} dV= P\left(Z = Z_{st}, Y_{OH} \le Y_{OH, c} \right) $ and its evolution over time is shown in \cref{fig:PEXT}. The excellent agreement between LES and DNS on the figure indicates that the timings of extinction and re-ignition as predicted by LES are accurate. The temporal evolution of the expected temperature conditioned on the stoichiometric mixture fraction in \cref{fig:CONDZST} corroborates the onset of extinction due to high dissipation and the subsequent  re-ignitionl at low dissipation. The  increase of temperature at final times is accompanied by  $Y_{CO2}$ production and $Y_{CO}$ consumption at later times, as observed  in \cref{fig:RZTC}.

\begin{figure}
    \centering

    \includegraphics[trim=0.0cm 1.25cm 0.75cm 0cm, clip=true, height=3.425cm]{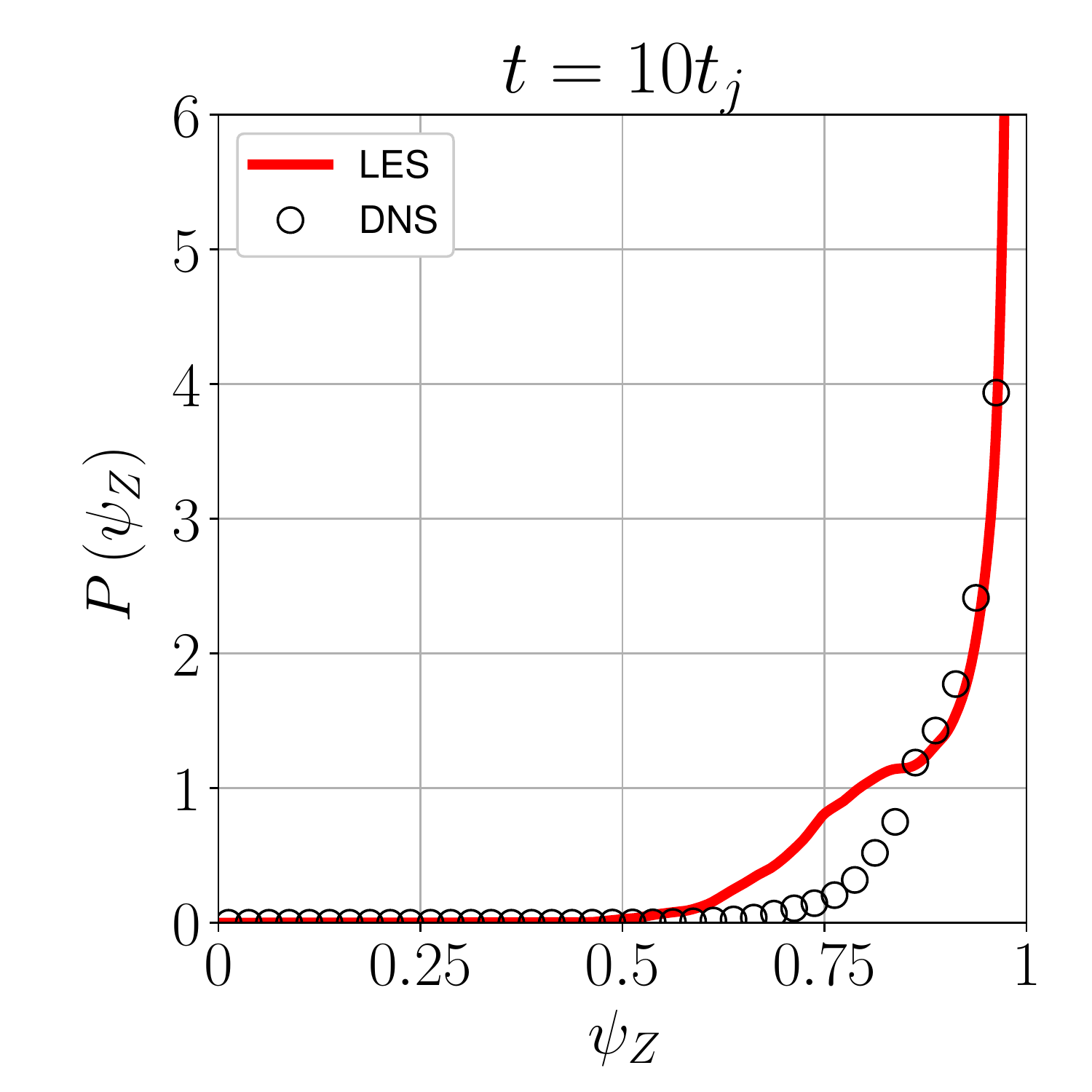} 
    \includegraphics[trim=2.8cm 1.25cm 0.75cm 0cm, clip=true, height=3.425cm]{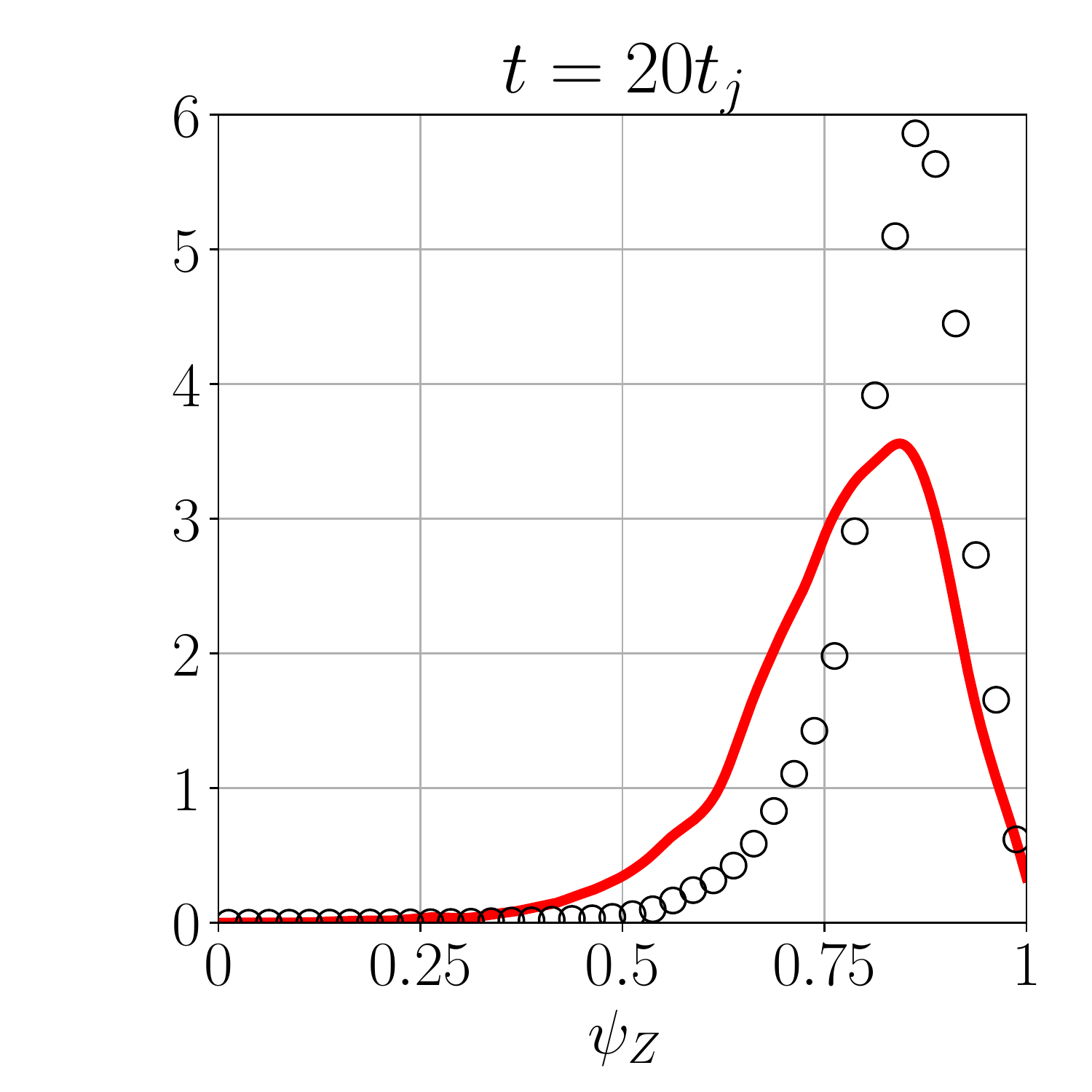}
    \includegraphics[trim=2.8cm 1.25cm 0.75cm 0cm, clip=true, height=3.425cm]{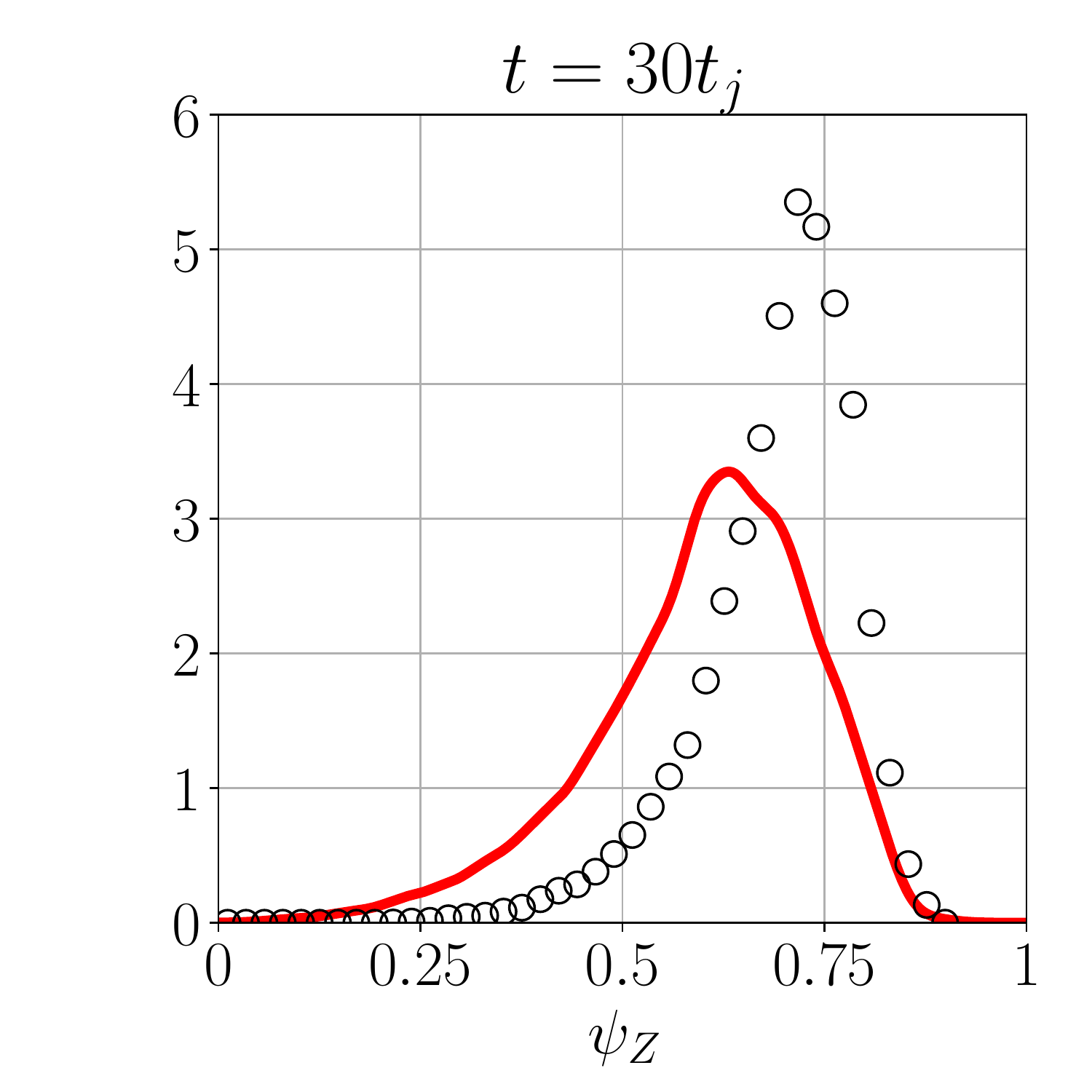}
    \includegraphics[trim=2.8cm 1.25cm 0.75cm 0cm, clip=true, height=3.425cm]{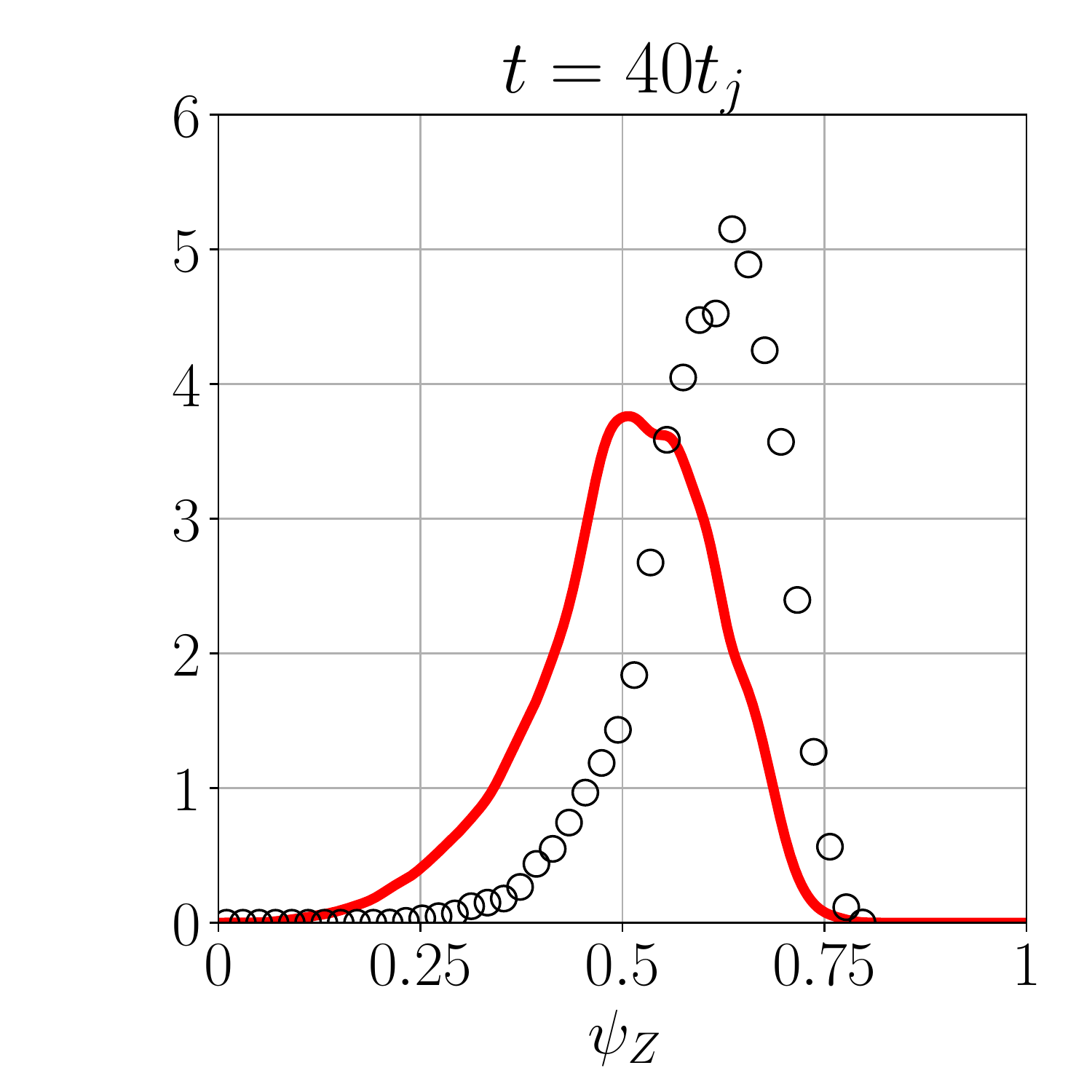}

    \vspace{-0.125cm}
    
    \includegraphics[trim=0.0cm 0cm 0.75cm 14cm, clip=true, width=3.52cm]{Figures/PDF_05000.pdf}
    \includegraphics[trim=2.8cm 0cm 0.75cm 14cm, clip=true, width=2.84cm]{Figures/PDF_05000.pdf}
    \includegraphics[trim=2.8cm 0cm 0.75cm 14cm, clip=true, width=2.84cm]{Figures/PDF_05000.pdf}
    \includegraphics[trim=2.8cm 0cm 0.75cm 14cm, clip=true, width=2.84cm]{Figures/PDF_05000.pdf}

    \caption{Probability density function of mixture fraction about $y=0$ plane. Lines and symbols denote the LES and DNS values, respectively.}
    \label{fig:PDF}
\end{figure}

\begin{figure}
    \centering
    \includegraphics[width=0.3\textwidth]{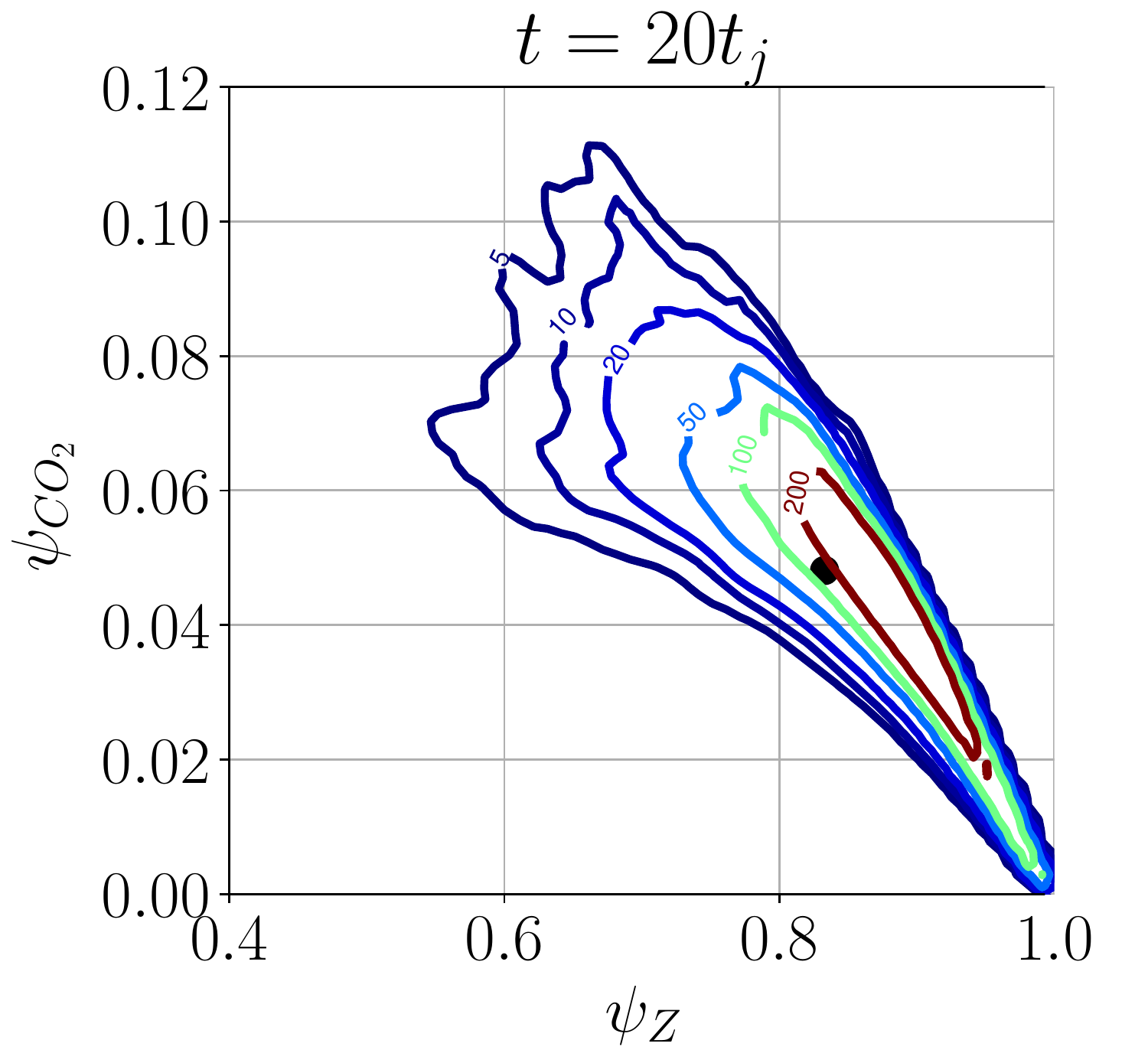}
    \includegraphics[width=0.3\textwidth]{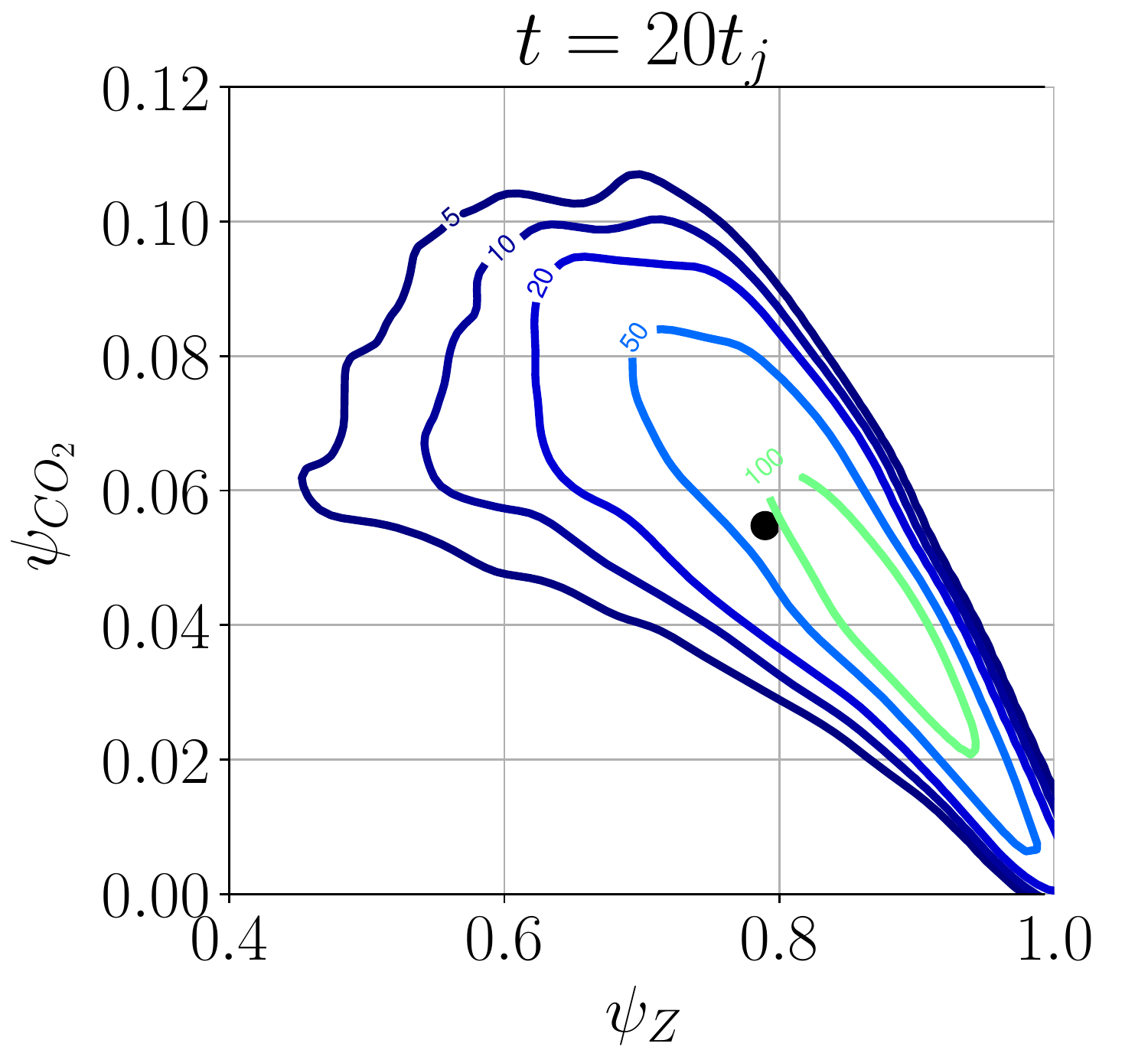}

    \includegraphics[width=0.3\textwidth]{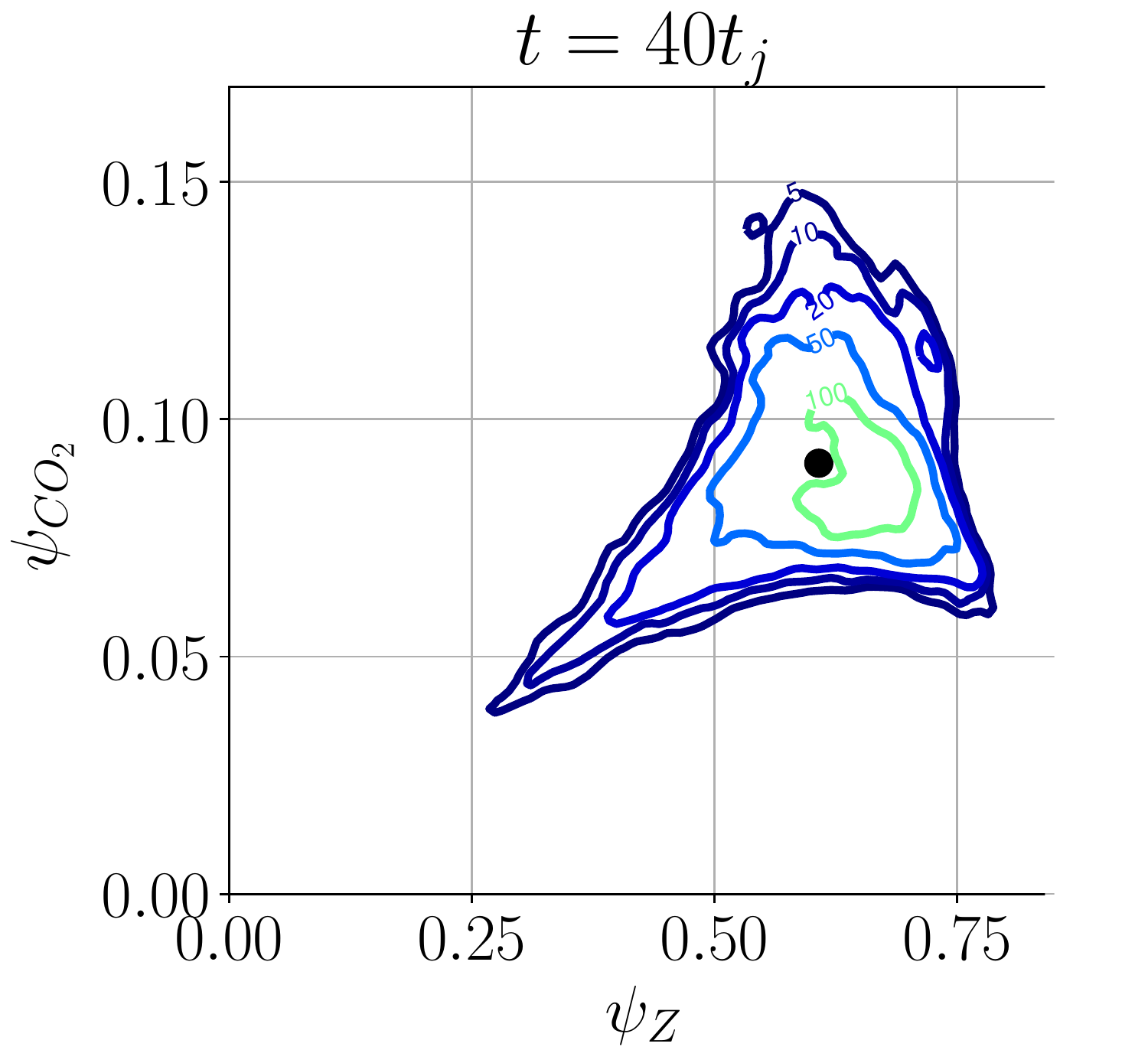}
    \includegraphics[width=0.3\textwidth]{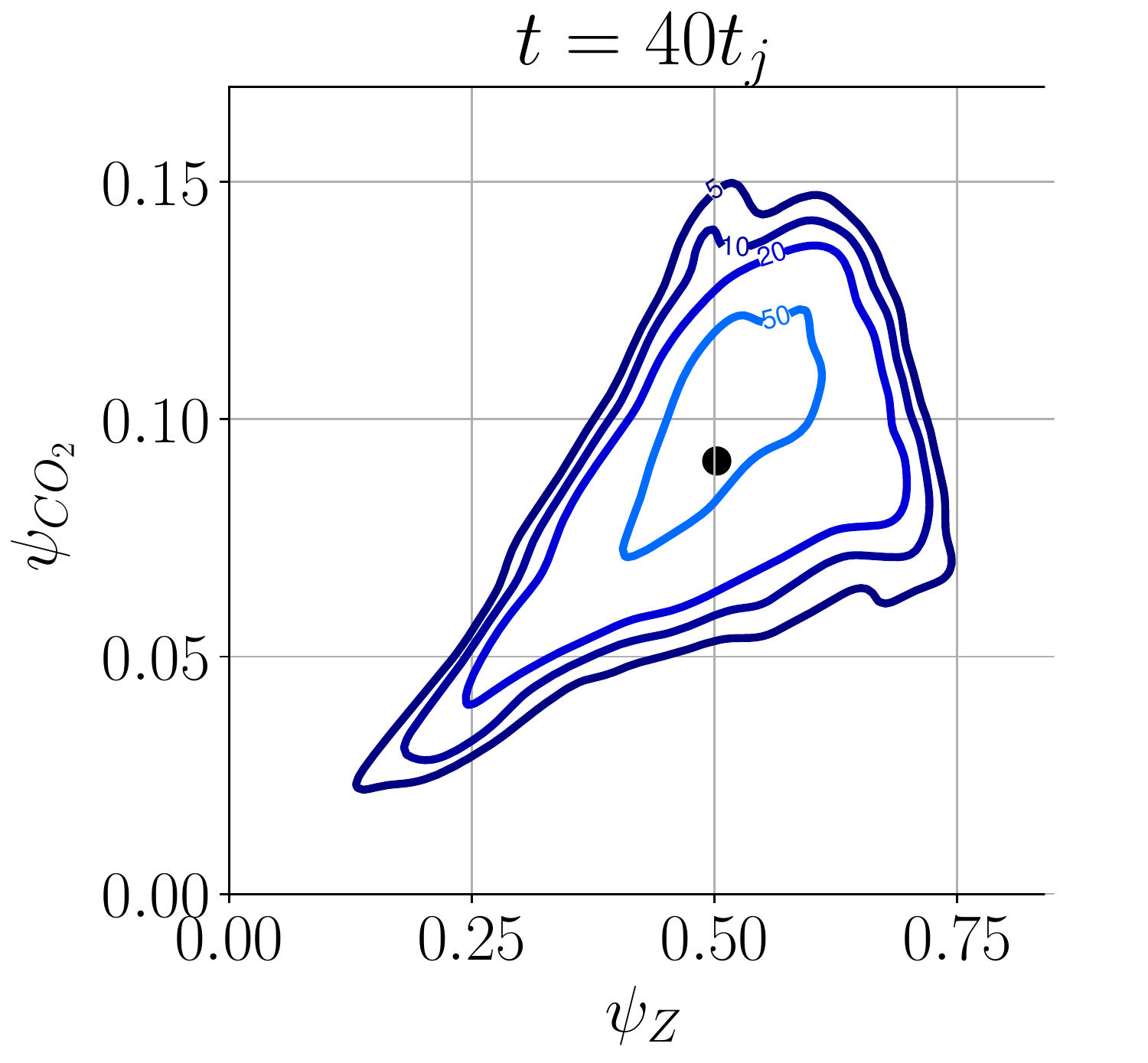}
    \caption{Joint probability density functions of mixture fraction and $Y_{CO_2}$ about  $y=0$ plane of DNS (left) and LES (right).}
    \label{fig:JPDF}
\end{figure}

\begin{figure}
    \centering
        \begin{subfigure}[b]{\textwidth}
        \centering
        \includegraphics[width=0.3\textwidth]{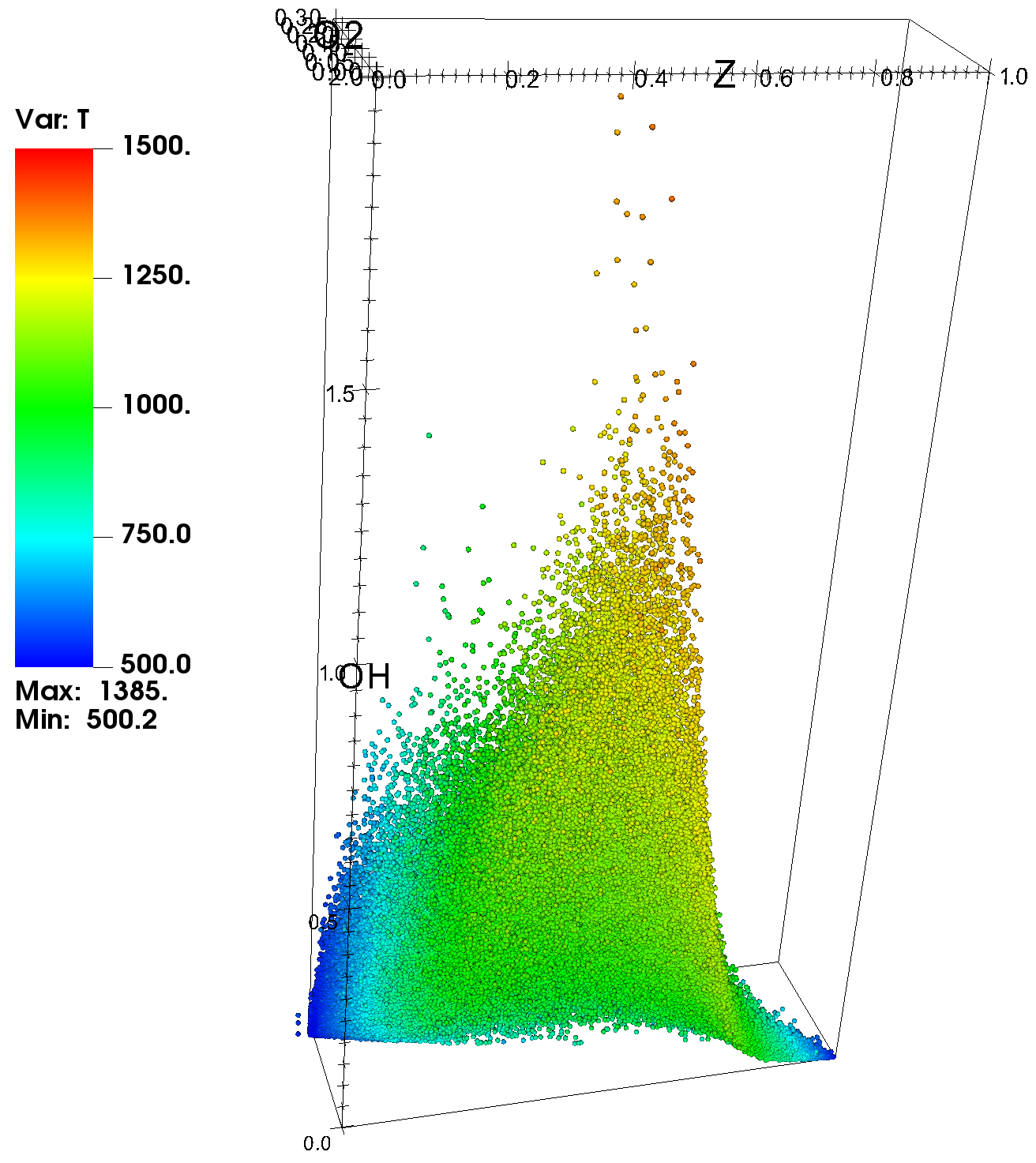}
        \includegraphics[width=0.3\textwidth]{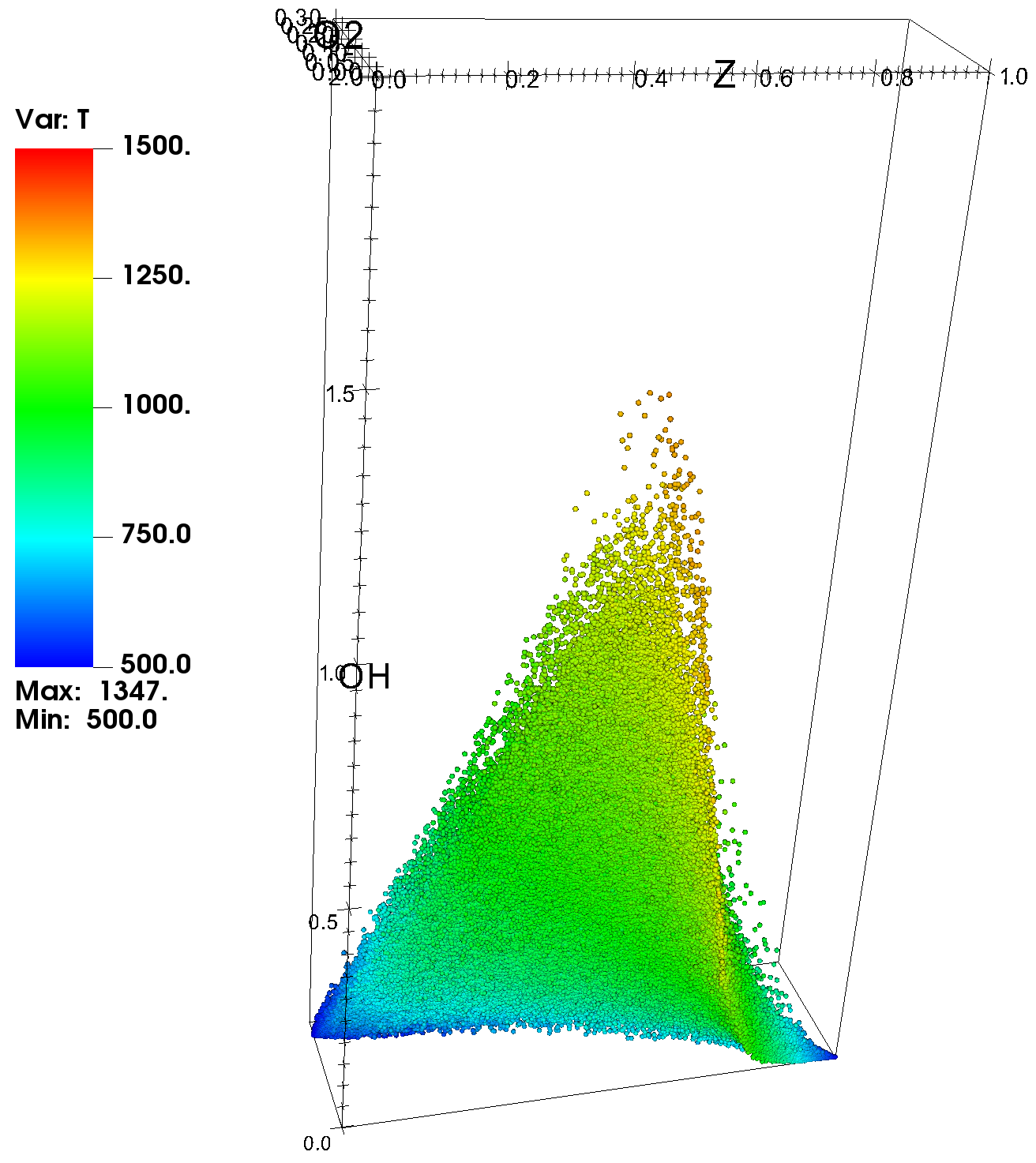}
        \caption{$t=20tj$}
        \label{fig:3DScatter_20}
    \end{subfigure}

    \begin{subfigure}[b]{\textwidth}
        \centering
        \includegraphics[width=0.3\textwidth]{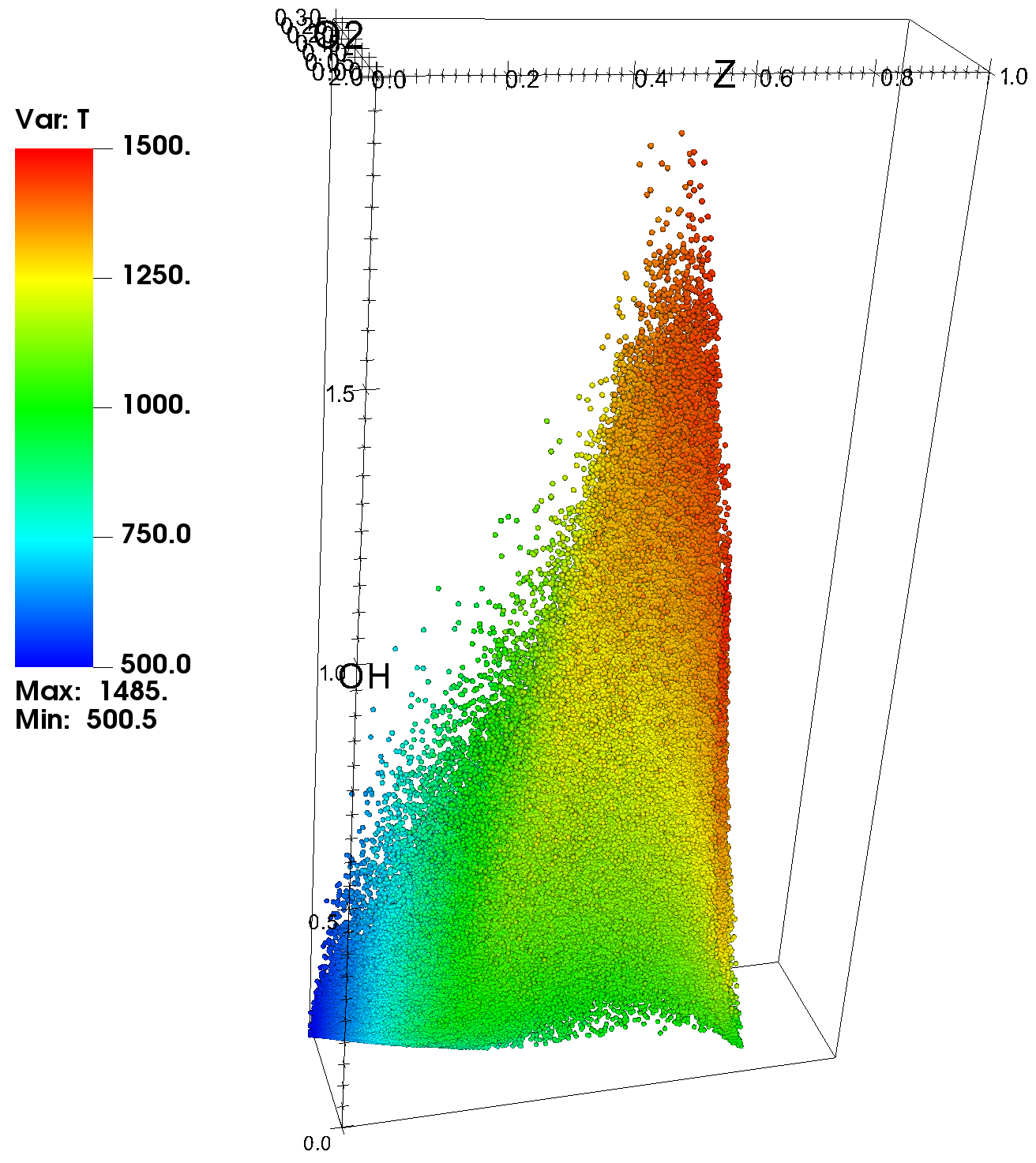}
        \includegraphics[width=0.3\textwidth]{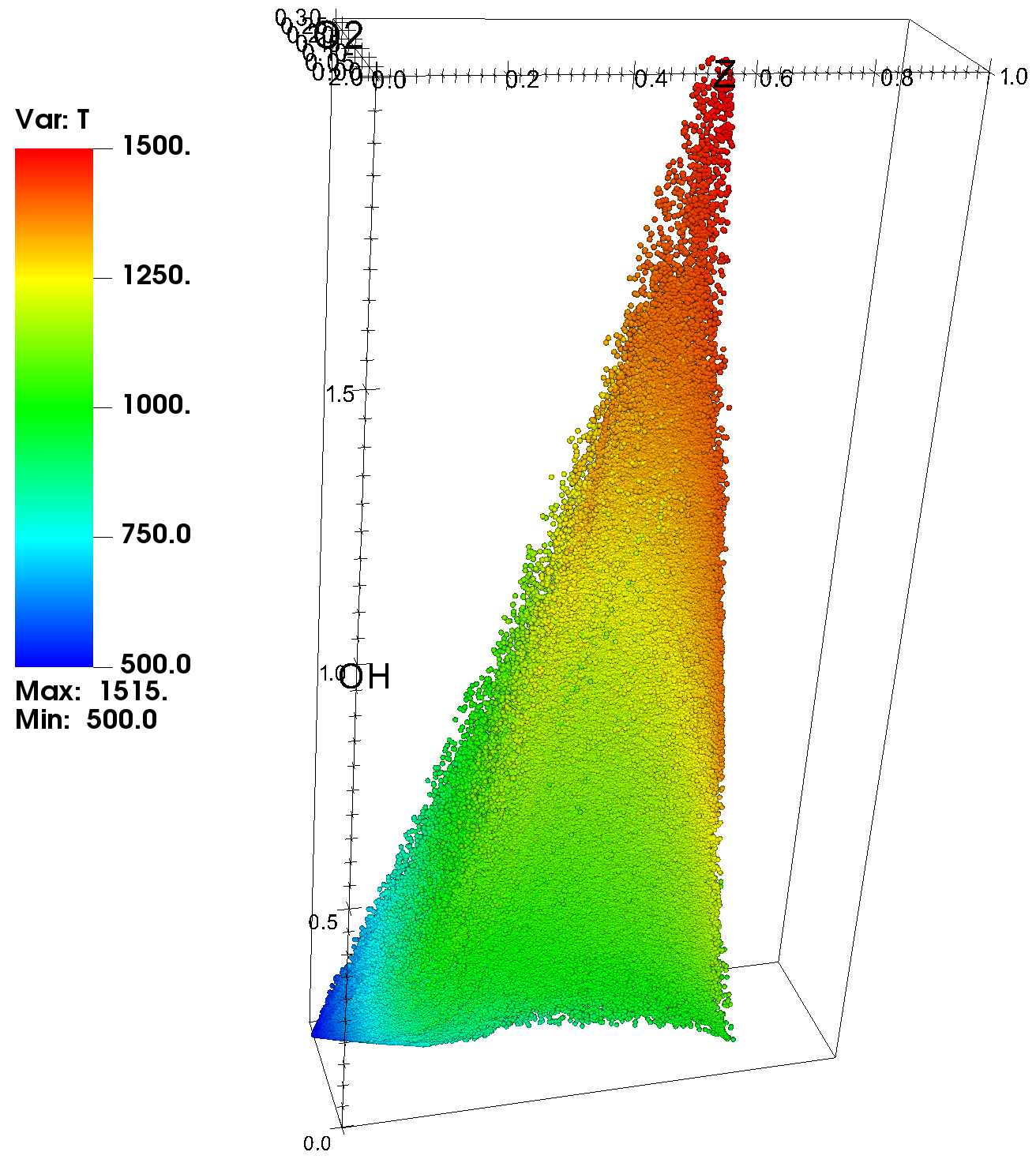}
        \caption{$t=40tj$}
        \label{fig:3DScatter_40}
    \end{subfigure}

    \caption{Scatter plot of mixture fraction $Z$, oxidant mass fraction $Y_{O_2}$, and hydroxyl radical mass fraction $Y_{OH} \times 1000$ colored by temperature of DNS (left) and LES (right).}
    \label{fig:3DScatter}
\end{figure}

A more comprehensive comparison with DNS is done by examination of the mixture fraction PDFs in  \cref{fig:PDF}.  In DNS these PDF generated by sampling of $N_{x, DNS} \times 8 \times N_{z, DNS}$ near the center-plane ($\left| y \right| < \Delta y $) of the jet (8 cross-stream planes)).  The LES generated PDFs are based on sampling of $N_{x} \times 2 \times N_{z}$ (2 cross-stream planes). While the two sets of PDFs are qualitatively the same, there are some small quantitative differences. The DNS generated PDFs tend to be concentrated near the higher mixture fraction values.  This is consistent with the observations made in \cref{fig:RZTC}, indicating a higher jet spreading rate in LES.  However, the width of the PDFs are the same,  consistent with the RMS values shown in \cref{fig:RZTC}.  To portray the dynamics of multi-scalar mixing and reaction, the joint PDFs of the scalar variables must be considered. The mixture fraction and the mass fraction of the CO\textsubscript{2} are considered, and the results are shown in  \cref{fig:JPDF}. In both cases, as the flow becomes fully turbulent at $t=40t_j$, the PDFs tend to have a multi-variate Gaussian distribution. In all cases, the  LES predicted PDFs are  in excellent agreement with those depicted by DNS.  Finally, to asses the LES predictions of the overall compositional structure, three-dimensional scatter plots of the mixture fraction, the mass fraction of oxidant O\textsubscript{2} and the mass fraction of hydroxyl radical OH colored by temperature are shown in Fig.\ \ref{fig:3DScatter}. Again, the  manifolds as predicted by LES-FDF are in very good agreements with those depicted by DNS.

\section{Conclusions}
\label{sec:conclusions}

Modeling of turbulence-combustion interactions has been the subject of broad investigations for over seventy years now \cite{HWH48}.   Large eddy simulation (LES) has been long recognized as a convenient means of capturing the unsteady evolution of turbulence  in both non-reacting and reactive flows \cite{Givi89}. The major issues associated with LES for prediction of practical turbulent combustion problems are: reliable modeling of the subgrid scale (SGS) quantities,  high fidelity solution of the modeled transport equations, and versatility in dealing with complex flames.  The filtered density function (FDF) \cite{Givi06,Haworth10,pope13,MF16,SRG20} has proven   particularly effective in resolving the first issue.  The present work makes a some progress in   dealing with the other two.  This progress is facilitated by developing a novel computational scheme by the merger of the PeleLM flow solver \cite{day2000numerical, Nonaka2012Deferred, Pazner2016HighOrder} and the Monte-Carlo (FDF) simulator.  The resulting computational scheme facilitates reliable and high fidelity simulation of turbulent combustion systems.  The novelty of the methodology, as developed, is its capability to capture the very intricate dynamics of turbulence-chemistry interactions.  This is  demonstrated by its implementation to conduct LES  of a CO/H\textsubscript{2} temporally developing jet flame. The results are assessed via detailed {\it a posteriori} comparative assessments against direct numerical simulation (DNS) data for the same flame \cite{Hawkes2007Scalar}.  Excellent   agreements are observed for  the temporal evolution of all of  the thermo-chemical variables, including the manifolds portraying the multi-scalar mixing.  The new methodology is shown to be particularly effective in capturing non-equilibrium turbulence-chemistry interactions.  This is demonstrated by capturing the  flame-extinction and its re-ignition as observed in DNS.  With its high fidelity and   computational affordability, the  new PeleLM-FDF simulator as developed here provides an excellent tool for computational simulations of complex turbulent combustion systems. 

At this point it is instructive to provide some suggestions for future work in continuation of this research:

\begin{enumerate}
    \item The hydrodynamic SGS closure adopted here is based on the  zero-order model of Vreman \cite{Vreman2004Eddy}.  This model has proven very effective for LES of many flows, including the one considered here. However, for more complex flows one may need to use more comprehensive SGS closures.  Therefore, the extension to include the velocity-FDF \cite{GGJP02,SDGP03,SGP07,SGP09} is encouraged.
    
    \item A very attractive feature of the PeleLM  is its adaptive gridding and mesh refinement strategy.  This feature is not utilized here because of the relative flow simplicity.  Future work is needed to refine  the MC strategy in conjunction with  AMR.  Some progress  in this regard has been reported \cite{Damasceno2018Simulation, Castro2021Implementation}.
    
    \item The PeleC code \cite{Sitaraman2021Adaptive} is the counterpart  of PeleLM for high speed flows.  It would be desirable to implement the FDF methodology in this code as well.  In doing so, the full self-contained form of the FDF \cite{NNGLP17} should be considered. 
    
    \item Resolution assessment in LES is of crucial importance. Several such studies have been conducted for other forms of LES-FDF \cite{DSMG07,NYSG10,SAMG20}, and is recommended for PeleLM-FDF.
    
    \item With its  flexibility and high fidelity, it is expected that the PeleLM-FDF methodology will be implemented for LES of a wide variety of other complex turbulent combustion systems.  
\end{enumerate}

 \section*{Acknowledgments}
We are grateful to Professor Evatt R.\ Hawkes of  the University of New South Wales for providing the DNS data as used for comparative studies here. We re indebted to Dr.\ Marcus Day of National Renewable Energy Laboratory, the original  developer of the PeleLM for excellent comments on the draft of this manuscript.  This work is sponsored by the National Science Foundation  under Grant CBET-2042918. Computational resources are provided by the University of Pittsburgh Center for Research Computing.

\newpage

\end{document}


\maketitle

\section{Appendix A.  Supplementary Data:}  Supplementary data associated with this work are available on online version of the manuscript.

\begin{figure}
    \centering
    \includegraphics[width=0.3\textwidth]{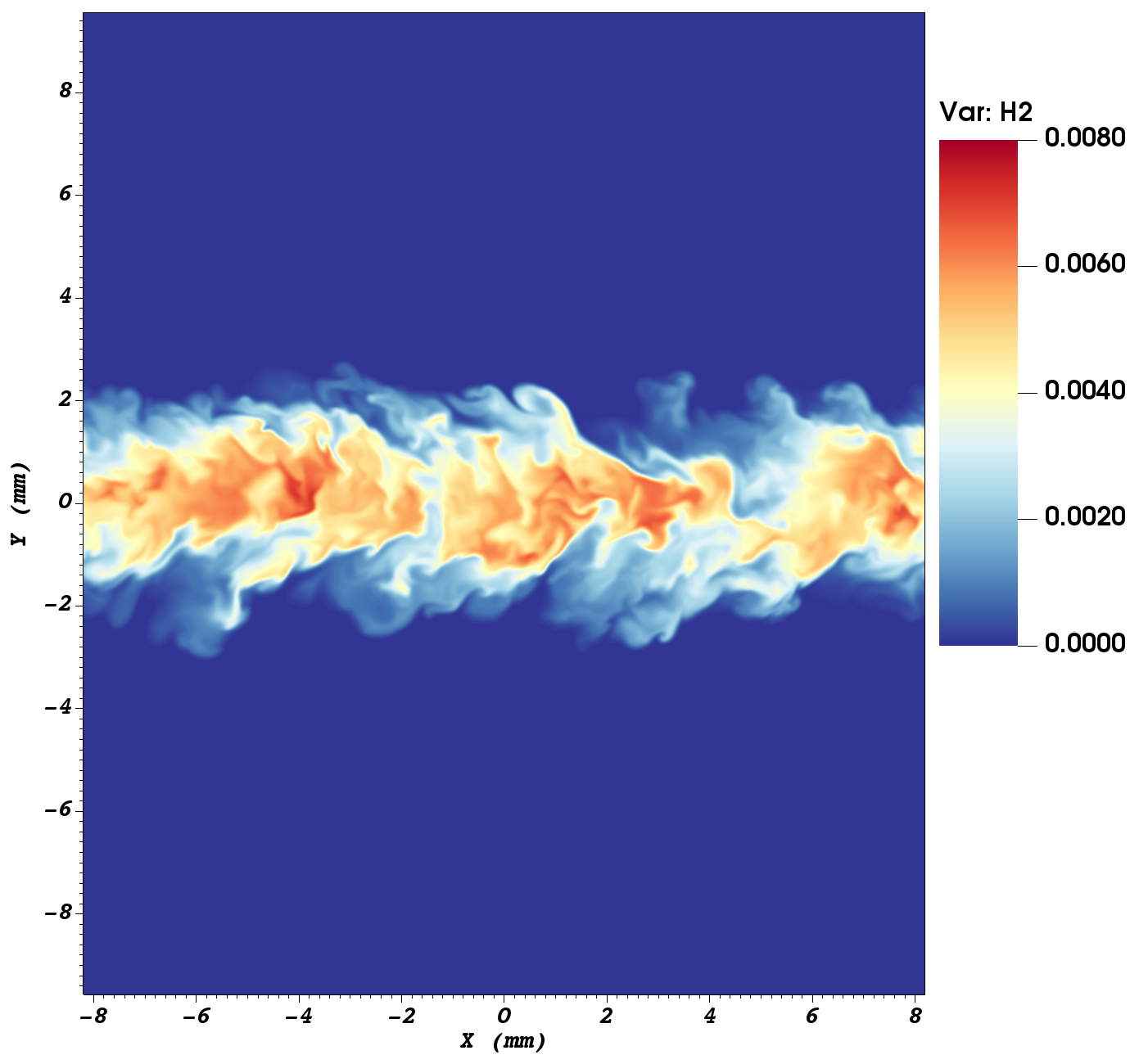}
    \includegraphics[width=0.3\textwidth]{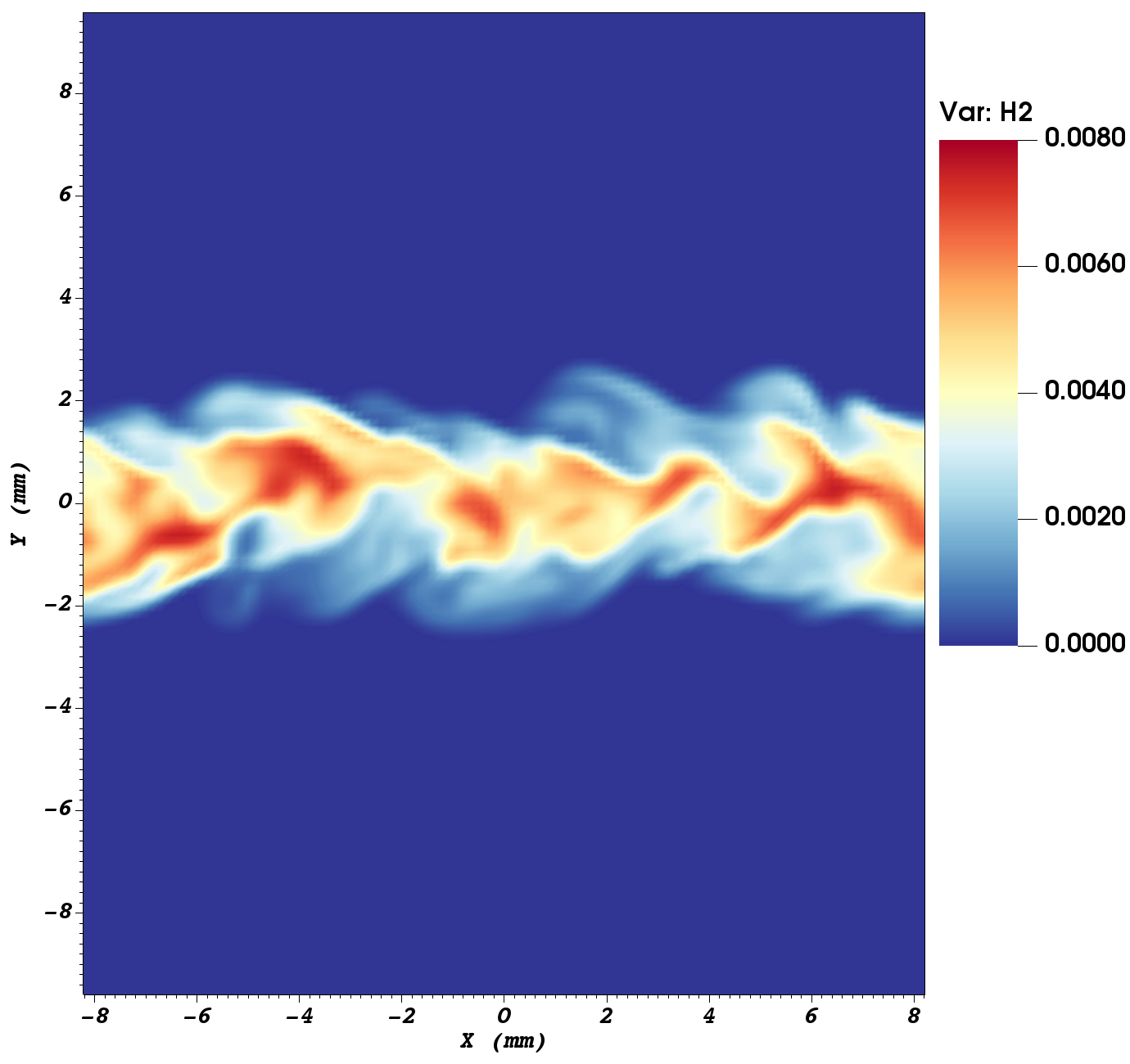}
    
    \includegraphics[width=0.3\textwidth]{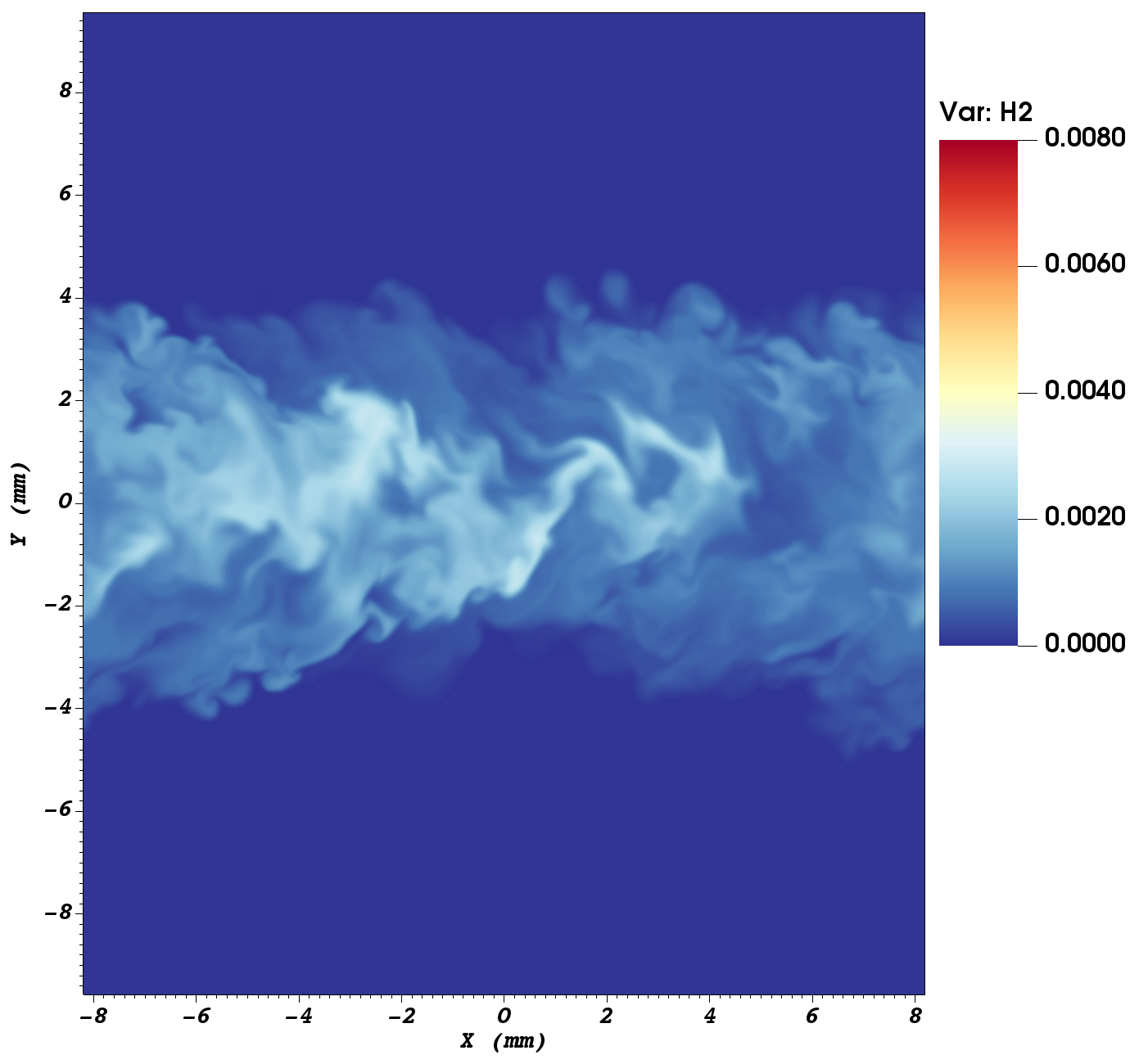}
    \includegraphics[width=0.3\textwidth]{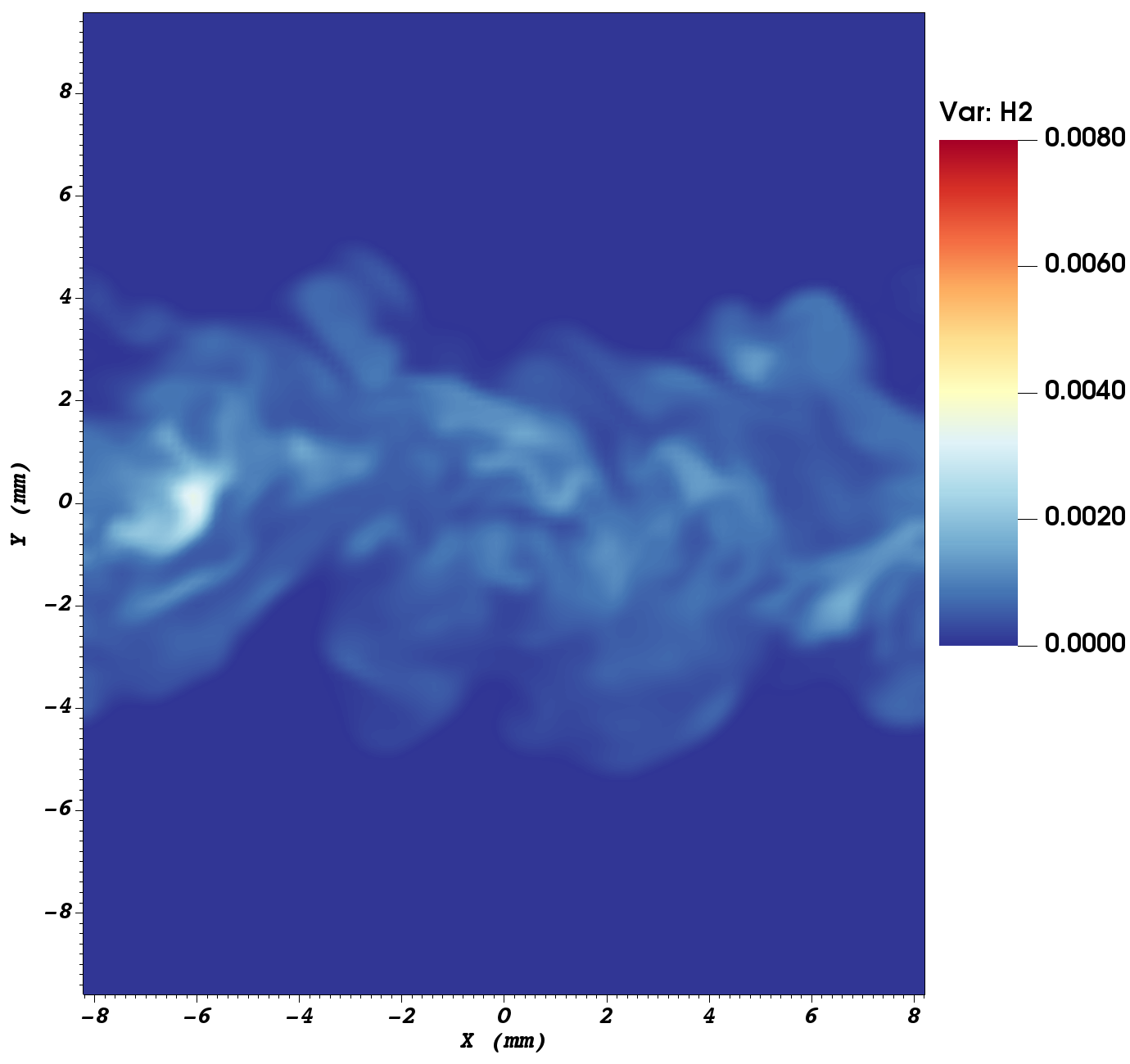}
    
    \caption{Instantaneous slice plots of H2 mass fraction at $t=20$ and $t=40$ obtained from DNS and LES-FDF.}
    \label{fig:H2Slices}
\end{figure}

\begin{figure}
    \centering
    \includegraphics[width=0.24\textwidth]{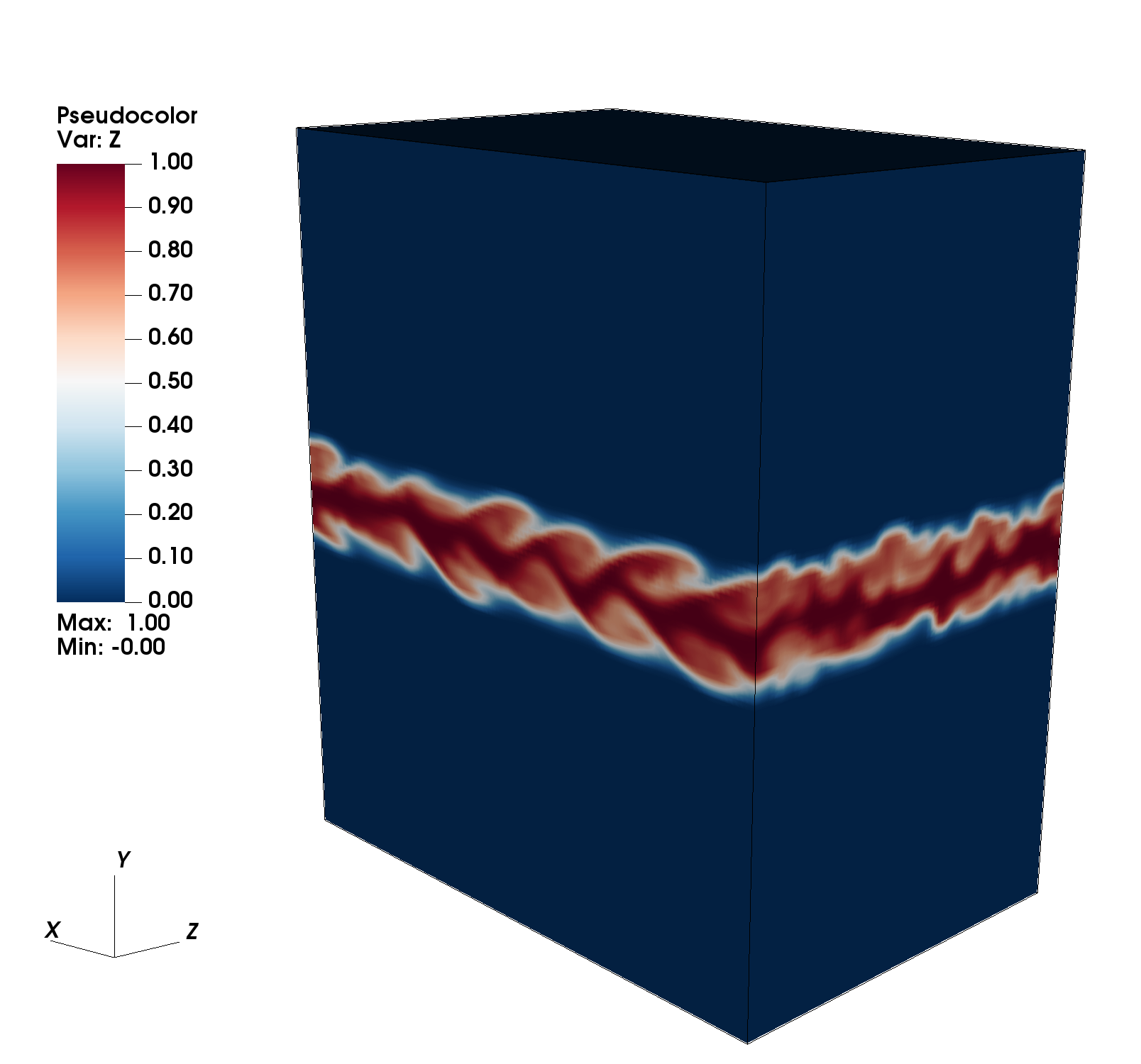}
    \includegraphics[width=0.24\textwidth]{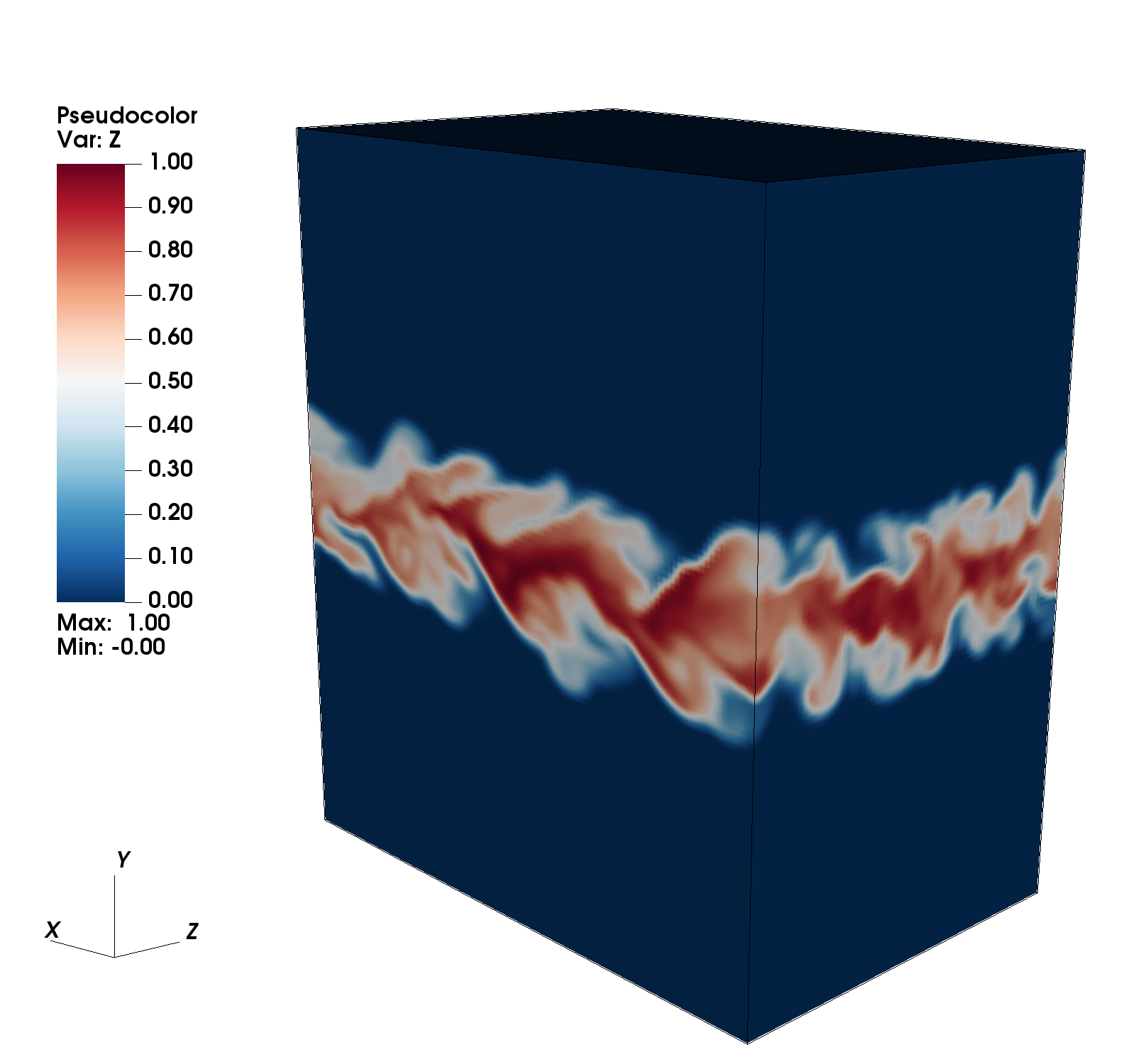}
    \includegraphics[width=0.24\textwidth]{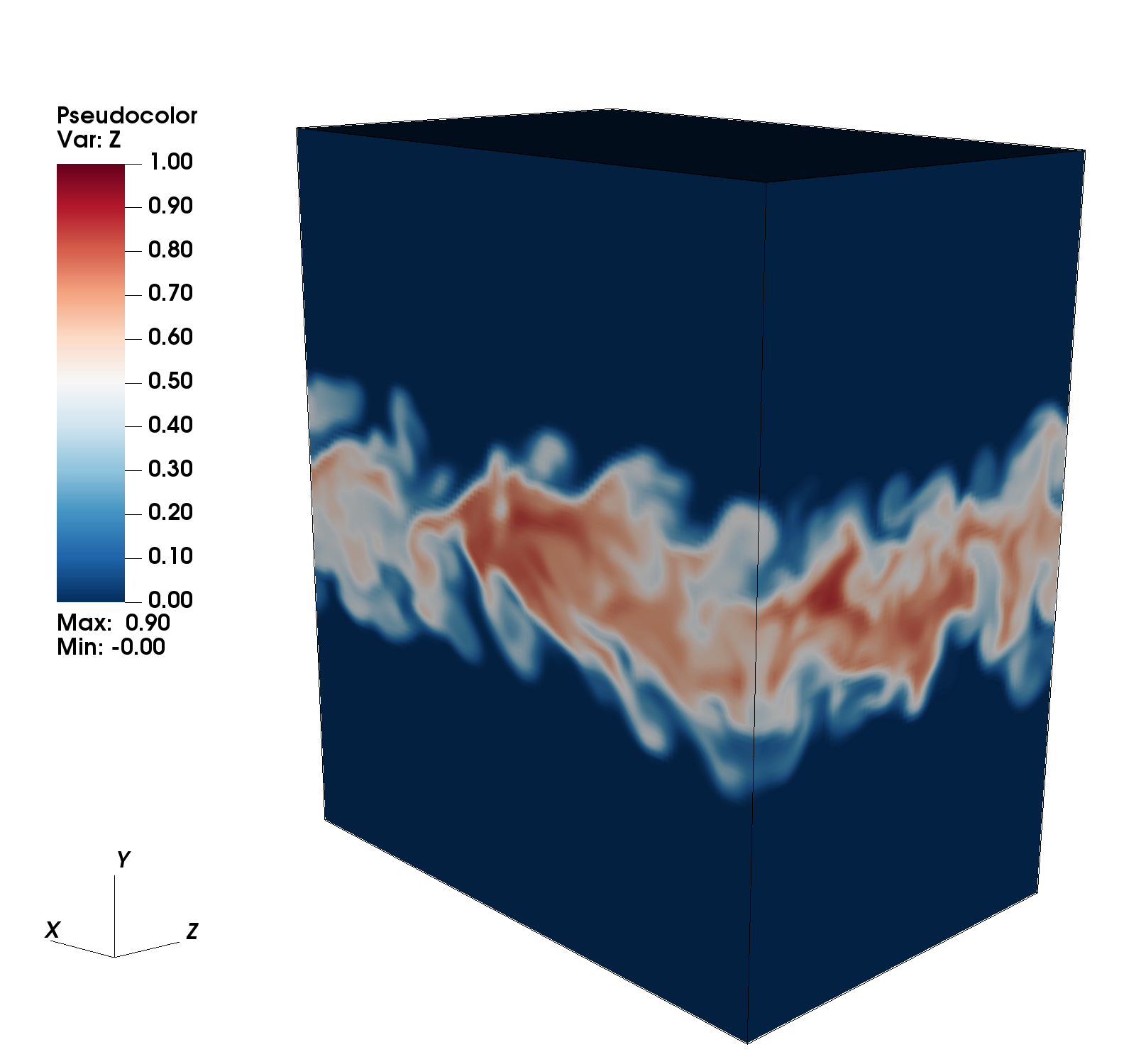}
    \includegraphics[width=0.24\textwidth]{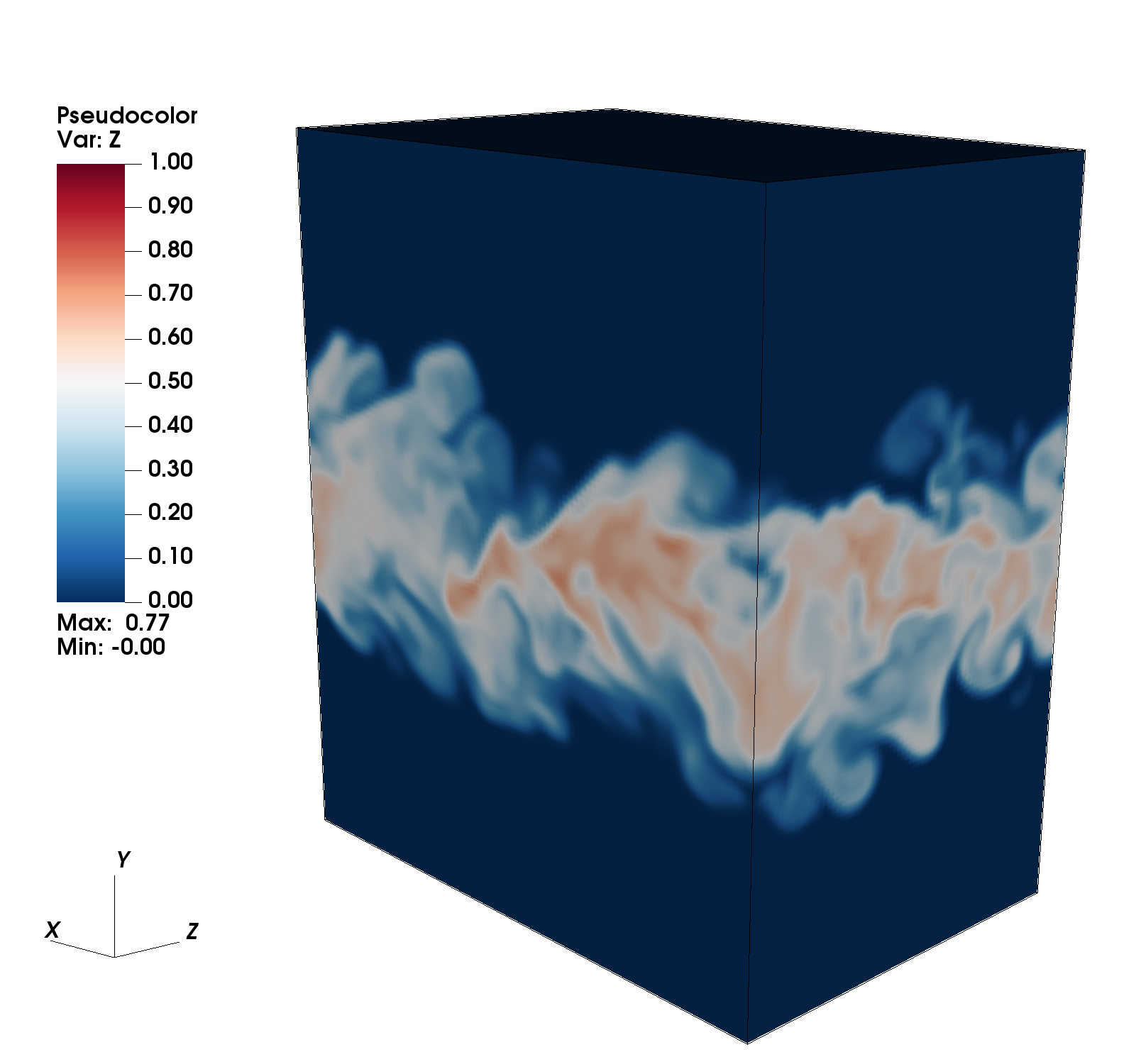}
    \caption{Temporal evolution of the mixture fraction.  (a) $t=10tj$, (b), $t=20tj$, (c), $t=30tj$, (d) $t=40tj$}
    \label{fig:SUPPLEMENTARY_Z}
\end{figure}

\begin{figure}
    \centering
    \includegraphics[width=0.24\textwidth]{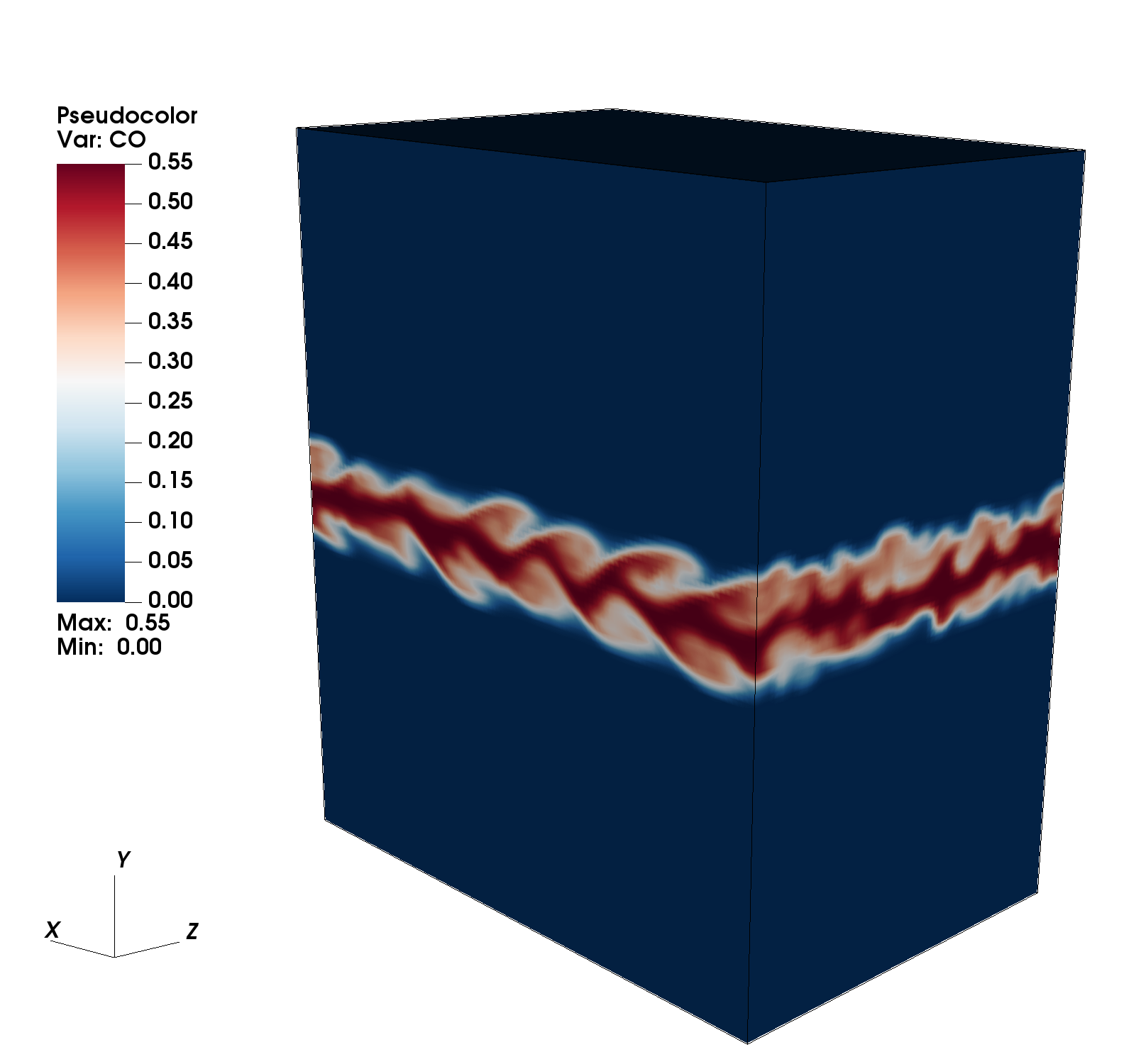}
    \includegraphics[width=0.24\textwidth]{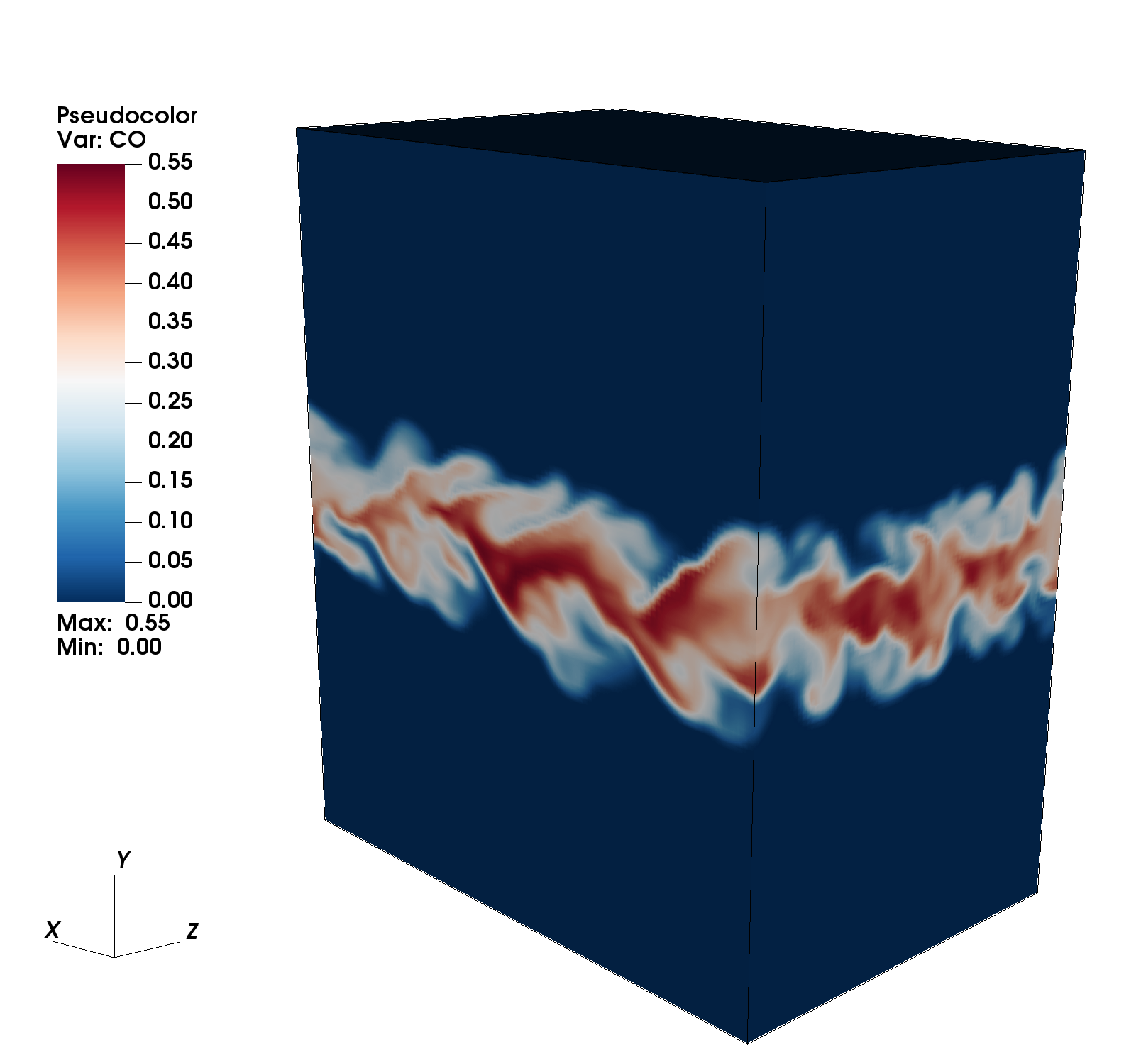}
    \includegraphics[width=0.24\textwidth]{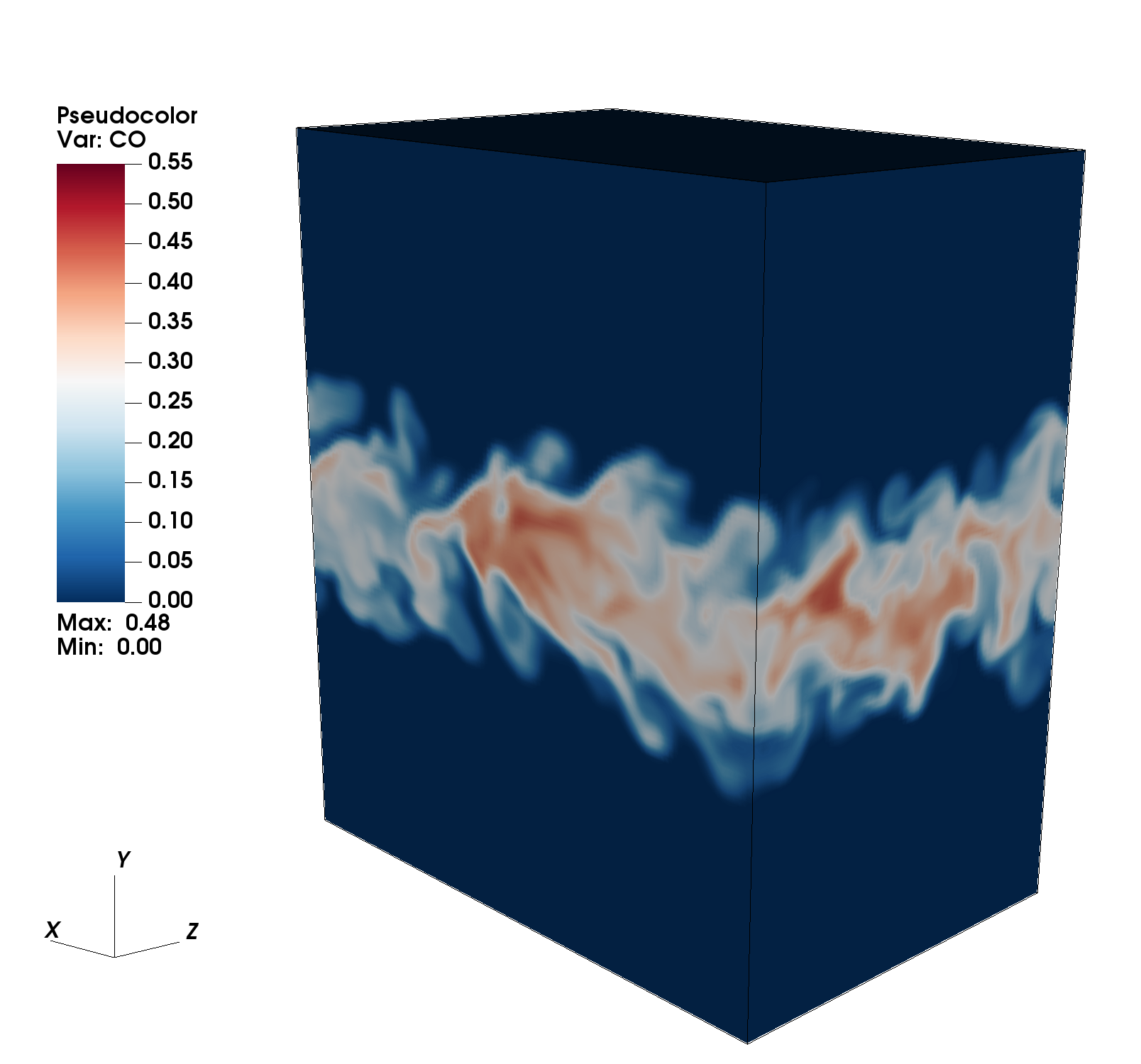}
    \includegraphics[width=0.24\textwidth]{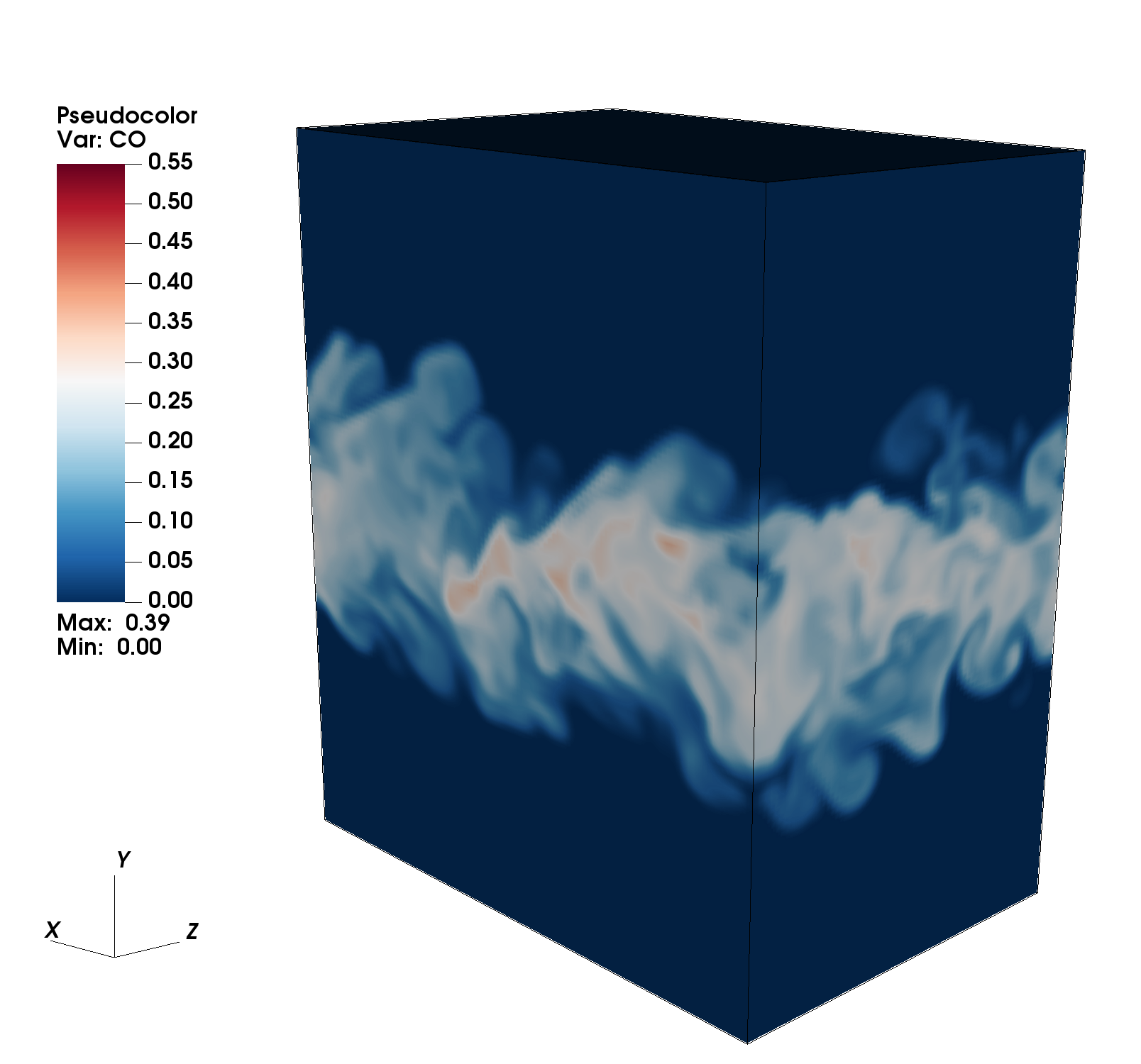}
    \caption{Temporal evolution of the CO.  (a) $t=10tj$, (b), $t=20tj$, (c), $t=30tj$, (d) $t=40tj$}
    \label{fig:SUPPLEMENTARY_CO}
\end{figure}

\begin{figure}
    \centering
    \includegraphics[width=0.24\textwidth]{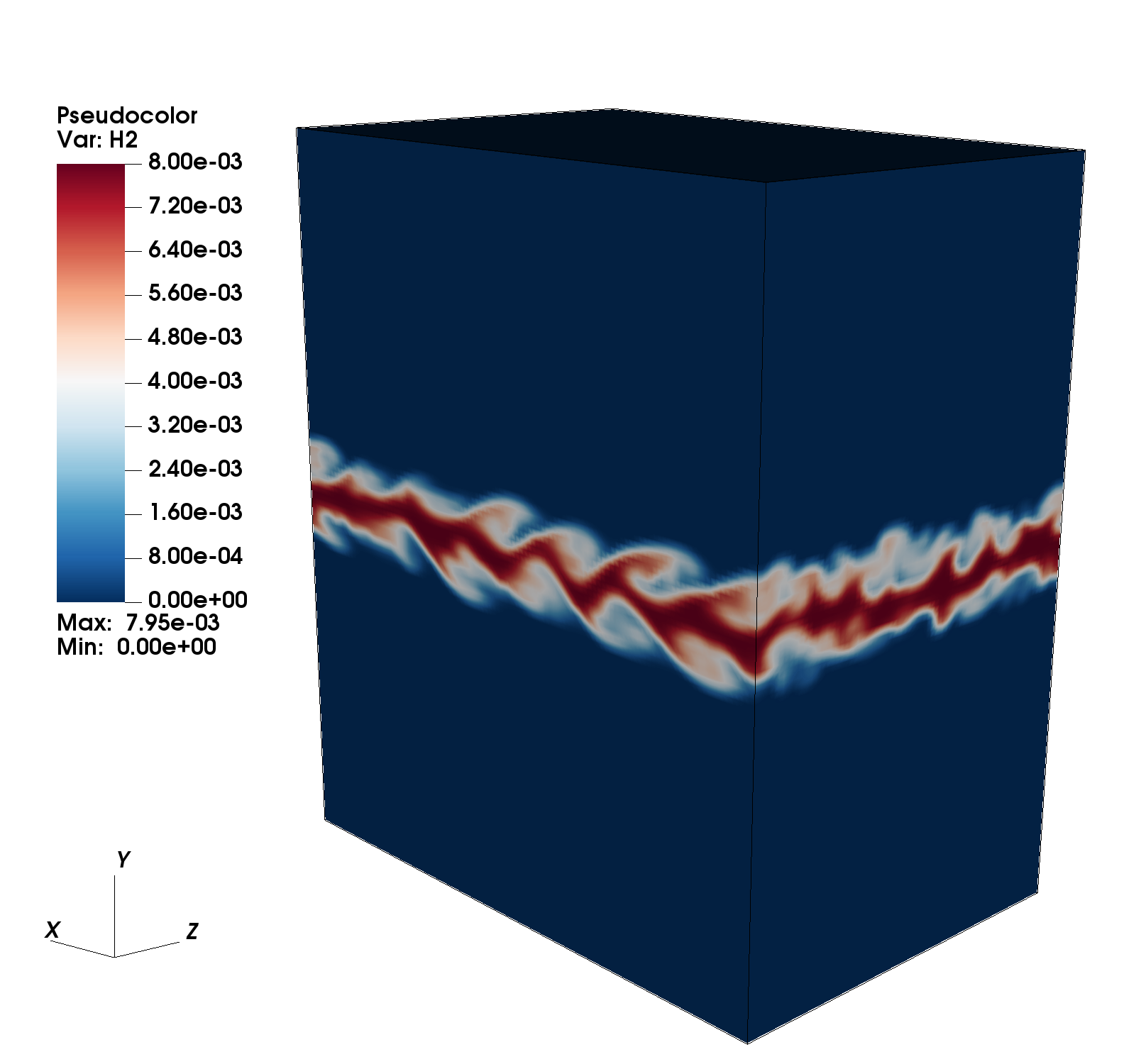}
    \includegraphics[width=0.24\textwidth]{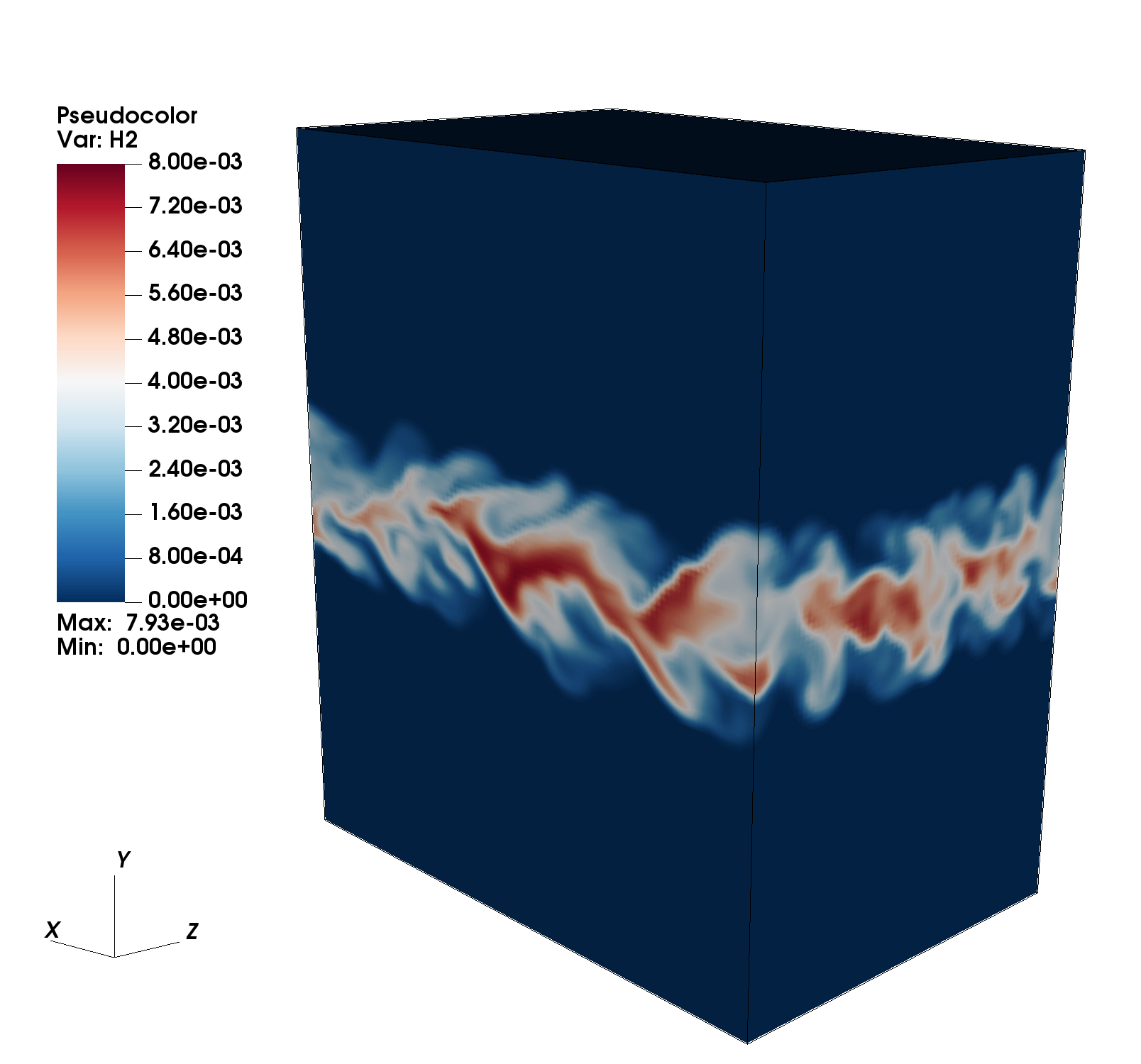}
    \includegraphics[width=0.24\textwidth]{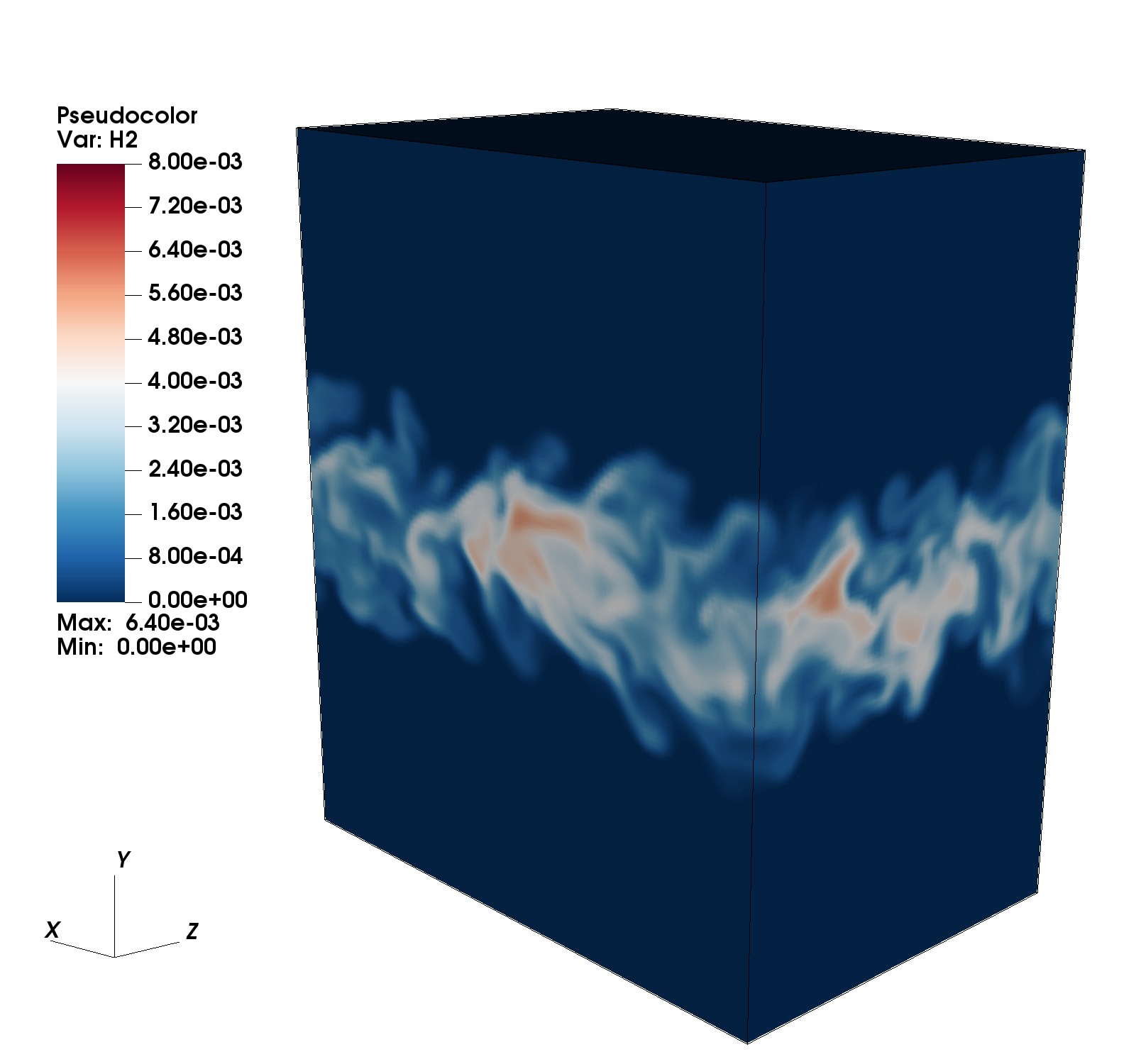}
    \includegraphics[width=0.24\textwidth]{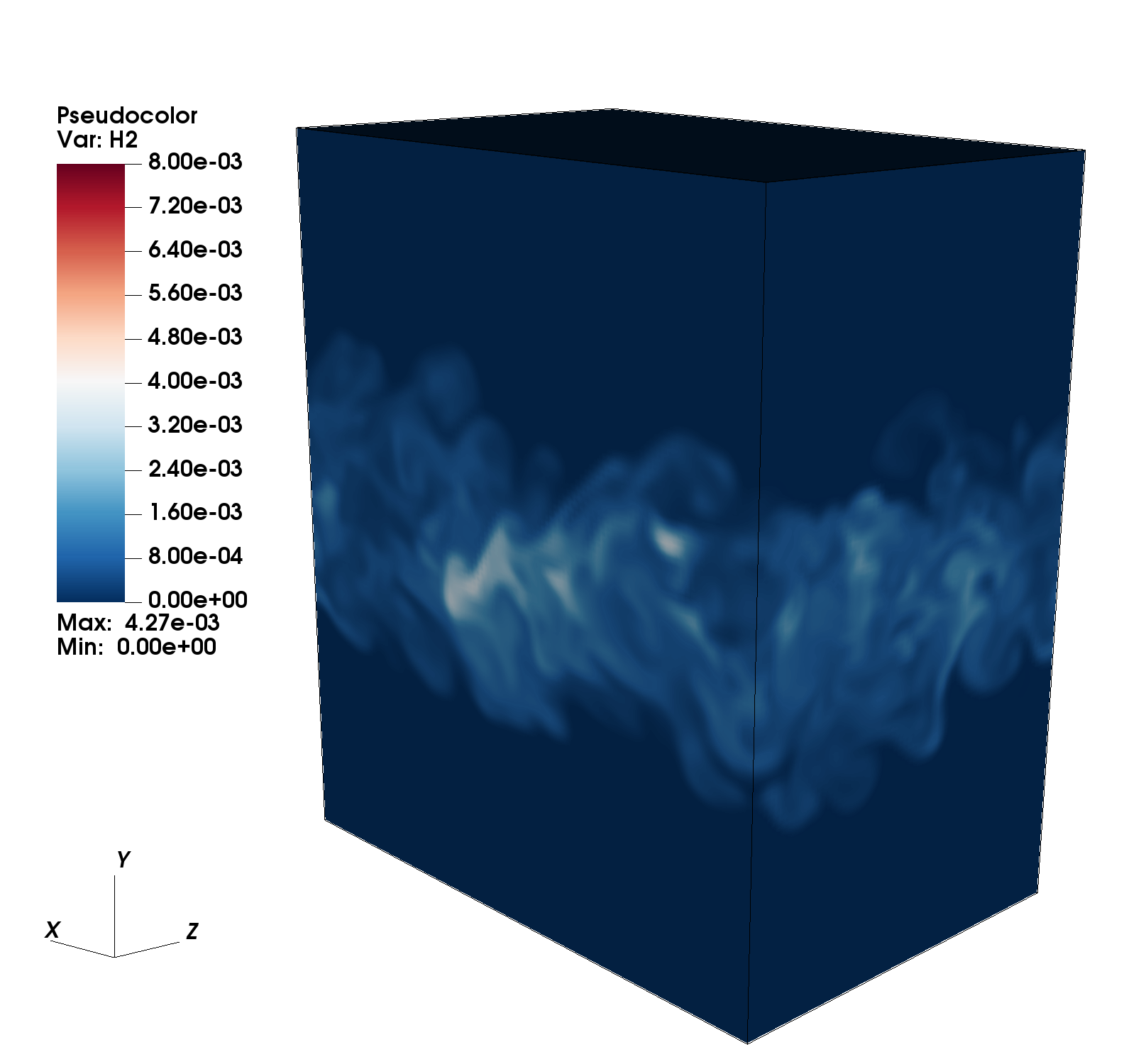}
    \caption{Temporal evolution of the H2.  (a) $t=10tj$, (b), $t=20tj$, (c), $t=30tj$, (d) $t=40tj$}
    \label{fig:SUPPLEMENTARY_H2}
\end{figure}

\begin{figure}
    \centering
    \includegraphics[width=0.24\textwidth]{Figures/O2_0000.png}
    \includegraphics[width=0.24\textwidth]{Figures/O2_0001.png}
    \includegraphics[width=0.24\textwidth]{Figures/O2_0002.png}
    \includegraphics[width=0.24\textwidth]{Figures/O2_0003.png}
    \caption{Temporal evolution of the O2.  (a) $t=10tj$, (b), $t=20tj$, (c), $t=30tj$, (d) $t=40tj$}
    \label{fig:SUPPLEMENTARY_O2}
\end{figure}

\begin{figure}
    \centering
    \includegraphics[width=0.24\textwidth]{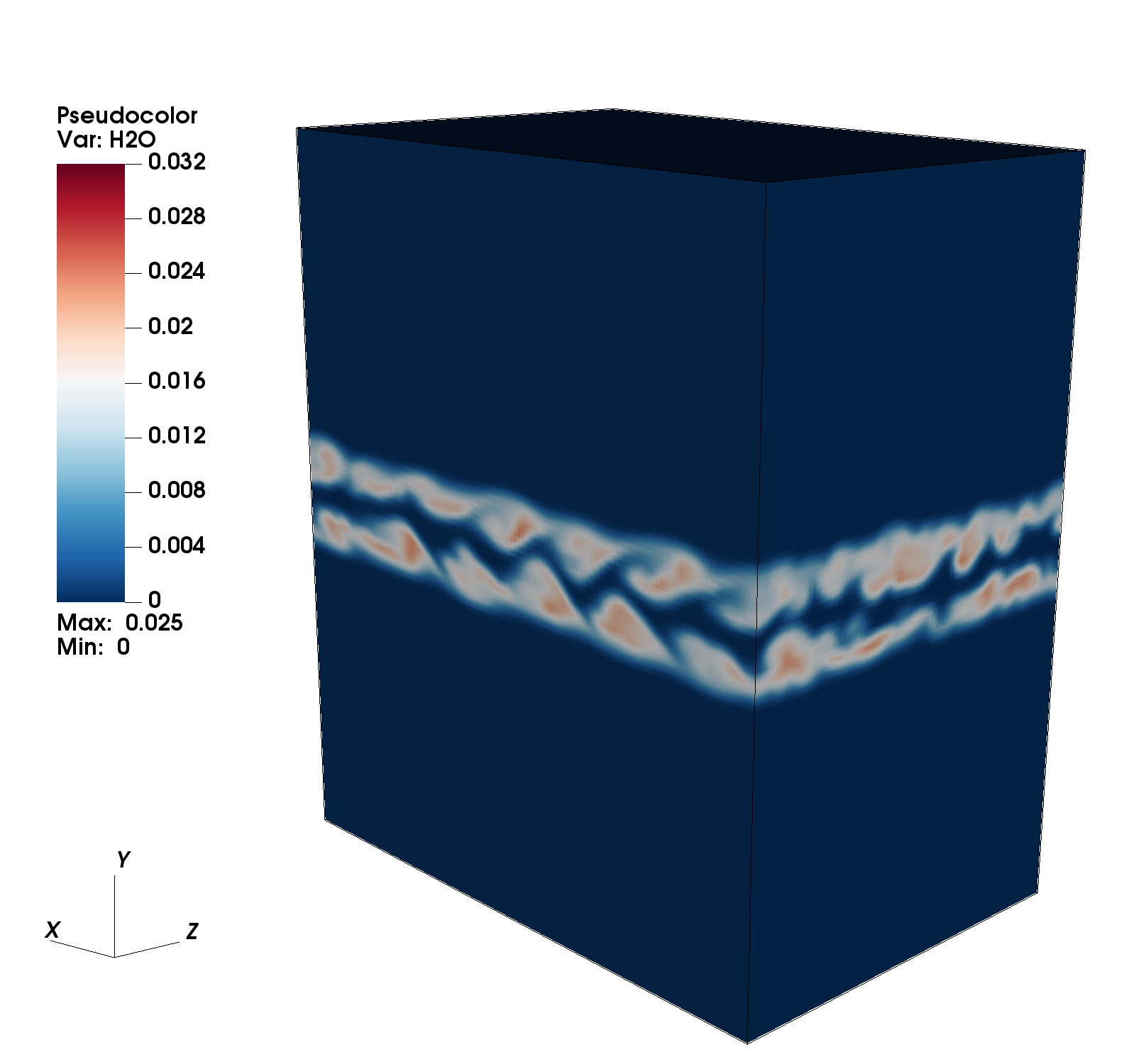}
    \includegraphics[width=0.24\textwidth]{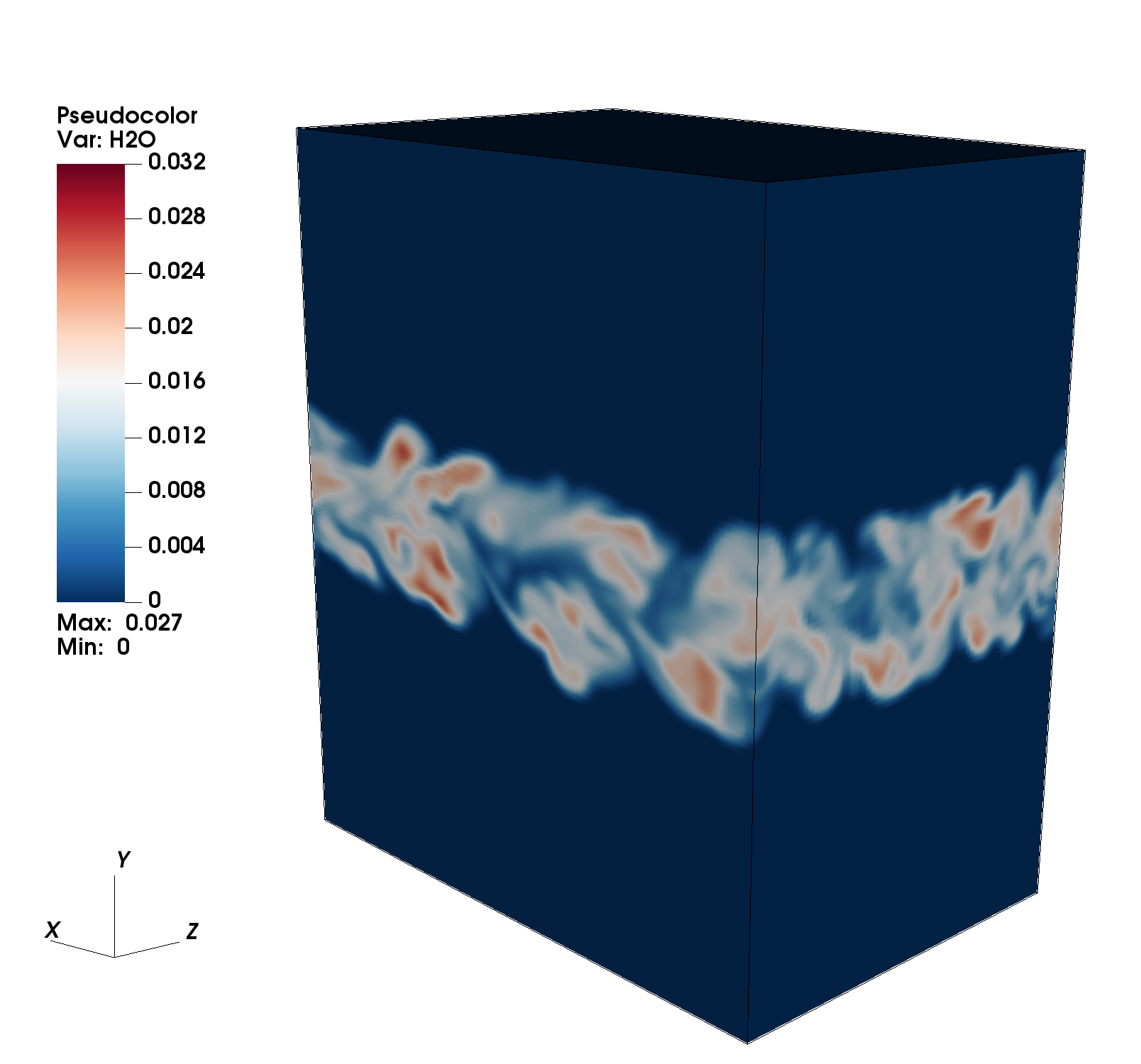}
    \includegraphics[width=0.24\textwidth]{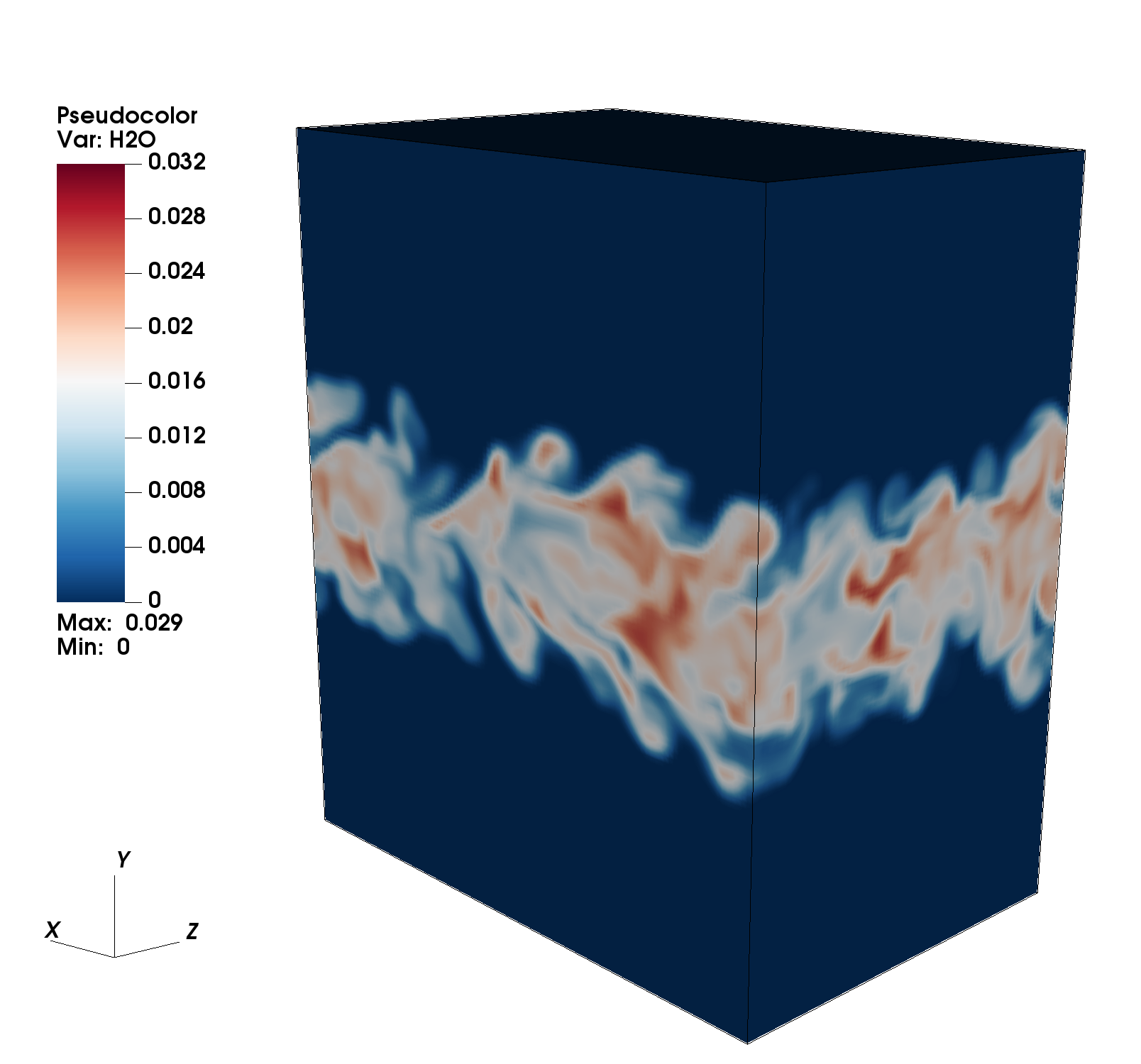}
    \includegraphics[width=0.24\textwidth]{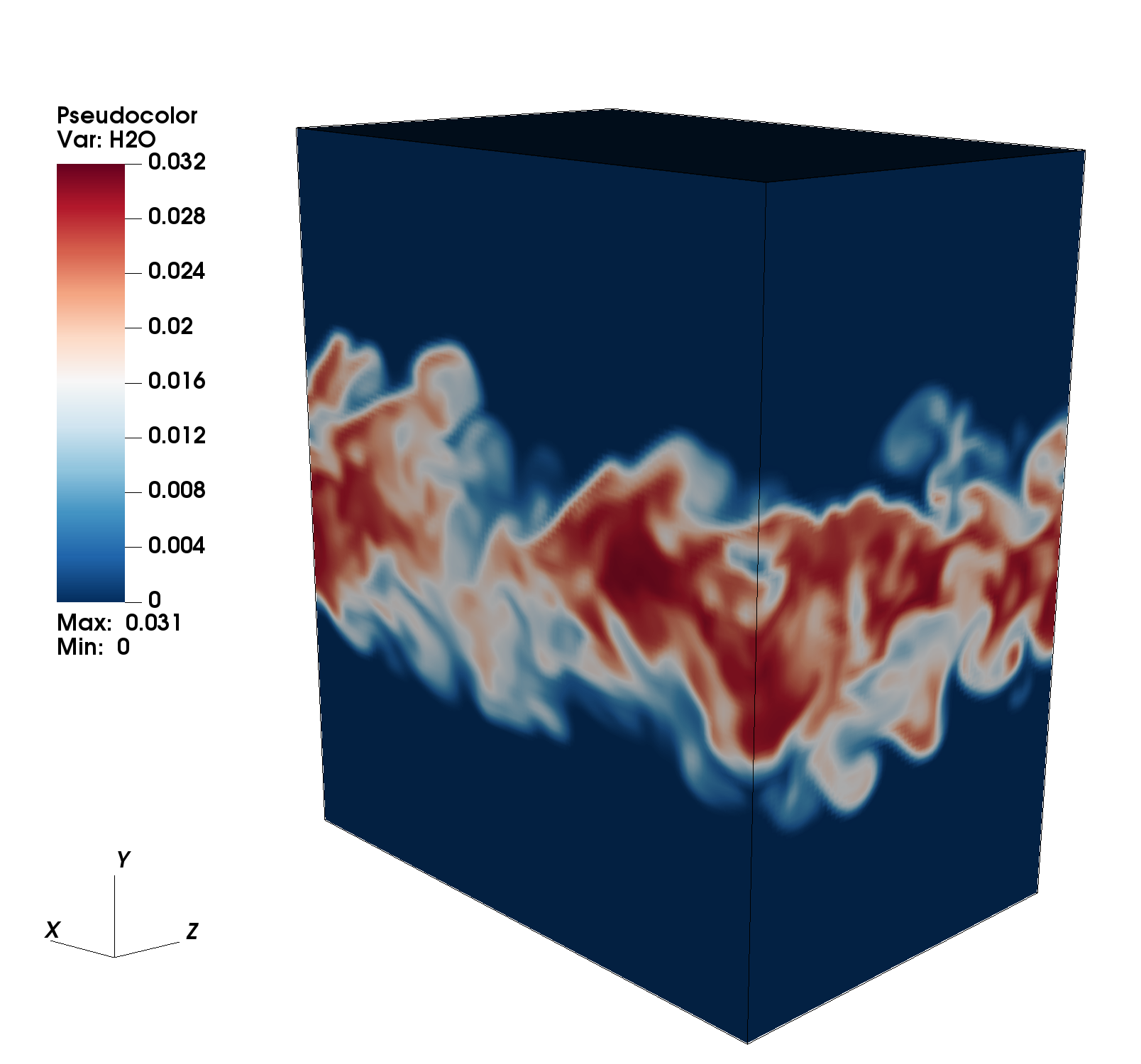}
    \caption{Temporal evolution of the H2O.  (a) $t=10tj$, (b), $t=20tj$, (c), $t=30tj$, (d) $t=40tj$}
    \label{fig:SUPPLEMENTARY_H2O}
\end{figure}

\begin{figure}
    \centering
    \includegraphics[width=0.24\textwidth]{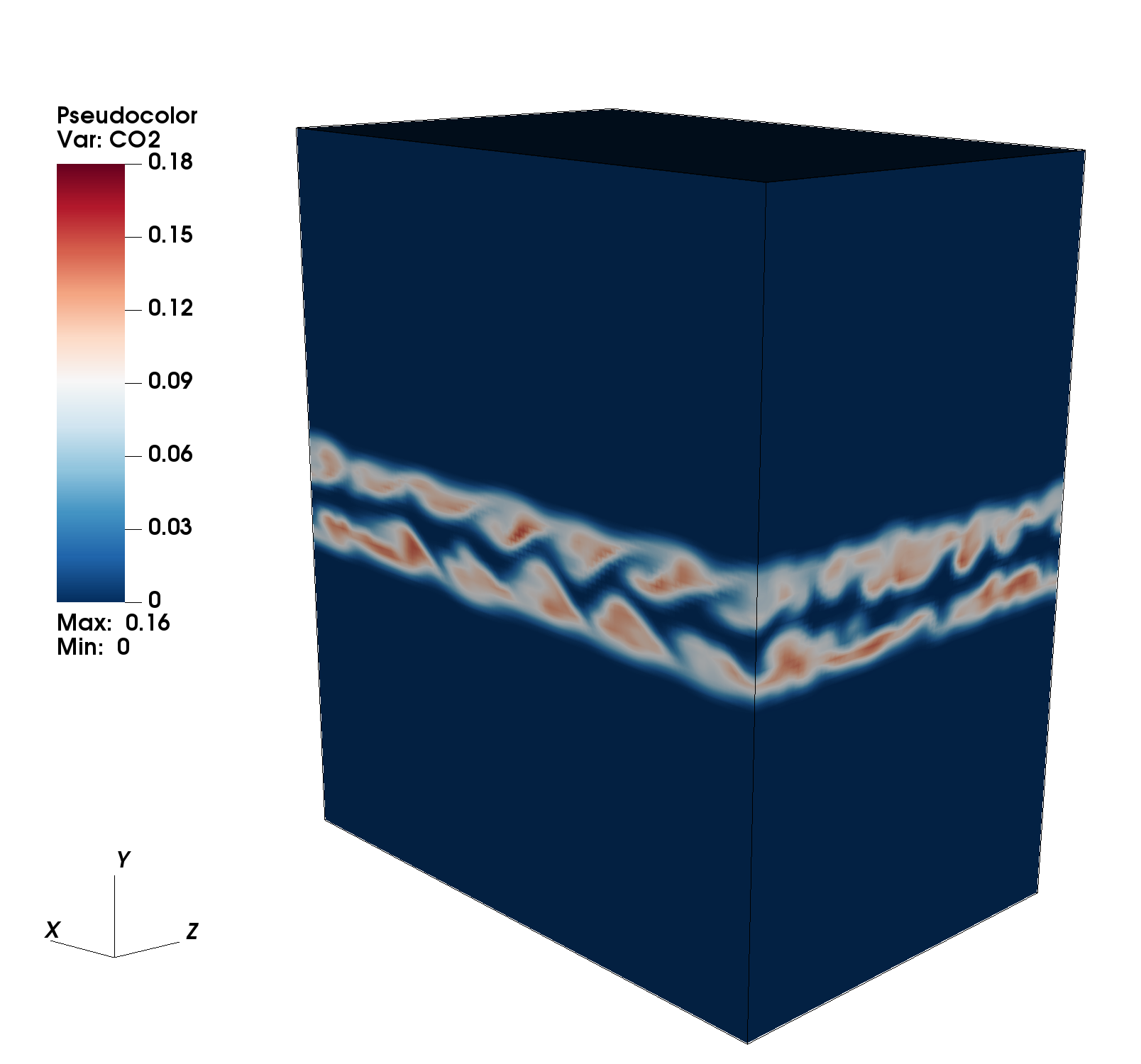}
    \includegraphics[width=0.24\textwidth]{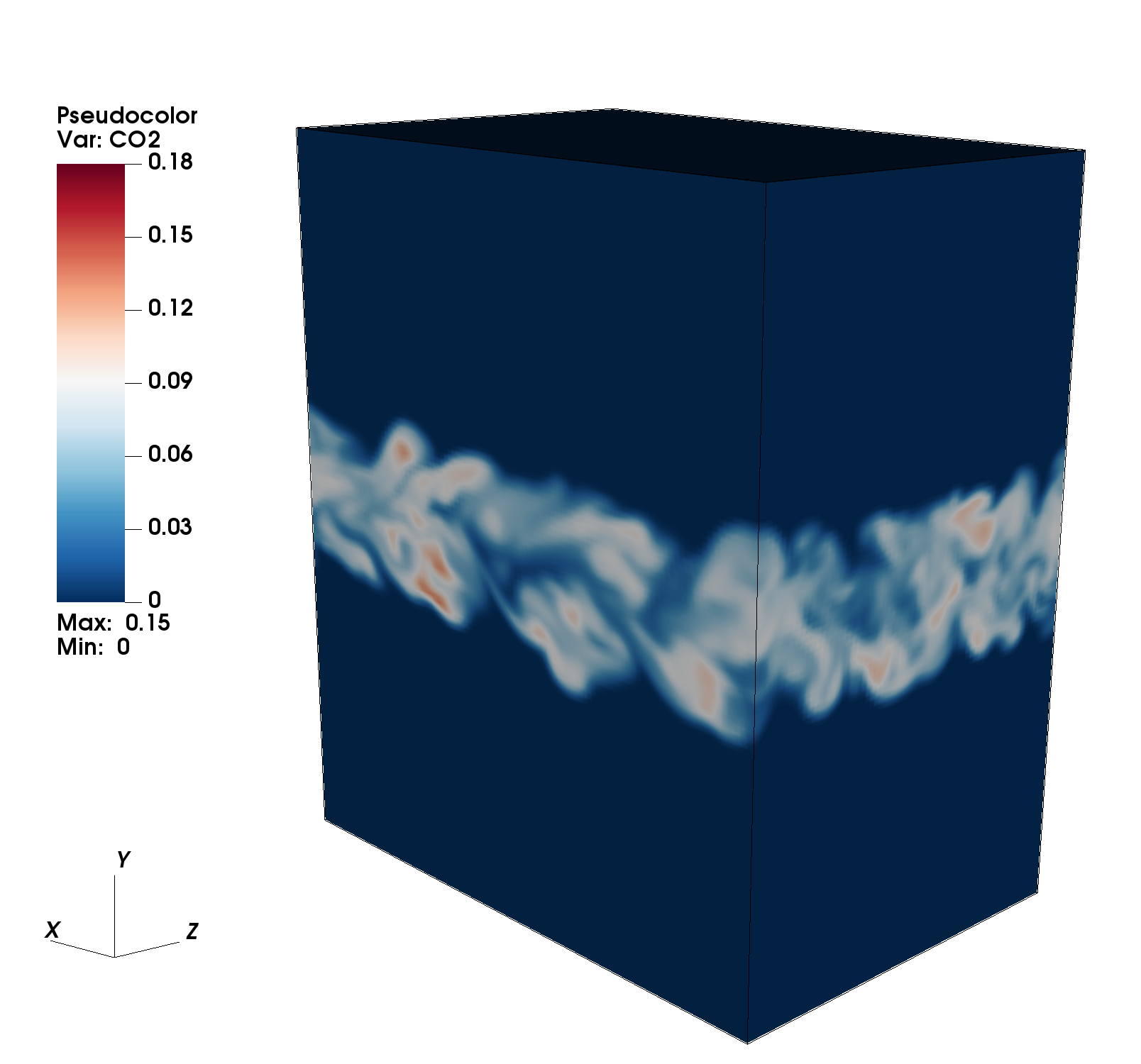}
    \includegraphics[width=0.24\textwidth]{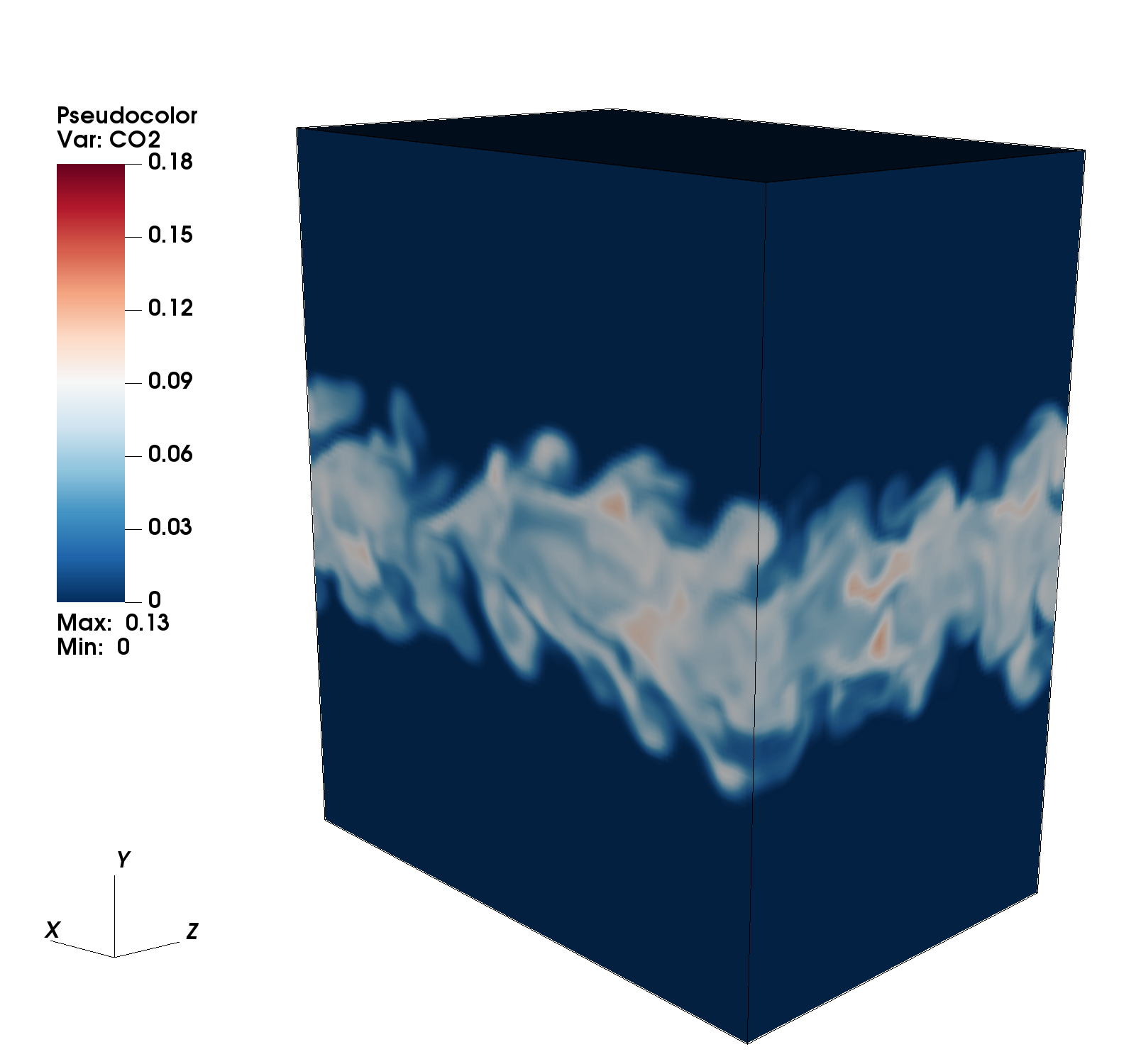}
    \includegraphics[width=0.24\textwidth]{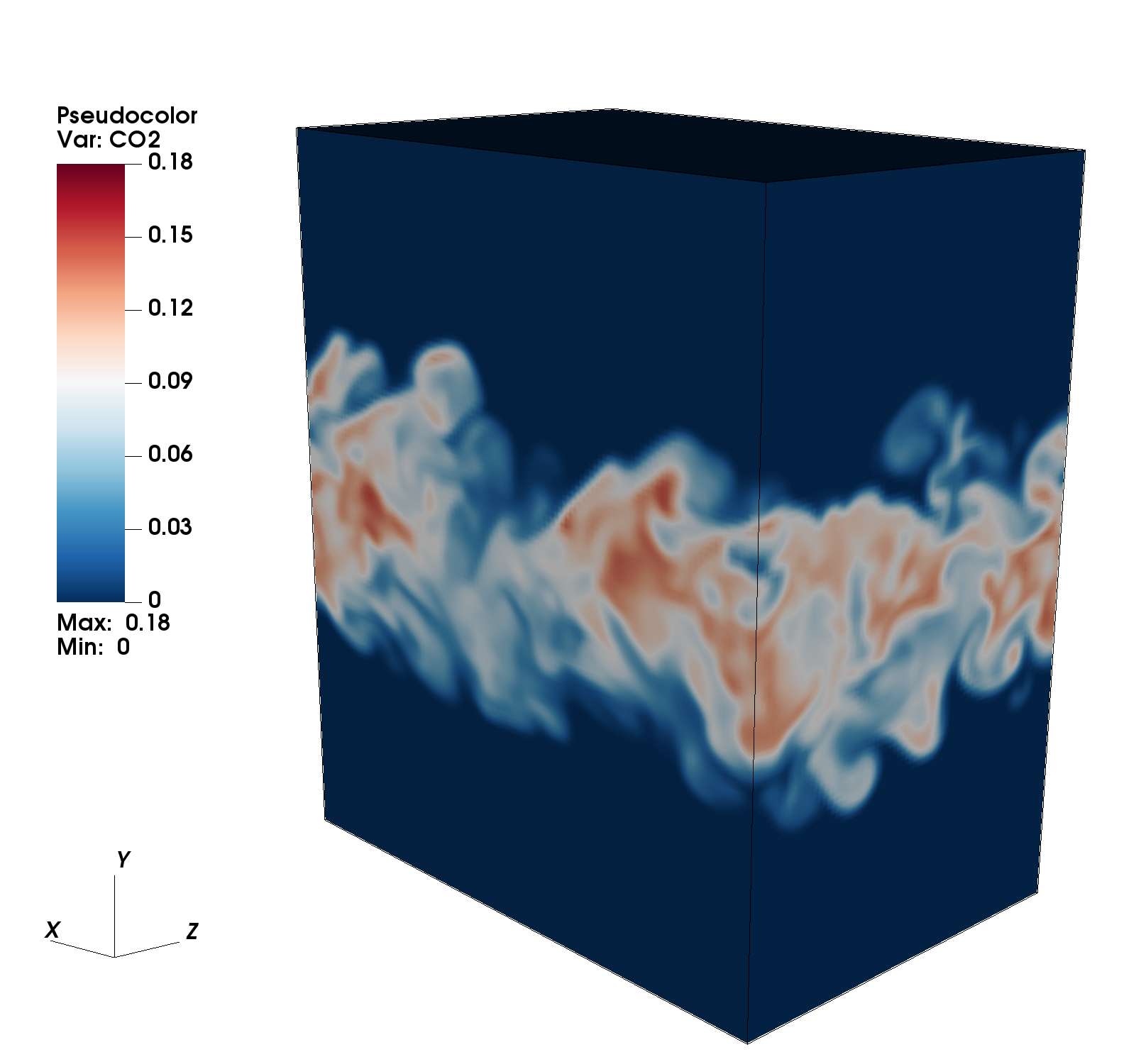}
    \caption{Temporal evolution of the CO2.  (a) $t=10tj$, (b), $t=20tj$, (c), $t=30tj$, (d) $t=40tj$}
    \label{fig:SUPPLEMENTARY_CO2}
\end{figure}

\begin{figure}
    \centering
    \includegraphics[width=0.24\textwidth]{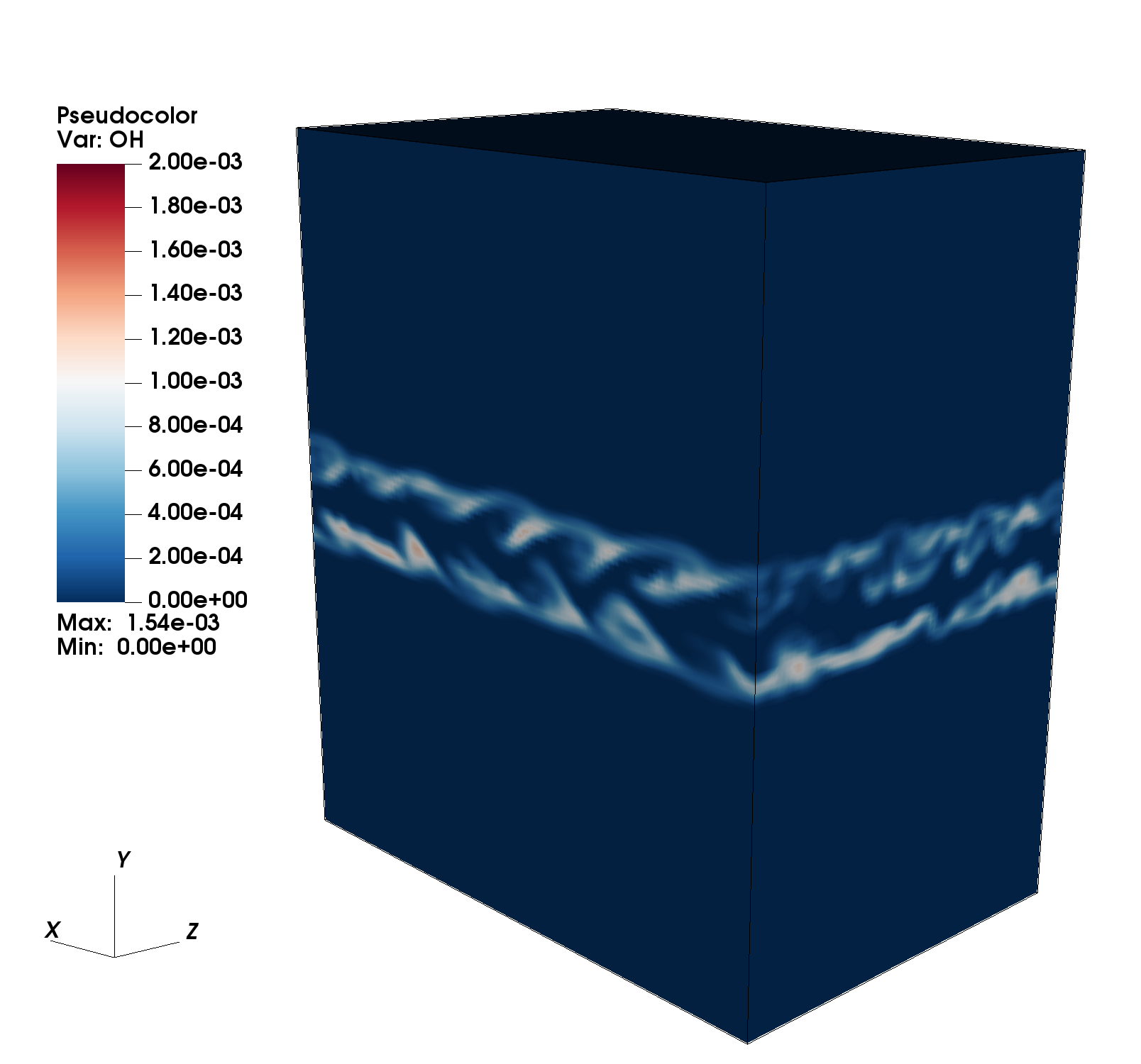}
    \includegraphics[width=0.24\textwidth]{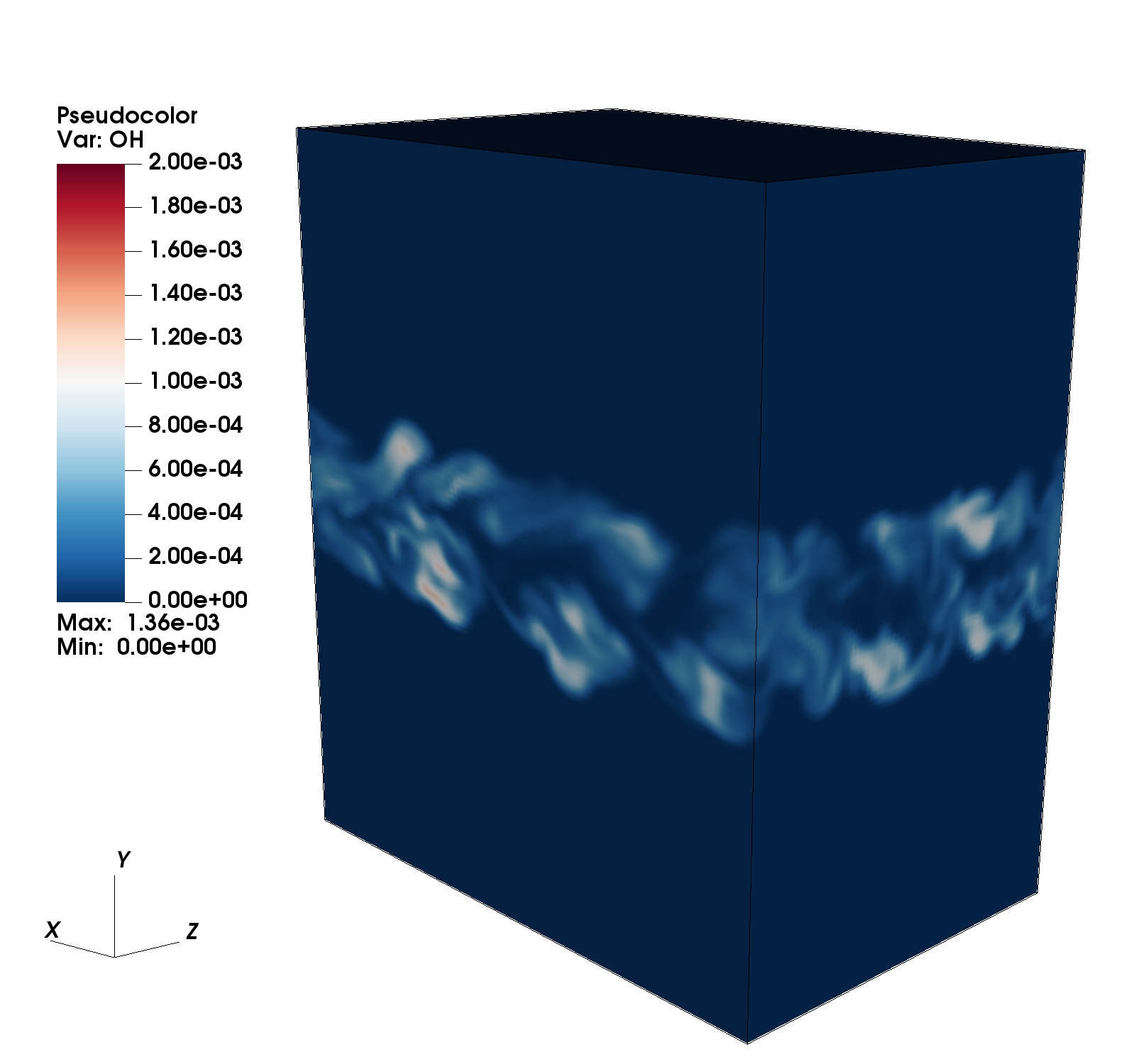}
    \includegraphics[width=0.24\textwidth]{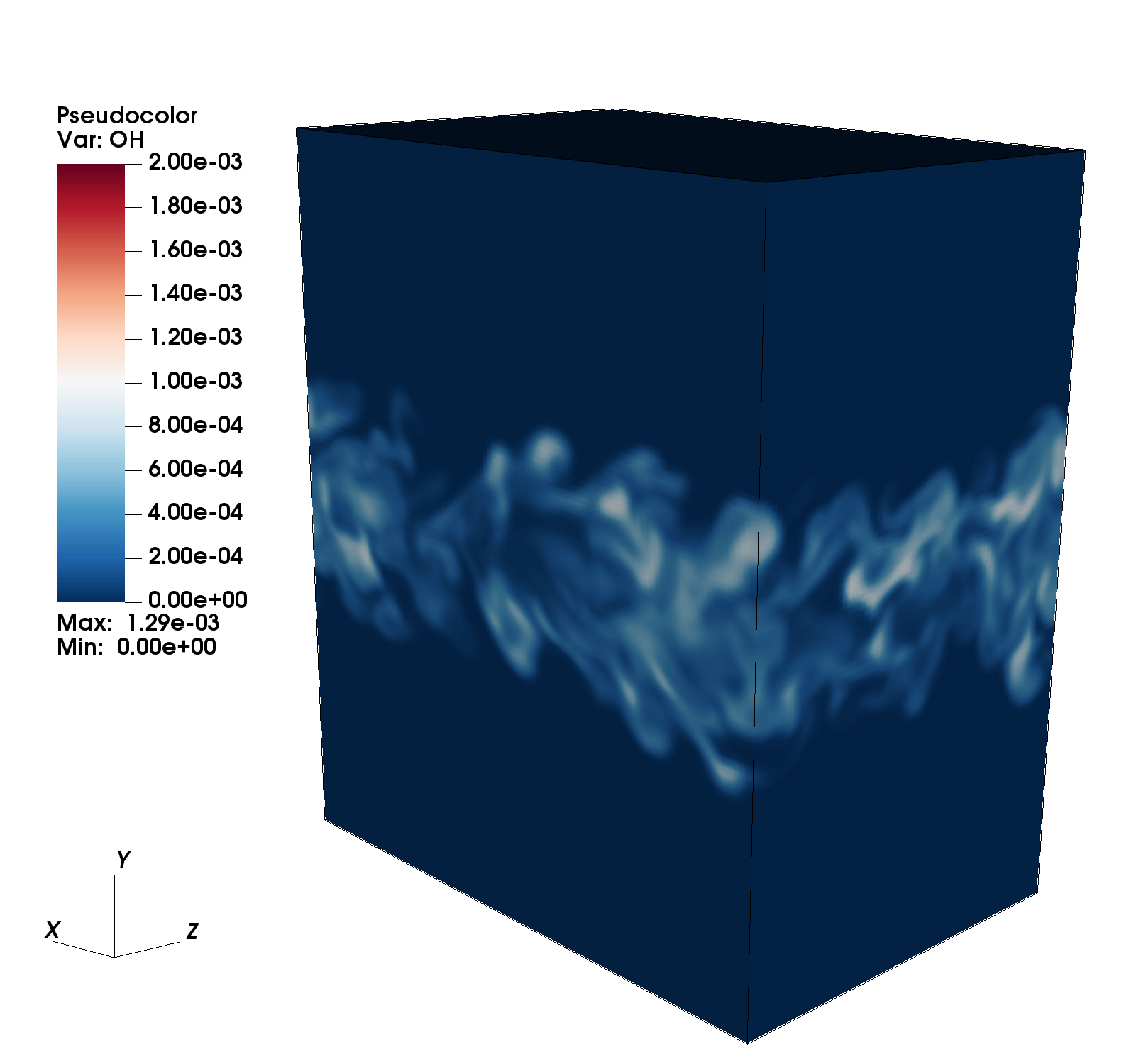}
    \includegraphics[width=0.24\textwidth]{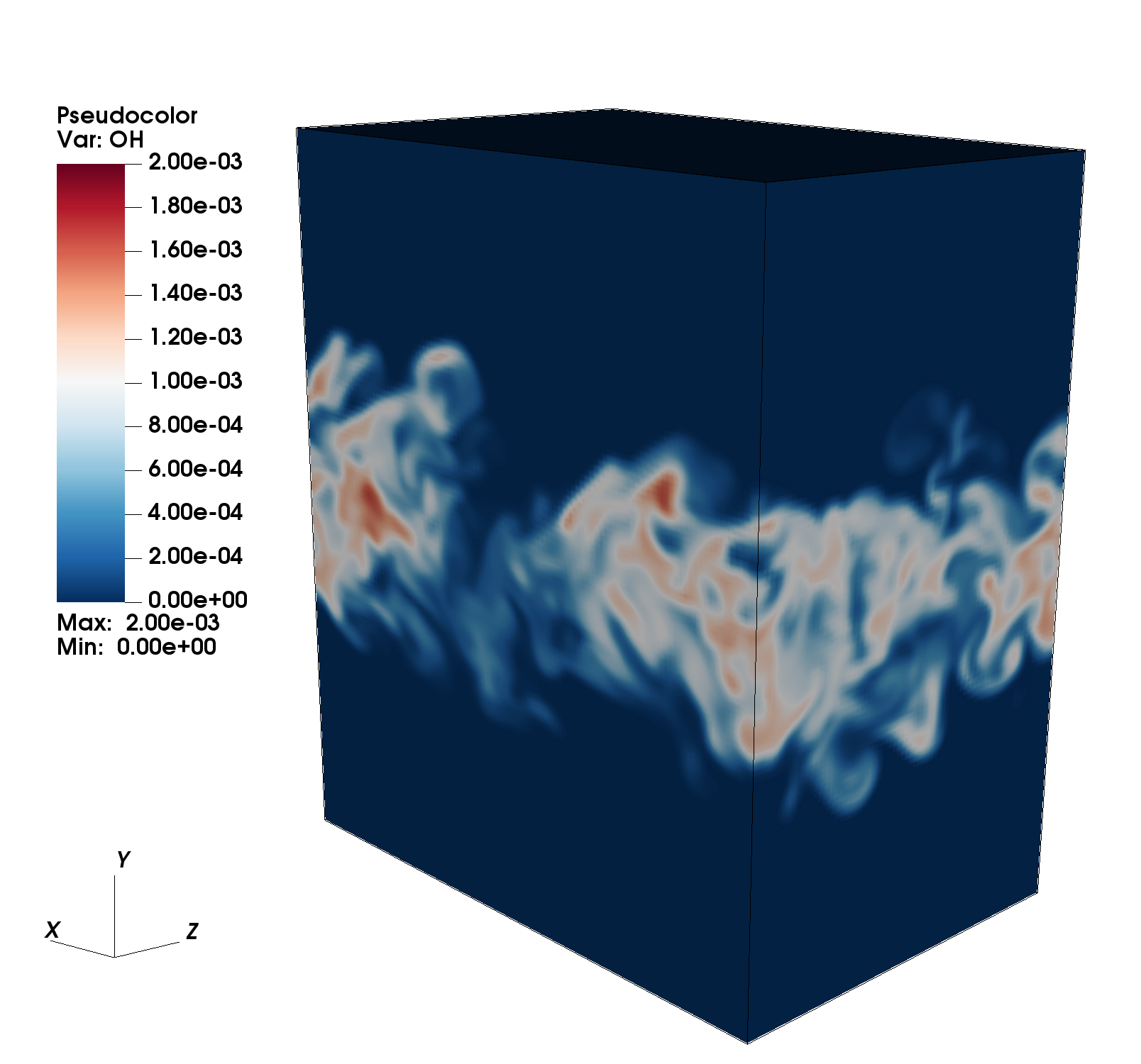}
    \caption{Temporal evolution of the OH.  (a) $t=10tj$, (b), $t=20tj$, (c), $t=30tj$, (d) $t=40tj$}
    \label{fig:SUPPLEMENTARY_OH}
\end{figure}

\begin{figure}
    \centering
    \includegraphics[width=0.2\textwidth]{Figures/bar_u_05000.eps}
    \includegraphics[width=0.2\textwidth]{Figures/bar_u_10000.eps}
    \includegraphics[width=0.2\textwidth]{Figures/bar_u_15000.eps}
    \includegraphics[width=0.2\textwidth]{Figures/bar_u_20000.eps}
    \caption{Reynolds-averaged and RMS values of velocity component $u$. Lines and symbols denote LES and  DNS results, respectively.}
    \label{fig:SUPPLEMENTARY_RU}
\end{figure}

\begin{figure}
    \centering
    \includegraphics[width=0.2\textwidth]{Figures/bar_v_05000.eps}
    \includegraphics[width=0.2\textwidth]{Figures/bar_v_10000.eps}
    \includegraphics[width=0.2\textwidth]{Figures/bar_v_15000.eps}
    \includegraphics[width=0.2\textwidth]{Figures/bar_v_20000.eps}
    \caption{Reynolds-averaged and RMS values of velocity component $v$. Lines and symbols denote LES and  DNS results, respectively.}
    \label{fig:SUPPLEMENTARY_RV}
\end{figure}

\begin{figure}
    \centering
    \includegraphics[width=0.2\textwidth]{Figures/bar_w_05000.eps}
    \includegraphics[width=0.2\textwidth]{Figures/bar_w_10000.eps}
    \includegraphics[width=0.2\textwidth]{Figures/bar_w_15000.eps}
    \includegraphics[width=0.2\textwidth]{Figures/bar_w_20000.eps}
    \caption{Reynolds-averaged and RMS values of velocity component $w$. Lines and symbols denote LES and  DNS results, respectively.}
    \label{fig:SUPPLEMENTARY_RW}
\end{figure}
